\documentclass[12pt,a4paper]{article}
\usepackage{jheppub}  
\usepackage{amssymb} 
\usepackage{amsmath}
\usepackage{mathtools}
\usepackage{amsfonts}    
\usepackage{dsfont}
\usepackage{pdfpages}
\usepackage{verbatim}
\usepackage{tensor}
\usepackage{mathrsfs}
\usepackage[mathscr]{euscript}
\usepackage{tikz}

\newcommand{\ba}{\begin{align}}

\newcommand{\be}{\begin{equation}}
\newcommand{\ee}{\end{equation}}
\def\bd{\begin{tikzpicture}}
\def\ed{\end{tikzpicture}}
\newcommand{\ket}[1]{\left| #1\right\rangle}

\allowdisplaybreaks[1]

\title{Strings on $\boldsymbol{\text{AdS}_3 \times \text{S}^3 \times \text{S}^3 \times \text{S}^1}$}

\author{Lorenz Eberhardt and Matthias R.\ Gaberdiel} 
\affiliation{Institut f\"ur Theoretische Physik, ETH Zurich, \\
\hspace*{0.3cm}CH-8093 Z\"urich, Switzerland}
\emailAdd{eberhardtl@itp.phys.ethz.ch, gaberdiel@itp.phys.ethz.ch}

\abstract{String theory on ${\rm AdS}_3 \times {\rm S}^3 \times {\rm S}^3 \times {\rm S}^1$ with pure NS-NS flux and minimal flux through one of the two ${\rm S}^3$'s is studied from a world-sheet perspective. It is shown that the spacetime spectrum, as well as the algebra of spectrum generating operators, matches precisely that of the symmetric orbifold of ${\rm S}^3\times \mathrm{S}^1$ in the large $N$ limit. This gives strong support for the proposal that these two descriptions are exactly dual to one another. }

\begin{document}

\maketitle

%make math in all titles bold
\makeatletter
\g@addto@macro\bfseries{\boldmath}
\makeatother
%end code

%%%%%%%%%%%%%%%%%%%%%%%%%%%%%%%%%%%%%%%%%%%%%%%%%%%%%%%%%%%%%%%
\section{Introduction} \label{sec:intro}
%%%%%%%%%%%%%%%%%%%%%%%%%%%%%%%%%%%%%%%%%%%%%%%%%%%%%%%%%%%%%%%

The AdS/CFT correspondence for the case of ${\rm AdS}_3$ is an interesting toy model that has the potential to allow for quantitative tests of the duality even in the stringy regime. Indeed, unlike the higher dimensional cousins, strings on AdS$_3$ with pure NS-NS flux have an exactly solvable worldsheet description in terms of a WZW model based on $\mathfrak{sl}(2,\mathds{R})$ \cite{Maldacena:2000hw,Maldacena:2000kv,Maldacena:2001km}. In a similar vein, the dual CFT is a $2$-dimensional conformal field theory for whose analysis also powerful tools exist. This opens the possibility that one may be able to identify an exact dual pair for which both sides of the duality are exactly solvable. Recently, very good evidence has been presented that this is indeed possible:  it was shown in \cite{Eberhardt:2018ouy}, following \cite{Gaberdiel:2018rqv} (see also \cite{Giribet:2018ada}), that the pure NS-NS background of ${\rm AdS}_3\times {\rm S}^3 \times \mathbb{T}^4$ with minimal flux ($k=1$) through AdS$_3$ (and ${\rm S}^3$) has exactly the same spacetime spectrum as the symmetric orbifold theory of $\mathbb{T}^4$ in the large $N$ limit. It was furthermore shown in \cite{Eberhardt:2019qcl} that also the algebraic structures, i.e.\ the operator algebra of the chiral fields, agree between the two descriptions. This gives very strong credence to the idea that these two descriptions are indeed exactly dual to one another.

Given the success of this correspondence, one may ask whether there are other examples of this kind. In this paper we show that a similar result holds for string theory on ${\rm AdS}_3\times {\rm S}^3 \times {\rm S}^3 \times {\rm S}^1$ with pure NS-NS flux: for the minimal value of the flux through one of the two ${\rm S}^3$, say $k^+=1$, the spacetime spectrum of string theory is equivalent to the symmetric orbifold of ${\cal S}_\kappa$ (in the large $N$ limit), where $\kappa$ is related to the flux through the other ${\rm S}^3$, $\kappa = k^- -1$. (Here $\mathcal{S}_\kappa$ denotes the supersymmetric WZW-model on $\mathrm{S}^3 \times \mathrm{S}^1$.) More specifically, we check this statement on the level of the spectrum (or partition function), and we also confirm that we reproduce the correct spacetime algebra from the worldsheet via a DDF construction following \cite{Giveon:1998ns,Elitzur:1998mm}. In particular, this gives in a sense a direct derivation of the proposed dual CFT of \cite{Eberhardt:2017pty}, see also \cite{deBoer:1999gea,Gukov:2004ym} for earlier attempts.

While the general philosophy of the approach is quite similar to what was done in \cite{Eberhardt:2018ouy,Eberhardt:2019qcl}, there are a number of technical difficulties one has to overcome. First, the relevant hybrid formalism (that we employ to make sense of the theory with $k^+=1$) had not been developed before, and we sketch its derivation. It involves the WZW model based on the superalgebra $\mathfrak{d}(2,1;\alpha)$, whose structure is somewhat more complicated than $\mathfrak{psu}(1,1|2)$ that was relevant for the case of $\mathbb{T}^4$ \cite{Eberhardt:2018ouy}. In particular, we need to work out the fusion rules and the characters from first principles. We also make a guess about the structure of the indecomposable representations (whose structure is again somewhat more complicated than for the case of $\mathfrak{psu}(1,1|2)_1$). Finally, we show that, following \cite{Eberhardt:2019qcl}, the analysis can also be done for $k^\pm >1$, in which case the resulting dual CFT is the symmetric orbifold of large ${\cal N}=4$ Liouville theory. 
\medskip

The paper is organised as follows. We begin by explaining our conventions for the description of string theory on 
${\rm AdS}_3\times {\rm S}^3 \times {\rm S}^3 \times {\rm S}^1$ in the RNS formalism in Section~\ref{sec:RNS}. In Section~\ref{subsec:reformulation} we explain how to rewrite these degrees of freedom in a hybrid-like formalism; the resulting theory then involves the WZW model on the superalgebra $\mathfrak{d}(2,1;\alpha)$, together with a free boson, two pairs of topologically twisted fermions and some ghosts, see eq.~(\ref{wstheory}). In Section~\ref{Sec:4} we then study the representation theory of $\mathfrak{d}(2,1;\alpha)$, and specialise to the case $k^+=1$ in Section~\ref{Sec:kp1}. We explain the structure of the representations, study their fusion rules, %see Section~\ref{sec:fusion}, 
and then make a proposal for the full worldsheet theory. The physical state condition is then studied in Section~\ref{sec:physical states in string theory}; we also explain there how the spacetime BPS states (and in particular the moduli) arise from the worldsheet perspective. In Section~\ref{sec:DDF} we then study the algebraic structure of the spacetime theory from the worldsheet, following \cite{Eberhardt:2019qcl}, and establish the result for general values of the flux. We end in Section~\ref{sec:discussion} with some discussion of our result. There are a large number of appendices to which some of the technical material has been relegated. In particular, we explain in Appendix~\ref{app:Wakimoto} the Wakimoto representation of $\mathfrak{d}(2,1;\alpha)_k$; we derive the characters  of $\mathfrak{d}(2,1;\alpha)$ at level $k^+=1$ as well as their modular properties in Appendix~\ref{app:characters}; we explain the free field realisation of $\mathfrak{d}(2,1;\alpha)$ for $k^+=k^-=1$ in Appendix~\ref{app:kappa=0}; and we study the indecomposable nature of the representations in Appendix~\ref{app:indecomposables}.

\section{The worldsheet theory} \label{sec:RNS}
In the RNS-formalism, the worldsheet theory is described by the WZW-model
\be 
\mathfrak{sl}(2,\mathds{R})_k^{(1)} \oplus \mathfrak{su}(2)_{k^+}^{(1)} \oplus \mathfrak{su}(2)_{k^-}^{(1)} \oplus \mathfrak{u}(1)^{(1)}\ ,
\ee
together with the usual superconformal ghost system. Here, $\mathfrak{g}^{(1)}_k$ denotes the $\mathcal{N}=1$ superconformal affine algebra of $\mathfrak{g}$ at level $k$. Criticality of the background requires the three levels to be related according to
\be 
\frac{1}{k}=\frac{1}{k^+}+\frac{1}{k^-}\ . \label{eq:criticality}
\ee
As is well known, the fermions of this algebra can be decoupled, leading to 
\begin{align}
\mathfrak{sl}(2,\mathds{R})_k^{(1)} \cong \mathfrak{sl}(2,\mathds{R})_{k+2} \oplus \text{3 free fermions}\ , \label{eq:free fermion decoupling sl2R}\\
\mathfrak{su}(2)_{k^\pm}^{(1)} \cong \mathfrak{su}(2)_{k^\pm-2} \oplus \text{3 free fermions}\ .\label{eq:free fermion decoupling su2}
\end{align}
We shall denote the decoupled currents by $\mathscr{J}^a$ and $\mathscr{K}^{(\pm)a}$ with levels $k+2$ and $k^\pm-2$, respectively. Here, $a\in \{+,-,3\}$ is an adjoint index of $\mathfrak{sl}(2,\mathds{R})$ or $\mathfrak{su}(2)$. Similarly, we denote the corresponding fermionic partners as $\psi^a$ and $\chi^{(\pm)a}$. Finally, the free boson of the $\text{S}^1$ will be denoted by $\partial \Phi$ and the corresponding fermion by $\lambda$. The relevant commutation relations are spelled out in Appendix~\ref{app:OPEs}.
The $\mathcal{N}=1$ superconformal currents on the worldsheet are defined by 
\begin{align}
T(z)&=\frac{1}{k}\Big(-\mathscr{J}^3\mathscr{J}^3+\tfrac{1}{2}\big(\mathscr{J}^+\mathscr{J}^-+\mathscr{J}^-\mathscr{J}^+\big)+\psi^3 \partial \psi^3-\tfrac{1}{2}\big(\psi^+\partial \psi^-+\psi^-\partial \psi^+\big)\Big) \nonumber\\
&\qquad+\frac{1}{k^+}\Big(\mathscr{K}^{(+)3}\mathscr{K}^{(+)3}+\tfrac{1}{2}\big(\mathscr{K}^{(+)+}\mathscr{K}^{(+)-}+\mathscr{K}^{(+)-}\mathscr{K}^{(+)+}\big)\nonumber\\
&\qquad\qquad\qquad-\chi^{(+)3}\partial \chi^{(+)3}-\tfrac{1}{2}\big(\chi^{(+)+}\partial\chi^{(+)-}+\chi^{(+)-}\partial \chi^{(+)+}\big)\Big)\nonumber\\
&\qquad+\frac{1}{k^-}\Big(\mathscr{K}^{(-)3}\mathscr{K}^{(-)3}+\tfrac{1}{2}\big(\mathscr{K}^{(-)+}\mathscr{K}^{(-)-}+\mathscr{K}^{(-)-}\mathscr{K}^{(-)+}\big)\Big)\nonumber\\
&\qquad\qquad\qquad-\chi^{(-)3}\partial \chi^{(-)3}-\tfrac{1}{2}\big(\chi^{(-)+}\partial\chi^{(-)-}+\chi^{(-)-}\partial \chi^{(-)+}\big)\Big)
\nonumber\\&\qquad
+\frac{1}{2}(\partial\Phi\partial\Phi)-\frac{1}{2}\big(\lambda\partial\lambda\big)\ , \label{eq:worldsheet Virasoro}\\
G(z)&=-\frac{1}{k}\Big(-\mathscr{J}^3\psi^3+\tfrac{1}{2}\big(\mathscr{J}^+\psi^-+\mathscr{J}^-\psi^+\big)-\tfrac{1}{k}(\psi^3\psi^+\psi^-)\Big) \nonumber\\
&\qquad-\frac{1}{k^+}\Big(\mathscr{K}^{(+)3}\chi^{(+)3}+\tfrac{1}{2}\big(\mathscr{K}^{(+)+}\chi^{(+)-}+\mathscr{K}^{(+)-}\chi^{(+)+}\big)+\tfrac{1}{k^+}(\chi^{(+)3}\chi^{(+)+}\chi^{(+)-})\Big) \nonumber\\
&\qquad-\frac{1}{k^-}\Big(\mathscr{K}^{(-)3}\chi^{(-)3}+\tfrac{1}{2}\big(\mathscr{K}^{(-)+}\chi^{(-)-}+\mathscr{K}^{(-)-}\chi^{(-)+}\big)+\tfrac{1}{k^-}(\chi^{(-)3}\chi^{(-)+}\chi^{(-)-})\Big) 
\nonumber\\ &\qquad
+\frac{1}{2} \big(\partial \Phi \lambda\big)\ . \label{eq:worldsheet supercurrent} 
\end{align}
The $\mathcal{N}=1$ superconformal structure on the worldsheet allows us to define the BRST charge as 
\be 
Q_\text{BRST}=\oint \mathrm{d}z\ \Big(c\big(T+\tfrac{1}{2}T_\text{gh}\big)+\gamma\big(G+\tfrac{1}{2}G_\text{gh}\big)\Big)\ . \label{eq:string BRST operator}
\ee
Here, $T^\text{gh}$ and $G^\text{gh}$ are the $\mathcal{N}=1$ generators of the superconformal ghost system; this consists of a $bc$ system with $\lambda=2$ and a $\beta\gamma$ system with $\lambda=\frac{3}{2}$,  whose OPE's we take to be (see also \cite{Eberhardt:2019qcl})
\be 
b(z)c(w) \sim \frac{1}{z-w}\ , \qquad \beta(z)\gamma(w) \sim -\frac{1}{z-w}\ .
\ee
In these conventions, the $\mathcal{N}=1$ superconformal algebra of the ghost system is then
\begin{align}
T_\text{gh}(z)&=-2b (\partial c) - (\partial b) c-\tfrac{3}{2}\hat{\beta} (\partial \hat{\gamma}) -\tfrac{1}{2} (\partial \hat{\beta}) \hat{\gamma}\ , \\
G_\text{gh}(z)&=(\partial \hat{\beta}) c+\tfrac{3}{2}\hat{\beta} (\partial c) -\tfrac{1}{2} b \hat{\gamma}\ ,
\end{align}
which realises the $\mathcal{N}=1$ superconformal algebra with central charge $c=-15$.

\subsection{Bosonisation}
In order to relate this description to the hybrid formalism it is convenient to bosonise the fermions as 
\begin{subequations}
\begin{align}
\partial H_1(z)&=\frac{1}{k} (\psi^+\psi^-)(z)\ ,  \qquad  & \partial H_2(z)&=\frac{1}{k^+} (\chi^{(+)+}\chi^{(+)-})(z)\ ,   \\
\partial H_3(z)&=\frac{2}{\sqrt{kk^+}}(\psi^3\chi^{(+)3})(z)\ , & \partial H_4(z)&=\frac{1}{k^-}(\chi^{(-)+}\chi^{(-)-})(z)\ , &  \\
\partial H_5(z)&=i \sqrt{\frac{2}{k^-}} (\lambda \chi^{(-)3})(z)\ . &
\end{align}
\end{subequations}
This bosonisation scheme reduces to that of \cite{Eberhardt:2019qcl} in the limit  $\gamma \to 1$, in which the geometry degenerates to $\mathrm{AdS}_3 \times \mathrm{S}^3 \times\mathbb{T}^4$. The bosons are normalised as
\be 
\partial H_i(z) \partial H_j(w) \sim \frac{\delta_{ij}}{(z-w)^2}\ .
\ee
We also choose the same bosonisation of the superconformal ghost system (the $\beta\gamma$ system) as there, i.e.\ we write 
\be \label{superconf}
\beta(z)=\mathrm{e}^{-\phi(z)+\chi(z)} \partial \chi(z)\ , \qquad \gamma=\mathrm{e}^{\phi(z)-\chi(z)}\ ,
\ee
where the two bosons $\phi(z)$ and $\chi(z)$ have background charge $Q_\phi=2$ and $Q_\chi=-1$, respectively,  and OPEs 
\be 
\phi(z) \phi(w) \sim -\log (z-w)\ , \qquad \chi(z)\chi(w) \sim \log(z-w)\ .
\ee
The energy-momentum tensor for the free-field representation then takes the form
\begin{align}
T&=T^\phi+T^\chi\ , \\
T^\phi&=-\tfrac{1}{2}(\partial\phi)^2+ \partial^2 \phi\ , \\
T^\chi&=\tfrac{1}{2}(\partial\chi)^2+\tfrac{1}{2} \partial^2 \chi\ .
\end{align}
Finally, the picture charge is defined as
\be 
Q_\text{pic}=\oint \mathrm{d}z\ \big(\partial \chi-\partial \phi\big)\ .
\ee

\section{The hybrid formalism} \label{subsec:reformulation}

Next we want to rewrite these degrees of freedom in a way that makes spacetime supersymmetry manifest. This can be done by passing to a $\mathfrak{d}(2,1;\alpha)_k$ WZW-model, thereby leading to the natural analogue of the `hybrid formalism' for this background; as far as we are aware, the hybrid formalism for ${\rm AdS}_3 \times {\rm S}^3 \times {\rm S}^3 \times {\rm S}^1$ has not been developed before. 

\subsection{Defining free field variables}
We start by defining the vertex operators, cf.\ \cite{Eberhardt:2019qcl}
\begin{subequations}
\begin{align} 
p^{\alpha\beta}&= \mathrm{e}^{\frac{1}{2}(\alpha H_1+\beta H_2+\alpha\beta H_3+H_4+H_5-\phi)}\ , \label{eq:definition p}\\
\theta^{\alpha\beta}&=\mathrm{e}^{\frac{1}{2}(\alpha H_1+\beta H_2-\alpha\beta H_3-H_4-H_5+\phi)}\ , \label{eq:definition theta}
\end{align}
\end{subequations}
which obey the free field OPEs
\be \label{toptwist}
p^{\alpha\beta}(z) \theta^{\gamma\delta}(w) \sim \frac{\epsilon^{\alpha\gamma} \epsilon^{\beta\delta}}{z-w}\ .
\ee
We have suppressed the cocycle factors in the expressions. These fields have conformal weight $1$ and $0$, and picture numbers $(-\tfrac{1}{2})$ and $(\tfrac{1}{2})$, respectively. The indices $\alpha$, $\beta \in \{+,-\}$ are spinor indices of $\mathfrak{sl}(2,\mathds{R})_k \oplus \mathfrak{su}(2)_{k^+}$. Note that we have explicitly broken the second $\mathfrak{su}(2)$ symmetry: $p^{\alpha\beta}$ carries charge $+\tfrac{1}{2}$ under $\mathfrak{su}(2)_{k^-}$, while that of $\theta^{\alpha\beta}$ is $-\tfrac{1}{2}$.  

In the case of $\mathrm{AdS}_3 \times \mathrm{S}^3 \times \mathbb{T}^4$, one can construct out of these fields the affine algebra $\mathfrak{psu}(1,1|2)_k$. Analogously, as we shall now explain, we can define a $\mathfrak{d}(2,1;\alpha)_k$ affine algebra in our case (and it will be part of the hybrid formulation). To start with, we define
\be 
S^{\alpha\beta +}=p^{\alpha\beta}-\frac{k^+}{2(k^++k^-)} (J^{(-)+} \theta^{\alpha\beta})\ ,
\ee
which define half of the supercurrents in $\mathfrak{d}(2,1;\alpha)_k$. They are also part of the Wakimoto construction of $\mathfrak{d}(2,1;\alpha)_k$ that is described in detail in Appendix~\ref{app:Wakimoto}, see eq.~(\ref{S-}).

\subsection{Remaining fields}

In order to construct the remaining fields of the hybrid formalism (and complete the construction of $\mathfrak{d}(2,1;\alpha)_k$) we now recall that the bosonic generators of $\mathfrak{d}(2,1;\alpha)_k$ form the Lie algebra $\mathfrak{sl}(2,\mathds{R})_{k} \oplus \mathfrak{su}(2)_{k^+} \oplus \mathfrak{su}(2)_{k^-}$.
The original boson $\partial\Phi$ corresponding to $\mathrm{S}^1$ commutes with $\mathfrak{d}(2,1;\alpha)_k$, and can be directly added to the theory. (It is naturally defined in the $(0)$-picture.) This accounts for all bosonic degrees of freedom. Furthermore, we have not used the $bc$ ghosts in our reformulation and they simply continue to be also part of the hybrid description. 

As regards the fermions, we can define four more fermions which commute with the $p^{\alpha\beta}$'s as well as with the $\theta^{\alpha\beta}$'s. As in \cite{Eberhardt:2019qcl}, they are given by
\be 
\mathrm{e}^{H_4-\phi+\chi}\ , \qquad
\mathrm{e}^{H_5-\phi+\chi}\ , \qquad
\mathrm{e}^{-H_4+\phi-\chi}\ , \qquad
\mathrm{e}^{-H_5+\phi-\chi}\ , \label{eq:remaining 4 fermions}
\ee
where we have also made use of the boson $\chi$ that was introduced in the bosonisation of the superconformal ghosts, see eq.~(\ref{superconf}).
%Let us construct the remaining fields of the hybrid formalism. So far, we have constructed $\mathfrak{d}(2,1;\alpha)_k$, which has nine bosonic fields and eight fermionic fields. Thus, we still need to define one more bosonic field and two fermionic fields as well as the ghosts.
%To start, the original boson $\partial\Phi$ describing $\mathrm{S}^1$ still commutes with $\mathfrak{d}(2,1;\alpha)_k$. It is defined in the $(0)$-picture and hence still constitutes a good field of the theory. 
%Let us discuss the fermionic generators. Analogously to the $\mathbb{T}^4$ case, we can define four more fermions which commute with the $p^{\alpha\beta}$'s as well as with the $\theta^{\alpha\beta}$'s. These are given by
%\be 
%\mathrm{e}^{H_4-\phi+\chi}\ , \qquad
%\mathrm{e}^{H_5-\phi+\chi}\ , \qquad
%\mathrm{e}^{-H_4+\phi-\chi}\ , \qquad
%\mathrm{e}^{-H_5+\phi-\chi}\ , \label{eq:remaining 4 fermions}
%\ee
%where we also made use of the remaining boson $\chi$, which was used in the bosonisation of the superconformal ghosts. 
The conformal weights of the first two fermions is one, while that of the last two fermions is zero; thus they define 2 pairs of topologically twisted fermions (i.e.\ two $bc$ systems of conformal weight 1 and 0).
%former two fermions have conformal weight equal to one, while the latter two equal to zero. We refer to such bosons as topologically twisted fermions (which are none other than a $bc$ system of conformal weight 1 and 0). 
Finally, we replace the other boson $\phi$ from the bosonisation of the superconformal ghosts by the combination 
\be 
\rho=2\phi-H_4-H_5-\chi\ ,
\ee
that commutes with all the above fields, and defines the new ghost field of the hybrid formalism. As in \cite{Eberhardt:2019qcl} one then checks that the central charge of all of these fields is equal to zero, i.e.\ that we have accounted for all degrees of freedom. Thus, we have reassembled the RNS degrees of freedom as 
\be\label{wstheory}
\begin{aligned}
\mathfrak{d}(2,1;\alpha)_k  \oplus \mathfrak{u}(1) & \oplus \text{2 pairs of topologically twisted fermions from eq.~(\ref{eq:remaining 4 fermions})} \\
& \oplus \text{$bc$ and $\rho$ ghosts}\ .
\end{aligned}
\ee

While this construction is fairly parallel to the case of $\mathbb{T}^4$, there is one important difference: the 
$\mathfrak{su}(2)_{k^-}$ currents that appear in $\mathfrak{d}(2,1;\alpha)_k$, see eqs.~(\ref{K-+}), (\ref{K-3}) and (\ref{K--}), 
 do \emph{not} correspond to the correct spacetime supersymmetry currents. One can repair this by redefining the generators of $\mathfrak{d}(2,1;\alpha)_k$ as 
\begin{align}
\tilde{K}^{(-)3} & := {K}^{(-)3}+\partial \phi-\partial H_5\ ,\\
\tilde{K}^{(-)-} & := {K}^{(-)-}-2(\partial \phi-\partial H_5) \hat{\gamma}\ , \\
\tilde{S}^{\alpha\beta-} & := {S}^{\alpha\beta-}+\frac{k^+}{k^-+k^-} (\partial \phi-\partial H_5)\theta^{\alpha\beta}\ ,
\end{align}
without changing the commutation relations of $\mathfrak{d}(2,1;\alpha)_k$. Here, $\hat{\gamma}$ is the free field appearing in the Wakimoto representation of $\mathfrak{sl}(2,\mathds{R})_{k^--2}$, see Appendix~\ref{app:Wakimoto} for details. However, 
this redefined $\mathfrak{d}(2,1;\alpha)_k$ algebra does not commute any longer with the remaining free fermions \eqref{eq:remaining 4 fermions}. 

We should mention that we can also express the physical state conditions in terms of these new variables, which entails rewriting the BRST operator \eqref{eq:string BRST operator}. The explicit expressions are quite complicated (as already in the $\mathbb{T}^4$ case \cite{Berkovits:1999im, Gaberdiel:2011vf}) but since we will  not need them for our purposes, we have not written them out explicitly.

\section{Representations of \texorpdfstring{$\boldsymbol{\mathfrak{d}(2,1;\alpha)}$}{d(2,1;alpha)}}\label{Sec:4}

For the following, it is important to understand representations of $\mathfrak{d}(2,1;\alpha)$ in detail. 
The bosonic subalgebra of $\mathfrak{d}(2,1;\alpha)$ is $\mathfrak{sl}(2,\mathds{R}) \oplus \mathfrak{su}(2)\oplus \mathfrak{su}(2)$. While the representations of $\mathfrak{su}(2)$ that appear are the familiar finite-dimensional spin $\ell$ representations, the representations of $\mathfrak{sl}(2,\mathds{R})$ that are relevant are either discrete lowest (or highest) weight representations that we denote by $\mathscr{D}^j_+$ (or $\mathscr{D}^j_-$ in the case of lowest weight); the other class of $\mathfrak{sl}(2,\mathds{R})$  representations that appear are the continuous representations that are neither highest nor lowest weight and that will be denoted by $\mathscr{C}^j_\lambda$. In either case, $j$ determines the value of the quadratic Casimir of $\mathfrak{sl}(2,\mathds{R})$
\be 
\mathcal{C}=-J_0^3J_0^3+\frac{1}{2}\big(J_0^+J_0^-+J_0^-J_0^+\big)\ ,
\ee
as $\mathcal{C}=-j(j-1)$, and in the case of the continuous representations $\lambda \in \mathds{R}/\mathds{Z}$ denotes the fractional part of the $J^3_0$-eigenvalues. Since the Casimir $\mathcal{C}$ is invariant under $j \to 1-j$, we may assume without loss of generality that $\mathrm{Re}(j)\geq 1/2$. More details about our conventions can be found in \cite{Eberhardt:2018ouy, Eberhardt:2019qcl}.

\subsection{Long representations}

Next we want to understand the structure of the representations of $\mathfrak{d}(2,1;\alpha)$. The fermionic generators of $\mathfrak{d}(2,1;\alpha)$ form a Clifford algebra, and the representations of  $\mathfrak{d}(2,1;\alpha)$ are thus generated from an irreducible representation of the bosonic subalgebra $\mathfrak{sl}(2,\mathds{R}) \oplus \mathfrak{su}(2) \oplus \mathfrak{su}(2)$ by the action of these fermionic modes. We shall mainly focus on the case where the representation with respect to $\mathfrak{sl}(2,\mathds{R})$ is a continuous representation,\footnote{The situation for the discrete representations is essentially identical.} and we shall label the representations of $\mathfrak{su}(2)$ by their dimension $\mathbf{m}^\pm$. A generic (long) multiplet decomposes then with respect to the bosonic subalgebra as 
\be 
\begin{tabular}{ccccccccc}
& $(\mathscr{C}^{j-\frac{1}{2}}_{\lambda+\frac{1}{2}},\mathbf{m}^+,\mathbf{m}^-)$ & \\
& $(\mathscr{C}^{j}_{\lambda},\mathbf{m^+\pm 1},\mathbf{m^-\pm 1})$ & \\
$(\mathscr{C}^{j+\frac{1}{2}}_{\lambda+\frac{1}{2}},\mathbf{m^+\pm 2},\mathbf{m^-})$ & $2 \cdot (\mathscr{C}^{j+\frac{1}{2}}_{\lambda+\frac{1}{2}},\mathbf{m^+},\mathbf{m^-})$ & $(\mathscr{C}^{j+\frac{1}{2}}_{\lambda+\frac{1}{2}},\mathbf{m^+},\mathbf{m^-\pm 2})$ \\
& $(\mathscr{C}^{j+1}_{\lambda},\mathbf{m^+\pm 1},\mathbf{m^-\pm 1})$ & \\
& $(\mathscr{C}^{j+\frac{3}{2}}_{\lambda+\frac{1}{2}},\mathbf{m^+},\mathbf{m^-})$ &
\end{tabular} \ . \label{eq:long representation}
\ee
%which corresponds to the Clifford module generated by the free action of half of the supercharges. The choice of $j$ is convenient. 
For small $\mathbf{m}^\pm$, additional shortenings occur; for example if $\mathbf{m}^+=1$ --- this case will be important below --- the representation shortens to 
\be 
\begin{tabular}{ccccccccc}
& $(\mathscr{C}^{j-\frac{1}{2}}_{\lambda+\frac{1}{2}},\mathbf{1},\mathbf{m^-})$ & \\
& $(\mathscr{C}^{j}_{\lambda},\mathbf{2},\mathbf{m^-\pm 1})$ & \\
$(\mathscr{C}^{j+\frac{1}{2}}_{\lambda+\frac{1}{2}},\mathbf{3},\mathbf{m^-})$ & $(\mathscr{C}^{j+\frac{1}{2}}_{\lambda+\frac{1}{2}},\mathbf{1},\mathbf{m^-})$ & $(\mathscr{C}^{j+\frac{1}{2}}_{\lambda+\frac{1}{2}},\mathbf{1},\mathbf{m^-\pm 2})$ \\
& $(\mathscr{C}^{j+1}_{\lambda},\mathbf{2},\mathbf{m^-\pm 1})$ & \\
& $(\mathscr{C}^{j+\frac{3}{2}}_{\lambda+\frac{1}{2}},\mathbf{1},\mathbf{m^-})$ &
\end{tabular} \ . \label{eq:long representation mp=1}
\ee
However, even in this case, the multiplet still contains a representation with $\mathbf{m^+} \ge \mathbf{3}$. 

In the following we shall mainly be interested in the $\mathfrak{d}(2,1;\alpha)$ representations that can appear as (Virasoro) highest weights of an affine $\mathfrak{d}(2,1;\alpha)$ representation at $k^+=1$. Then, because of the usual representation theory of $\mathfrak{su}(2)_1$, see also the analogous discussion in \cite{Eberhardt:2018ouy}, only $\mathfrak{d}(2,1;\alpha)$ representations with $\mathbf{m^+} \leq \mathbf{2}$ are allowed. The above argument therefore shows that only `short' representations of $\mathfrak{d}(2,1;\alpha)$ are then possible. 

% representation, because of the appearance of the representation $(\mathscr{C}^{j+1}_\lambda,\mathbf{m^++2},\mathbf{m^-})$ in the third line. We can already anticipate that any long multiplet of this kind \textit{cannot} appear as a ground state representation of the $\mathfrak{d}(2,1;\alpha)_k$ WZW model at $k^+=1$, since this would in particular imply that that $\mathfrak{su}(2)_{k^+}=\mathfrak{su}(2)_1$ would have at least a spin 1 representation as ground state representation. However, it is well-known that a $\mathfrak{su}(2)_k$ WZW-model has representation whose spin is bounded by $\tfrac{k}{2}$, hence long representations cannot appear.
%The way out of this impasse is to consider short representations.

\subsection{Short representations}

We have analysed systematically the (short) representations with  $\mathbf{m^+} \leq \mathbf{2}$. The analysis is fairly parallel to the case discussed in detail in \cite{Eberhardt:2018ouy}, and up to relabelling, the only  representations with this property have the form 
%
%Here, we systematically determine all possible short representations of $\mathfrak{d}(2,1;\alpha)$ which do not involve a $\mathbf{m^+} \ge \mathbf{3}$ representation. These coming from suitable shortenings of the long representations with $\mathbf{m^+}=\mathbf{1}$ or $\mathbf{m^+}=\mathbf{2}$ as their highest weight state.
%
%We start now with the long representation \eqref{eq:long representation mp=1} and ask for a possible shortenings. We want to demand that the representation $(\mathscr{C}^{j+\frac{1}{2}}_{\lambda+\frac{1}{2}},\mathbf{3},\mathbf{m^-})$ is absent. This will be only the case of also already one of the two representations in the second line are absent. This then yields the following two potential short multiplets:
%\be 
%\begin{tabular}{ccc}
%$(\mathscr{C}^{j-\frac{1}{2}}_{\lambda+\frac{1}{2}},\mathbf{1},\mathbf{m})$ & & \\
%& $(\mathscr{C}^{j}_{\lambda},\mathbf{2},\mathbf{m\pm 1})$ & \\
%& & $(\mathscr{C}^{j+\frac{1}{2}}_{\lambda+\frac{1}{2}},\mathbf{1},\mathbf{m\pm 2})$ \\
%\end{tabular} \ .
%\ee
%The two choices are actually equivalent, since we can turn the picture around and redefine $j$ and $\mathbf{m}$.\footnote{Recall for this that $\mathscr{C}^j_\lambda$ and $\mathscr{C}^{1-j}_\lambda$ describe equivalent $\mathfrak{sl}(2,\mathds{R})$ representations.} Similarly, also starting with the $\mathbf{m^+}=\mathbf{2}$ long multiplet and analysing shortening conditions does not yields new possibilities. We are thus led to the conclusion that there is at most one short multiplet, which we can conveniently parametrise as 
\be 
\begin{tabular}{ccc}
& $(\mathscr{C}^{j}_\lambda,\mathbf{2},\mathbf{m})$ & \\
$(\mathscr{C}^{j-\frac{1}{2}}_{\lambda+\frac{1}{2}},\mathbf{1},\mathbf{m+1})$ & & $(\mathscr{C}^{j+\frac{1}{2}}_{\lambda+\frac{1}{2}},\mathbf{1},\mathbf{m-1})$ \ , 
\end{tabular} \label{eq:d21alpha short multiplet}
\ee
where $j$ (with $\mathfrak{Re}(j)\geq 1/2$) will be determined momentarily. Note that if the multiplet was a discrete multiplet (i.e.\ if the continuous representations $\mathscr{C}^j_\lambda$ were replaced by the discrete representations $\mathscr{D}^j_+$), we could easily determine the relevant shortening condition: it requires that the lowest weight state, i.e.\ the state in $(\mathscr{D}^{j-1/2}_+,\mathbf{1},\mathbf{m+1})$ is BPS, and hence saturates the familiar BPS bound \cite{Gunaydin:1988re, Gukov:2004ym}, which in the above parametrisation (see also \cite{Gaberdiel:2013vva, Eberhardt:2017fsi}) takes the form 
\be 
j=(1-\gamma)\big(\ell+\tfrac{1}{2}\big)+\tfrac{1}{2}\ . \label{eq:j shortening condition}
\ee
Here we have defined $\gamma = \frac{\alpha}{1+\alpha}$, and expressed the $\mathfrak{su}(2)$-representation via its spin, $\mathbf{m}=2\ell+1$. 
%Note that since $\mathscr{C}^j_\lambda \cong \mathscr{C}^{1-j}_\lambda$, we can also reinterpret the result as follows:
%\be 
%\begin{tabular}{ccc}
%& $(\mathscr{C}^{j}_\lambda,\mathbf{2},\mathbf{m})$ & \\
%$(\mathscr{C}^{j+\frac{1}{2}}_{\lambda+\frac{1}{2}},\mathbf{1},\mathbf{m+1})$ & & $(\mathscr{C}^{j-\frac{1}{2}}_{\lambda+\frac{1}{2}},\mathbf{1},\mathbf{m-1})$
%\end{tabular} \label{eq:d21alpha short multiplet reinterpreted}
%\ee
%where now
%\be 
%j=-(1-\gamma)\big(\ell+\tfrac{1}{2}\big)+\tfrac{1}{2}\ .
%\ee
%To fix this ambiguity, we require $j >\tfrac{1}{2}$.

The same result is also true in the continuous case, as we shall now explain. One way of seeing this is to decompose the $\mathfrak{d}(2,1;\alpha)$-Casimir into its bosonic and its fermionic pieces
\begin{align}
\mathcal{C}^{\mathfrak{d}(2,1;\alpha)}&=\mathcal{C}_\text{bos}^{\mathfrak{d}(2,1;\alpha)}+\mathcal{C}_\text{ferm}^{\mathfrak{d}(2,1;\alpha)}\ , \\
\mathcal{C}_\text{bos}^{\mathfrak{d}(2,1;\alpha)}&=\mathcal{C}^{\mathfrak{sl}(2,\mathds{R})}+\gamma\mathcal{C}^{\mathfrak{su}(2)_+}+(1-\gamma)\mathcal{C}^{\mathfrak{su}(2)_-}\ , \\
\mathcal{C}_\text{ferm}^{\mathfrak{d}(2,1;\alpha)}&=-\frac{1}{2} \epsilon_{\alpha\mu} \epsilon_{\beta\nu}\epsilon_{\gamma \rho} S_0^{\alpha\beta\gamma} S_0^{\mu\nu\rho}\ .
\end{align}
The fermionic Casimir can be computed explicitly on the representations of the bosonic subalgebra with the result\footnote{One can also work this out on the third representation $(\mathscr{C}^{j+\frac{1}{2}}_\lambda,\mathbf{1},\mathbf{m-1})$, but the analysis is more complicated in that case.}
\be 
\left. \mathcal{C}_\text{ferm}^{\mathfrak{d}(2,1;\alpha)}\right|_{\big(\mathscr{C}^j_\lambda,\mathbf{2},\mathbf{m}\big)}=-\gamma\ , \qquad 
\left. \mathcal{C}_\text{ferm}^{\mathfrak{d}(2,1;\alpha)}\right|_{\big(\mathscr{C}^{j-\frac{1}{2}}_{\lambda+\frac{1}{2}},\mathbf{1},\mathbf{m+1}\big)}=-(1-\gamma)(2\ell+1)\ .
\ee
On these two representations of the bosonic subalgebra, the full $\mathfrak{d}(2,1;\alpha)$ Casimir therefore takes the values 
\begin{align}
\left.\mathcal{C}^{\mathfrak{d}(2,1;\alpha)}\right|_{(\mathscr{C}^j_\lambda,\mathbf{2},\mathbf{m})}&= -j(j-1)+\frac{3\gamma}{4}+(1-\gamma)\ell(\ell+1)-\gamma\ , \\
\left.\mathcal{C}^{\mathfrak{d}(2,1;\alpha)}\right|_{(\mathscr{C}^{j-\frac{1}{2}}_\lambda,\mathbf{1},\mathbf{m+1})}&= -\big(j-\tfrac{1}{2}\big)\big(j-\tfrac{3}{2}\big)+(1-\gamma)\big(\ell+\tfrac{1}{2}\big)\big(\ell+\tfrac{3}{2}\big)\nonumber\\
&\qquad\qquad\qquad-(1-\gamma)(2\ell+1)\ .
\end{align}
Demanding the two expressions to be equal reproduces then \eqref{eq:j shortening condition}, in which case the Casimir simplifies to 
\be 
\mathcal{C}^{\mathfrak{d}(2,1;\alpha)}=\gamma(1-\gamma) \big(\ell+\tfrac{1}{2}\big)^2\ .
\ee
%Thus, we have defined the short representations on the worldsheet, which depend on $\lambda$ and $\ell\in \tfrac{1}{2}\mathds{Z}_{\ge 0}$. 
We should mention that for the minimal value of $\ell=0$, the third term in (\ref{eq:d21alpha short multiplet}) is absent, i.e.\ the representation is ultrashort, and takes the form 
\be \label{ultrashort}
(\mathscr{C}^j_\lambda,\mathbf{2},\mathbf{1})\oplus (\mathscr{C}^{j-\frac{1}{2}}_{\lambda+\frac{1}{2}},\mathbf{1},\mathbf{2})
\ee
with $j=1-\tfrac{\gamma}{2}$.

In the limit $\gamma \to 1$, $\mathfrak{d}(2,1;\alpha)$ degenerates to $\mathfrak{psu}(1,1|2)$ and the second $\mathfrak{su}(2)$ becomes an outer automorphism. Then the short representation reduces as 
\be 
\setlength{\tabcolsep}{-5pt}
\begin{tabular}{ccc}
& $(\mathscr{C}^{j}_\lambda,\mathbf{2},\mathbf{m})$ & \\
$(\mathscr{C}^{j+\frac{1}{2}}_{\lambda+\frac{1}{2}},\mathbf{1},\mathbf{m+1})$ & & $(\mathscr{C}^{j-\frac{1}{2}}_{\lambda+\frac{1}{2}},\mathbf{1},\mathbf{m-1})$
\end{tabular} 
\setlength{\tabcolsep}{6pt}
\ \ \overset{\gamma \to 1}{\longrightarrow} m \times \left((\mathscr{C}^{\frac{1}{2}}_\lambda,\mathbf{2}) \oplus 2 \cdot (\mathscr{C}^0_{\lambda+\frac{1}{2}},\mathbf{1})\right)\ .
\ee
The expression in the bracket on the right hand side is the short representation of $\mathfrak{psu}(1,1|2)$ that was discussed in \cite{Eberhardt:2018ouy}. Similarly, the shortening condition (\ref{eq:j shortening condition}) just becomes $j=\frac{1}{2}$ in this limit, again in agreement with  \cite{Eberhardt:2018ouy}.

The continuous representations $\mathscr{C}^j_\lambda$ of $\mathfrak{sl}(2,\mathds{R})$ are indecomposable for $j = \lambda$ and $1-j=\lambda$, i.e.\ for $\lambda = \pm j\, (\text{mod}\, \mathds{Z})$. The same property carries, of course, also through to the $\mathfrak{d}(2,1;\alpha)$ representations. For the above short representations this becomes
\be 
\lambda=\pm j=\pm \lambda_\ell\ , \qquad \lambda_\ell=(1-\gamma)(\ell+\tfrac{1}{2})+\tfrac{1}{2}\ . \label{eq:lambdaell}
\ee
In each case, there is a discrete subrepresentation, and we can define the continuous representation such that the discrete subrepresentation is either highest of lowest weight. Altogether there are therefore four cases.

In addition there is another degeneration that occurs if the Casimir of a $\mathfrak{sl}(2,\mathds{R})$ representation vanishes, since it contains then the trivial representation as a subrepresentation. There are two ways in which this may occur in \eqref{eq:d21alpha short multiplet}. First, we can formally set $\mathbf{m}=\mathbf{0}$, in which case we just keep the left-hand-factor
\be
(\mathscr{C}^{0}_{\lambda+\frac{1}{2}},\mathbf{1},\mathbf{1})  \ , 
\ee
where we have used that $\mathbf{m}=\mathbf{0}$ leads to $\ell=-\frac{1}{2}$ and hence to $j=\frac{1}{2}$ in (\ref{eq:j shortening condition}). This then contains the trivial representation for $\lambda=\frac{1}{2}$. 
The other case arises for 
\be 
\mathbf{m}=\frac{1}{1-\gamma} \in \mathds{Z}_{\ge 0}\ ,\label{eq:short module existence condition}
\ee
since then (\ref{eq:j shortening condition}) leads to $j=1$. (This is obviously only possible provided that $(1-\gamma)^{-1}$ is an integer.) 
In this case the middle representation in \eqref{eq:d21alpha short multiplet} can contain the trivial $\mathfrak{sl}(2,\mathds{R})$ representation, and then the two other terms will be absent. Thus, we conclude that also
\be 
\left(\mathscr{C}_\lambda^1,\mathbf{2},\mathbf{m}=\frac{1}{1-\gamma}\right) \label{eq:Lp ground state representation}
\ee
is a consistent multiplet. Note that there is no analogue of this in the limit $\gamma\rightarrow 1$.

%Interestingly, this multiplet \textit{does not} exist in the limit $\gamma \to 1$, where we recover $\mathfrak{psu}(1,1|2)$. 

\section{The \texorpdfstring{$\boldsymbol{\mathfrak{d}(2,1;\alpha)}$}{d(2,1;alpha)} WZW-model at \texorpdfstring{$\boldsymbol{k^+=1}$}{k+=1}}\label{Sec:kp1}

In the following we shall concentrate on the WZW-model based on $\mathfrak{d}(2,1;\alpha)$ with $k^+=1$. We shall set $k^-=\kappa+1$ with $\kappa \in \mathds{Z}_{\ge 0}$, as this will be convenient below. With this choice of parameters, we then have
\be \label{para}
k=\gamma=\frac{\kappa+1}{\kappa+2}\ \qquad \hbox{so that} \qquad (1-\gamma) = \frac{1}{\kappa+2} \ ,
\ee
see \eqref{eq:criticality}.

\subsection{The affine short representations}

As we have explained in the previous section, the only allowed ground state representations are the short 
continuous representations of $\mathfrak{d}(2,1;\alpha)$ of eq.~(\ref{eq:d21alpha short multiplet}). 
We will denote the resulting affine representations by $\mathscr{F}^\ell_\lambda$, where $\ell=2\mathbf{m}+1$. Since the $\mathfrak{su}(2)_{k^-}$ ground state representations must have spin less or equal to $\tfrac{1}{2}k^-$, $\ell$  is allowed to take only the values $\ell \in \{0,\tfrac{1}{2},\dots,\tfrac{\kappa}{2}\}$ --- note that the multiplet  (\ref{eq:d21alpha short multiplet}) also contains a representation with ${\bf m + 1}$. This bound was noted in the discrete case already in \cite{Gunaydin:1988re}. 

As we have explained above, the ground state representations of $\mathscr{F}^\ell_\lambda$ become indecomposable for $\lambda=\pm\lambda_\ell$ \eqref{eq:lambdaell} and the same is, of course, also true for the affine representations. We shall denote the corresponding discrete subrepresentations by 
\be
\mathscr{G}_{>,\pm}^\ell \subset  \mathscr{F}^\ell_{\lambda_\ell} \ , \qquad 
\mathscr{G}_{<,\pm}^\ell \subset  \mathscr{F}^\ell_{-\lambda_\ell} \ , 
\ee
where $\pm$ refers to whether the representation is lowest weight $(+)$, i.e.\ runs to the right, or whether it is highest weight $(-)$, i.e.\ runs to the left. Note that for $\gamma\neq 0,1$, the parameter $\lambda_\ell \not\in \tfrac{1}{2}\mathds{Z}$, and hence $\lambda_\ell$ and $-\lambda_\ell$ never differ by an integer (and hence never define the same representation).

%$\mathscr{F}^\ell_{\lambda_\ell}$ contains the discrete version of the module, which we shall denote by $\mathscr{G}_{>,+}^\ell$, $\mathscr{F}^\ell_{-\lambda_\ell}$ contains the lowest weight discrete version $\mathscr{G}_{<,+}^\ell$. Here, the + means that this is a lowest weight representation, i.e.~runs to the right. The signs $>$ and $<$ indicate whether they are submodules of $\mathscr{F}^\ell_{\lambda_\ell}$ or $\mathscr{F}^\ell_{-\lambda_\ell}$. Furthermore, there are the corresponding highest weight representations, which we denote by $\mathscr{G}_{>,-}^\ell$ and $\mathscr{G}_{<,-}^\ell$. Note that it is impossible that $\lambda_\ell \in \tfrac{1}{2}\mathds{Z}$, thus the cases cannot occur simultaneously. 

The other representations that will be relevant for us is the vacuum representation $\mathscr{L}$ of $\mathfrak{d}(2,1;\alpha)_\kappa$ --- this is the affine representation based on the trivial representation of 
$\mathfrak{d}(2,1;\alpha)$ --- as well as the representation $\mathscr{L}'$, whose ground state representation is 
\eqref{eq:Lp ground state representation}. Note that, because of (\ref{para}), $(1-\gamma)^{-1} = \kappa+2$ is an integer, and hence this representation exists for all $\kappa$. As we shall see below, see eq.~\eqref{eq:spectral flow identification c},  $\mathscr{L}'$ arises naturally by applying the joint spectral flow in the two affine $\mathfrak{su}(2)$'s to the vacuum representation.

Thus, up to now, we have the following irreducible modules of $\mathfrak{d}(2,1;\alpha)_k$ for $k^+=1$
\begin{align}
\mathscr{F}^\ell_\lambda\ , \quad\mathscr{G}^\ell_{<,\pm}\ , \quad\mathscr{G}^\ell_{>,\pm}\ , \quad
\mathscr{L}\ ,  \quad \mathscr{L}'\ , \label{eq:representations without spectral flow}
\end{align}
where $\ell$ runs over $\ell \in \{0,\tfrac{1}{2},\dots,\tfrac{\kappa}{2}\}$ and $\lambda \in \mathds{R}/\mathds{Z}$ with $\lambda\ne \pm \lambda_\ell$.

\subsection{Spectral flow} \label{subsec:spectral flow}

For the following it will be important that $\mathfrak{d}(2,1;\alpha)_k$ possesses a spectral flow automorphism $\sigma$. On the bosonic subalgebra $\mathfrak{sl}(2,\mathds{R})_k \oplus \mathfrak{su}(2)_{k^+} \oplus \mathfrak{su}(2)_{k^-}$, we define it to act by a simultaneous spectral flow on $\mathfrak{sl}(2,\mathds{R})_k \oplus \mathfrak{su}(2)_{k^+}$, 
\begin{subequations}
\begin{align}
\sigma^w (J^3_m)&=J^3_m+\tfrac{kw}{2}\delta_{m,0}\ , \label{eq:spectral flow a}\\
\sigma^w(J^\pm_m)&=J^\pm_{m\mp w}\ , \label{eq:spectral flow b}\\
\sigma^w(K^{(+)3}_m)&=K^{(+)3}_m+\tfrac{k^+ w}{2}\delta_{m,0}\ , \label{eq:spectral flow c}\\
\sigma^w(K^{(+)\pm}_m)&=K^{(+)\pm}_{m \pm w}\ , \label{eq:spectral flow d}\\
\sigma^w(K^{(-)a}_m)&=K^{(-)a}_m\ , \label{eq:spectral flow e}\\
\sigma^w(S^{\alpha\beta\gamma}_m)&=S^{\alpha\beta\gamma}_{m+\frac{1}{2}w(\beta-\alpha)}\ .\label{eq:spectral flow f}
\end{align}
\end{subequations}
In particular, this spectral flow keeps the supercharges integer moded. As we will see below, see also \cite{Maldacena:2000hw}, these spectrally flowed representations will have to be included in order to get a well-defined worldsheet theory; we therefore need to extend \eqref{eq:representations without spectral flow} by their spectrally flowed images. 

We should mention that we have made an artificial choice in flowing in $\mathfrak{su}(2)_{k^+}$, and not in $\mathfrak{su}(2)_{k^-}$. This is reflected by the existence of another spectral flow $\rho$, which flows simultaneously in the two $\mathfrak{su}(2)$'s. This spectral flow does not generate any new representations, and it satisfies $\rho^2=1$.\footnote{This is to say, $\rho^2$ is an inner automorphism that maps each representation to itself. However, $\rho^2$ does not act trivially on the individual states.}  

Since spectral flow maps representations to representations, there are in fact a number of identifications. In particular, we have
\begin{subequations}
\begin{align} 
\rho(\mathscr{F}^\ell_\lambda)&=\mathscr{F}^{\frac{\kappa}{2}-\ell}_{\lambda+\frac{1}{2}}\ , &
\rho(\mathscr{G}_{>,\pm}^\ell)&=\mathscr{G}_{<,\pm}^{\frac{\kappa}{2}-\ell}\ , &
\rho(\mathscr{L})&=\mathscr{L}'\ , \label{eq:spectral flow identification a}\\
\sigma(\mathscr{L}) &\cong \mathscr{G}_{<,+}^0\ ,& \sigma^{-1}(\mathscr{L}) &\cong \mathscr{G}_{>,-}^0\ , \label{eq:spectral flow identification b} \\
\sigma(\mathscr{L}') &\cong \mathscr{G}_{>,+}^{\frac{\kappa}{2}}\ ,& \sigma^{-1}(\mathscr{L}') &\cong \mathscr{G}_{<,-}^{\frac{\kappa}{2}}\ , \label{eq:spectral flow identification c} \\
\sigma(\mathscr{G}^{\ell+\frac{1}{2}}_{>,-})&\cong\mathscr{G}^{\ell}_{>,+}\ , & \sigma^{-1}(\mathscr{G}^{\ell+\frac{1}{2}}_{<,+})&\cong\mathscr{G}^\ell_{<,-}\ . \label{eq:spectral flow identification d}
\end{align}
\end{subequations}
Finally, as in the case studied in \cite{Eberhardt:2018ouy}, the CFT is actually logarithmic, and one also needs to consider indecomposable representations. We have already seen that for $\lambda=\pm \lambda_\ell$ the module $\mathscr{F}^\ell_\lambda$ becomes indecomposable and contains a discrete subrepresentation. As it turns out --- this is typical for logarithmic CFTs --- $\mathscr{F}^\ell_\lambda$ itself does not appear in the spectrum of the theory, but it is instead part of an even larger indecomposable module. While these indecomposable modules lead to many technical complications, most of our results are largely unaffected by this subtlety, see also \cite{Troost:2011fd,Gaberdiel:2011vf}. We have therefore relegated the analysis of these indecomposable representations to Appendix~\ref{app:indecomposables}.

%While the emergence of the indecomposable representations leads to many technical complications, it will turn out that the resulting physical spectrum is largely unaffected by this subtlety, see also \cite{Troost:2011fd,Gaberdiel:2011vf}. Furthermore, for many considerations (in particular, for the analysis of the partition function) we may work with the so-called 
%\textit{Grothendieck ring} of modules, where modules related by short exact sequences are identified, i.e.\ 
%\be \label{Grothendieck}
%\mathscr{C} \sim \mathscr{A} \oplus \mathscr{B} \quad \Longleftrightarrow \quad 0 \longrightarrow \mathscr{A} \longrightarrow \mathscr{C} \longrightarrow \mathscr{B} \longrightarrow 0\ .
%\ee
%This equivalence relation therefore forgets the indecomposablity of modules. 

\subsection{The fusion rules}\label{sec:fusion}

Next we want to describe the fusion rules of the model. For the case of $\mathfrak{psu}(1,1|2)_1$ that was discussed in \cite{Eberhardt:2018ouy}, there exists a free field realisation from which the fusion rules can be deduced. We are not aware of such a free-field representation in the present case, except for $\kappa=0$; this free-field realisation for $\kappa=0$ is discussed in Appendix~\ref{app:kappa=0}.

We therefore have to resort to other methods. In particular, we can use a continuum version of the Verlinde formula to determine the typical fusion rules, i.e.\ those that do not involve indecomposable representations.\footnote{For the case of $\mathfrak{psu}(1,1|2)_1$, this was also done in \cite{Eberhardt:2018ouy}.}
%
%We are considering the set of modules 
%\be 
%\sigma^w(\mathscr{F}^\ell_\lambda)\ , \qquad w \in \mathds{Z}, \ \ell \in \{0,\tfrac{1}{2},\dots,\tfrac{\kappa}{2}\}, \ \lambda \in \mathds{R}/\mathds{Z}\ .
%\ee
%We have seen that the modules can become indecomposable for $\lambda=\pm \lambda_\ell$. This constitutes a measure zero set in the set of all representations. We now analyse the fusion rules in the typical case (i.e.\ when $\lambda=\pm \lambda_\ell$ is not satisfied for any of the three modules appearing in the fusion product).
%
 The calculation is somewhat lengthy, see Appendix~\ref{app:characters}, but it leads to the simple result
\begin{align}
\mathscr{F}^{\ell_1}_{\lambda_1} \times \mathscr{F}^{\ell_2}_{\lambda_2}&\cong\bigoplus_{\ell_3=0}^{\frac{\kappa}{2}} N^{\ell_3}_{\ell_1\ell_2}\Big(\sigma\big(\mathscr{F}^{\ell_3}_{\lambda_1+\lambda_2-\frac{\gamma}{2}}\big) \oplus \mathscr{F}^{\ell_3+\frac{1}{2}}_{\lambda_1+\lambda_2+\frac{1}{2}}\oplus \mathscr{F}^{\ell_3-\frac{1}{2}}_{\lambda_1+\lambda_2+\frac{1}{2}}\oplus\sigma^{-1}\big(\mathscr{F}^{\ell_3}_{\lambda_1+\lambda_2+\frac{\gamma}{2}}\big)\Big)\ . \label{fusionrules} 
\end{align}
Here, $N_{\ell_1\ell_2}^{\ell_3}$ are the $\mathfrak{su}(2)_\kappa$ fusion rules, and, by definition, $\mathscr{F}^{-\frac{1}{2}}_\lambda$ and $\mathscr{F}^{\frac{\kappa+1}{2}}_\lambda$ are considered to be zero. Since the Verlinde formula is blind to indecomposability issues, it is conceivable that some modules on the right hand side are actually part of a bigger indecomposable module.
In fact, if $\lambda_1+\lambda_2+\tfrac{1}{2}=\pm\lambda_{\ell_3\pm \frac{1}{2}}$ for some $\ell_3$, then we expect indecomposable modules to appear. While it is difficult to 
derive this for general $\kappa$,  we can use our knowledge from the free-field realisation at $\kappa=0$, see  Appendix~\ref{app:kappa=0}, and from the $\mathfrak{psu}(1,1|2)_1$ analysis (which arises for $\kappa \to \infty$) to make a reasonable guess for the indecomposable structure in general. This is also described in Appendix~\ref{app:indecomposables}. 
%In any case, it is satisfying that the typical fusion rules close which is a good indicator that we have indeed found all $\mathfrak{d}(2,1;\alpha)_k$ modules for $k^+=1$.

As in  \cite{Eberhardt:2018ouy}, the fusion rules are compatible with spectral flow,
\be 
\sigma^{w_1}\big(\mathscr{F}^{\ell_1}_{\lambda_1}\big) \times \sigma^{w_2}\big(\mathscr{F}^{\ell_2}_{\lambda_2}\big) \cong \sigma^{w_1+w_2}\big(\mathscr{F}^{\ell_1}_{\lambda_1} \times\mathscr{F}^{\ell_2}_{\lambda_2}\big)\ ,
\ee
and they reduce to the ones for $\mathfrak{psu}(1,1|2)_1$ in the limit $\kappa \to \infty$. In that limit, 
$\mathfrak{su}(2)_{\kappa+1}$ becomes an outer automorphism, and we therefore get from (\ref{fusionrules})
\begin{align} 
\mathscr{F}^0_{\lambda_1} \times \mathscr{F}^0_{\lambda_2} &\cong \sigma\big(\mathscr{F}^0_{\lambda_1+\lambda_2+\frac{1}{2}}\big)\oplus \mathscr{F}^{\frac{1}{2}}_{\lambda_1+\lambda_2+\frac{1}{2}} \oplus\sigma^{-1}\big(\mathscr{F}^0_{\lambda_1+\lambda_2+\frac{1}{2}}\big) \\
&\cong\sigma\big(\mathscr{F}^0_{\lambda_1+\lambda_2+\frac{1}{2}}\big)\oplus 2\cdot \mathscr{F}^{0}_{\lambda_1+\lambda_2+\frac{1}{2}} \oplus\sigma^{-1}\big(\mathscr{F}^0_{\lambda_1+\lambda_2+\frac{1}{2}}\big)\ ,
\end{align}
where the isomorphism breaks the outer automorphism $\mathfrak{su}(2)$; this then reproduces eq.~(4.17) of 
 \cite{Eberhardt:2018ouy}. As a second cross-check, we notice that they reduce, for $\kappa=0$, to 
\be 
\mathscr{F}^0_{\lambda_1} \times \mathscr{F}^0_{\lambda_2} \cong \sigma\big(\mathscr{F}^0_{\lambda_1+\lambda_2-\frac{1}{4}}\big) \oplus\sigma^{-1}\big(\mathscr{F}^0_{\lambda_1+\lambda_2+\frac{1}{4}}\big)\ ,
\ee
thereby reproducing the special case derived in Appendix~\ref{app:kappa=0} from the free field realisation.

%This shows that we have defined a complete set of modules closing under fusion.

\subsection{The Hilbert space and modular invariance}

With these preparations at hand, we can now write down the complete worldsheet spectrum. It takes the form 
%The simplest modular invariant spectrum is given by the diagonal modular invariant with respect to spectral flow,
\be 
\mathcal{H}=\bigoplus_{w \in \mathds{Z}} \bigoplus_{\ell=0,\, \bar{\ell}=0}^{\frac{\kappa}{2}} M_{\ell\bar{\ell}}\hspace{.3cm}\boldsymbol{\oplus} \hspace{-.576cm} \int\limits_{[0,1)} \mathrm{d}\lambda \ \sigma^w\big(\mathscr{F}^\ell_\lambda\big) \otimes \overline{\sigma^w\big(\mathscr{F}^\ell_\lambda\big)}\ ,\label{eq:Hilbert space Grothendieck}
\ee
where $M_{\ell\bar{\ell}}$ is any $\mathfrak{su}(2)_\kappa$ modular invariant. In Appendix~\ref{app:characters}, we determine the $S$-matrix for the modular transformations of the characters, see eq.~(\ref{eq:S matrix})
\be \label{Smatrix}
S_{(w,\lambda,\ell),(w',\lambda',\ell')}=-i\, \mathrm{sgn}(\mathrm{Re}(\tau)) \, \mathrm{e}^{2\pi i\big(w'\lambda+w\lambda'-\frac{w w'}{2(\kappa+2)}\big)} S^{\mathfrak{su}(2)}_{\ell \ell'}\ ,
\ee 
where $S^{\mathfrak{su}(2)}_{\ell\ell'}$ is the standard modular S-matrix of $\mathfrak{su}(2)_\kappa$. The $S$-matrix in (\ref{Smatrix}) is formally unitary, and hence the spectrum (\ref{eq:Hilbert space Grothendieck}) 
is (formally) modular invariant. This is true for any modular invariant of $\mathfrak{su}(2)_\kappa$, since the $S$-matrix is of tensor product form, i.e.\ the $\ell$-dependence only appears in $\mathfrak{su}(2)_\kappa$ $S$-matrix, which factors out from the rest, see eq.~(\ref{Smatrix}).

In writing down (\ref{eq:Hilbert space Grothendieck}) we have ignored the subtlety that the fusion rules require us to consider also some indecomposable modules. There is a general recipe for how to deal with this issue that was already explained in some detail in \cite{Eberhardt:2018ouy}; we have sketched some aspects of this in 
Appendix~\ref{app:indecomposables}. 
%But since this is a measure zero set, this does not influence modular invariance. 

%Thus, we have shown that the $\mathfrak{d}(2,1;\alpha)_k$ WZW-model at $k^+=1$ can be consistently defined and gives a local CFT. 

In Appendix~\ref{app:characters}, we have also derived the characters of the spectrally flowed representation $\sigma^w\big(\mathscr{F}^\ell_\lambda\big)$, which take the form
\begin{multline} 
\mathrm{ch}\big[\sigma^w\big(\mathscr{F}^\ell_\lambda\big)\big](t,u,v;\tau)=q^{\frac{w^2}{4(\kappa+2)}}x^{\frac{\kappa+1}{2(\kappa+2)}w} y^{\frac{w}{2}}\\
\times\sum_{n \in \mathds{Z}} \mathrm{e}^{2\pi i (\lambda+\frac{1}{2}) n} \, \delta(t-w\tau-n) \, \frac{\vartheta_2\big(\frac{t+u+v}{2};\tau\big)\vartheta_2 \big(\frac{t+u-v}{2};\tau\big)}{\eta(\tau)^4} \, \chi^{(\ell)}_\kappa(v;\tau)\ . \label{eq:continuous representation character main text} 
\end{multline}
Here, $u$, $v$ and $t$ are the chemical potentials of $\mathfrak{su}(2)_{1}$, $\mathfrak{su}(2)_{\kappa+1}$, and 
$\mathfrak{sl}(2,\mathds{R})_k$, respectively, which we write as 
\be 
q=\mathrm{e}^{2\pi i \tau}\ , \qquad x=\mathrm{e}^{2\pi i t}\ , \qquad y=\mathrm{e}^{2\pi i u}\ ,\qquad z=\mathrm{e}^{2\pi i v}\ .
\ee
We have also included a $(-1)^{\mathrm{F}}$ factor in the character, and $\chi^{(\ell)}_\kappa(v;\tau)$ is the $\mathfrak{su}(2)_\kappa$ affine character, for more details see Appendix~\ref{subapp:characters}. 
At this point, the appearance of $\mathfrak{su}(2)_\kappa$ is somewhat mysterious, since we started out with $\mathfrak{su}(2)_1 \oplus \mathfrak{su}(2)_{\kappa+1}\subset \mathfrak{d}(2,1;\alpha)_k$. However, its appearance is very natural from a spacetime perspective since the dual theory is expected \cite{Eberhardt:2017pty} to be the symmetric orbifold of $\mathcal{S}_\kappa$, which also contains a $\mathfrak{su}(2)_\kappa$ algebra; this will be explained in more detail in Section~\ref{sec:physical states in string theory}. 

We should also draw attention to the delta function which appears in the character. As in the case discussed in 
\cite{Eberhardt:2018ouy}, it means that the character localises on solutions which map the worldsheet torus (with modular parameter $\tau$) \textit{holomorphically} to the boundary torus (with modular parameter $t$).
This is the hallmark of a topological string theory and hence suggests that also $\mathrm{AdS}_3 \times \mathrm{S}^3 \times \mathrm{S}^3$ becomes essentially topological at $k^+=1$.

\section{Physical states in string theory} \label{sec:physical states in string theory}

Now we are ready to compute the full string theory spectrum of our theory. As we shall see, it will turn out to equal the partition function of the symmetric orbifold of $\mathcal{S}_\kappa$, nicely confirming the prediction of \cite{Eberhardt:2017pty}, see also \cite{Elitzur:1998mm}. Here $\mathcal{S}_\kappa$ is the $\mathcal{N}=1$ supersymmetric WZW-model on $\mathrm{S}^3 \times \mathrm{S}^1$ (with $\kappa$ units of flux through the $\mathrm{S}^3$), which exhibits in fact large $\mathcal{N}=(4,4)$ supersymmetry. 

\subsection{The theory \texorpdfstring{$\mathcal{S}_\kappa$}{Skappa} and its symmetric orbifold}
Let us begin by reviewing briefly the $\mathcal{S}_\kappa$ theory  \cite{Sevrin:1988ew, Gukov:2004ym, Eberhardt:2017pty}. The $\mathcal{S}_\kappa$ theory is defined by 
\be 
\mathfrak{su}(2)_{\kappa+2}^{(1)} \oplus \mathfrak{u}(1)^{(1)} \cong \mathfrak{su}(2)_{\kappa} \oplus \mathfrak{u}(1) \oplus \text{4 free fermions}\ ,
\ee
and possesses large $\mathcal{N}=(4,4)$ superconformal symmetry whose R-symmetry group is $\mathfrak{su}(2)_{\kappa+1} \oplus \mathfrak{su}(2)_1 \oplus \mathfrak{u}(1)$. Some background material about the large $\mathcal{N}=4$ superconformal algebra can be found in \cite{Gukov:2004ym, Gaberdiel:2013vva, Eberhardt:2017fsi}. 

For the comparison with the worldsheet answer, we will need the partition function of the $\mathcal{S}_\kappa$ theory, which is explicitly given (in the NS sector) as 
\be \label{Skappachar}
Z_{\mathcal{S}_\kappa}^{\text{NS}}(u,v;t)= \left|\frac{\vartheta_3\big(\tfrac{u+v}{2};t\big)\vartheta_3\big(\tfrac{u-v}{2};t\big)}{\eta(t)^3}\right|^2 Z_{\mathfrak{su}(2)_\kappa}(v;t)\, \Theta(\tau)\ .
\ee
Here, $\Theta(\tau)$ is the momentum-winding sum of the free boson, $u$ and $v$ are the chemical potentials for $\mathfrak{su}(2)_{\kappa+1}$ and $\mathfrak{su}(2)_1$, respectively,\footnote{To keep the notation simple, we have not introduced a chemical potential for the $\mathfrak{u}(1)$ factor. It is straightforward to include it and in fact the analysis of this paper carries through directly.} while  $t$ is the modular parameter, and 
\be 
Z_{\mathfrak{su}(2)_\kappa}(v;t)=\sum_{\ell=0}^\kappa M_{\ell\bar{\ell}}\, \chi^{(\ell)}_\kappa(v;t)\, \overline{\chi^{(\ell)}_\kappa(v;t)}
\ee
is the partition function of $\mathfrak{su}(2)_\kappa$. The central charge of this theory equals
\be 
c=\frac{6(\kappa+1)}{\kappa+2}\ ,\label{eq:Skappa central charge}
\ee
and the formula in the R-sector is obtained upon replacing $\vartheta_3$ by $\vartheta_2$. 

Given the partition function of the seed theory, it is straightforward to work out the partition function of the $N$-fold symmetric product \cite{Dijkgraaf:1996xw, Maldacena:1999bp, Eberhardt:2017pty}, and the partition function of the single particle states equals
\begin{multline} 
Z_{\mathrm{Sym}^N(\mathcal{S}_\kappa)}(u,v;t)=x^{-\frac{Nc}{24}}\bar{x}^{-\frac{Nc}{24}}\Bigg(\sum_{w=1\text{ odd}}^N x^{\frac{cw}{24}}\bar{x}^{\frac{cw}{24}}Z_{\mathcal{S}_\kappa}^{\text{NS}'}\big(u,v;\tfrac{t}{w}\big)\\
+\sum_{w=1\text{ even}}^N x^{\frac{cw}{24}}\bar{x}^{\frac{cw}{24}}Z_{\mathcal{S}_\kappa}^{\text{R}'}\big(u,v;\tfrac{t}{w}\big)\Bigg)\ . \label{eq:symmetric product Skappa partition function}
\end{multline}
Here $'$ denotes the orbifold projection, which ensures that only states with $h-\bar{h}\in \mathds{Z}$ are kept (resp.\ $h-\bar{h}\in  \mathds{Z}+\tfrac{1}{2}$ for fermions in the NS-sector). Since we are interested in the large $N$ limit, we will strip off the prefactor $x^{-\frac{Nc}{24}}\bar{x}^{-\frac{Nc}{24}}$; in the holographic setting, it corresponds to the divergent vacuum contribution.

\subsection{Adding the remaining matter and ghost fields}

Now we want to reproduce this answer from our worldsheet description. Recall that the complete worldsheet theory has in addition to $\mathfrak{d}(2,1;\alpha)_k$ an additional $\mathfrak{u}(1)$ current, four topologically twisted fermions, as well as the $bc$ and $\rho$ ghost system, see eq.~(\ref{wstheory}). The additional fields are all free, so it is a trivial matter to compute their partition functions. For the free bosons describing $\mathrm{S}^1$, we have
\be 
Z_{\text{S}^1}(\tau)=\frac{\Theta(\tau)}{|\eta(\tau)|^2}\ ,
\ee
where $\Theta(\tau)$ is the momentum-winding sum. We have already accounted for eight fermions on the worldsheet (since we constructed $\mathfrak{d}(2,1;\alpha)$ out of 8 fermions). So there should not be any additional fermionic contributions to the partition function, and indeed the $\rho$ ghost cancels the four topologically twisted fermions, as was discussed in \cite{Eberhardt:2018ouy}. Finally, the bosonic ghosts remove two neutral oscillators. Thus the full partition function of the worldsheet theory is simply obtained by multiplying the partition function of $\mathfrak{d}(2,1;\alpha)_k$ with $\Theta(\tau) \cdot |\eta(\tau)|^2$.

\subsection{The mass shell condition}
Finally, we need to impose the mass shell condition on the worldsheet, i.e.\ we need to demand that $L_0=0$. For this it is convenient to rewrite the delta function in \eqref{eq:continuous representation character main text} as an infinite sum --- this is in fact how the delta function was obtained in the first place --- so that the character reads
\begin{multline} 
\mathrm{ch}\big[\sigma^w\big(\mathscr{F}^\ell_\lambda\big)\big](t,u,v;\tau)=q^{\frac{w^2}{4(\kappa+2)}}x^{\frac{\kappa+1}{2(\kappa+2)}w} y^{\frac{w}{2}}\\
\times\sum_{m \in \mathrm{Z}+\lambda+\frac{1}{2}} x^m q^{-mw} \frac{\vartheta_2\big(\frac{t+u+v}{2};\tau\big)\vartheta_2\big(\frac{t+u-v}{2};\tau\big)}{\eta(\tau)^4} \chi^{(\ell)}_\kappa(v;\tau)\ . \label{eq:continuous representation character main text rewritten}
\end{multline}
Imposing the mass shell condition now amounts to solving
\be \label{massshell}
\frac{w^2}{4(\kappa+2)}-mw+h_\text{osc}=0\qquad\Rightarrow\qquad m=\frac{w}{4(\kappa+2)}+\frac{h_\text{osc}}{w}\ ,
\ee
where $h_\text{osc}$ is the conformal weight coming from the oscillator part (i.e.\ the theta-functions, the eta-functions and the affine $\mathfrak{su}(2)_\kappa$ character). Thus one term in the infinite sum of (\ref{eq:continuous representation character main text rewritten}) is picked out, for a specific choice of $\lambda$ (which is thereby also fixed). We correspondingly solve the mass shell condition for the right-movers. Since $\lambda$ is the same for both left- and right-movers, this imposes the additional condition
\be \label{leftright}
h_\text{osc}-\bar{h}_\text{osc}\equiv 0 \bmod w\ .
\ee
In terms of the character, imposing the two mass shall conditions can thus be implemented by removing the infinite sum, replacing $\tau \to \tfrac{t}{w}$, including the appropriate prefactor (coming from the first term in  (\ref{massshell})), and imposing the constraint (\ref{leftright}). Using the theta-function identities
\be 
\vartheta_2\big(\tfrac{t+u\pm v}{2};\tfrac{t}{w}\big)=x^{-\frac{w}{8}} y^{-\frac{w}{4}} z^{\mp \frac{w}{4}}
\begin{cases}
\vartheta_2\big(\tfrac{u\pm v}{2};\tfrac{t}{w}\big)\ , \qquad w\text{ even}\ , \\
\vartheta_3\big(\tfrac{u\pm v}{2};\tfrac{t}{w}\big)\ , \qquad w\text{ odd}\ .
\end{cases}
\ee
the partition function of the physical spectrum can thus be written as 
\be 
Z_\text{string}(u,v;t)=\sum_{w=1\text{ odd}}^\infty x^{\frac{cw}{24}}\bar{x}^{\frac{cw}{24}}Z_{\mathcal{S}_\kappa}^{\text{NS}'}\big(u,v;\tfrac{t}{w}\big)
+\sum_{w=1\text{ even}}^\infty x^{\frac{cw}{24}}\bar{x}^{\frac{cw}{24}}Z_{\mathcal{S}_\kappa}^{\text{R}'}\big(u,v;\tfrac{t}{w}\big)\ , \label{eq:full string partition function}
\ee
where $c$ is given by \eqref{eq:Skappa central charge}.
This then agrees precisely with the large $N$ limit of \eqref{eq:symmetric product Skappa partition function}. We note in passing that this works for any modular invariant of $\mathfrak{su}(2)_\kappa$.

We should mention that we have restricted the calculation here to the $w \ge 1$ sector. It is easy to see that there are no physical states in  the $w=0$ sector, while the states from the $w \le -1$ sector have the interpretation of out-states in the dual CFT \cite{Maldacena:2001km,Eberhardt:2018ouy}, and hence should not be included in the partition function. 

\subsection{The BPS spectrum}

It is instructive to understand how the BPS spectrum arises from the worldsheet. Recall that the single-particle BPS spectrum of the 
symmetric orbifold of $\mathcal{S}_\kappa$ is \cite{Gukov:2004ym,Eberhardt:2017pty}\footnote{There are some additional BPS states in the $N$-twisted sector, which disappear in the large $N$ limit \cite{Eberhardt:2017pty}. We therefore do not consider them here.} 
\be 
\bigoplus_{\ell=0}^{\frac{cN}{12}} [h=\ell,\ell^+=\ell,\ell^-=\ell,u=0] \otimes \overline{[h=\ell,\ell^+=\ell,\ell^-=\ell,u=0]}\ . \label{eq:BPS spectrum AdS3xS3xS3xS1}
\ee
Here, $[h=h_\text{BPS}(\ell^+,\ell^-,u),\ell^+,\ell^-,u]$ denotes a large $\mathcal{N}=4$ BPS multiplet in the representation $(\ell^+,\ell^-,u)$ of the R-symmetry  algebra $\mathfrak{su}(2) \oplus \mathfrak{su}(2) \oplus \mathfrak{u}(1)$. This BPS spectrum also agrees with the supergravity BPS spectrum for $\mathrm{AdS}_3 \times \mathrm{S}^3 \times \mathrm{S}^3 \times \mathrm{S}^1$ \cite{Eberhardt:2017fsi,Baggio:2017kza}.

The different states in eq.~(\ref{eq:BPS spectrum AdS3xS3xS3xS1}) arise as follows. There 
is a BPS representation in every $w$-twisted sector, provided that $w \not\in (\kappa+2)\mathds{Z}$. In order to describe it, we write
\be 
w=m(\kappa+2)+2\ell+1
\ee
for some $m \in \mathds{Z}$ and $\ell \in \{0,\tfrac{1}{2},\dots, \tfrac{\kappa}{2}\}$; this is possible since $w$ is not divisible by $(\kappa+2)$. Then we consider the $(2\ell+m(\kappa+1),m)$-fold spectral flow of the ground state representation of spin $(0,\ell)$ of $\mathfrak{su}(2)_1 \oplus \mathfrak{su}(2)_{\kappa+1}$. This gives a state in the $w$ twisted sector which is indeed BPS. It was furthermore shown in \cite{Eberhardt:2017pty} that all BPS states arise in this manner.

This structure can be directly translated to the worldsheet: 
%It is the $k^+=1$ analogue of the worldsheet BPS spectrum, which was also discussed in detail in \cite{Eberhardt:2017pty} for the case $k^\pm \ge 2$. 
BPS states come from the representations
\be 
\sigma^{m(\kappa+2)+2\ell+1}\big(\mathscr{F}^\ell_{\lambda_\ell}\big)\quad \text{($m$ even)}\ , \qquad   \hbox{and} \qquad \sigma^{m(\kappa+2)+2\ell+1}\big(\mathscr{F}^{\frac{\kappa}{2}-\ell}_{-\lambda_{\frac{\kappa}{2}-\ell}}\big) \quad \text{($m$ odd)}\ .
\ee
To see this, we first recall that the $m$-fold spectral flow on $\mathfrak{su}(2)_\kappa$ maps the spin-$\ell$ representation back to itself if $m$ is even, and to $\tfrac{\kappa}{2}-\ell$ if $m$ is odd; the resulting state therefore sits in the correct representation of $\mathfrak{su}(2)_{\kappa+1}$. This leaves us with determining the $\lambda$-parameters, which can be computed by requiring that the $\mathfrak{sl}(2,\mathds{R})$ weights agree with the BPS bound up to an integer. We see that we obtain precisely the values at which the modules become indecomposable. (Strictly speaking, we should therefore replace $\mathscr{F}^\ell_{\lambda_\ell}$ by its indecomposable analogue $\mathscr{T}^{\ell+\frac{1}{2}}_>$ and $\mathscr{F}^{\frac{\kappa}{2}-\ell}_{-\lambda_{\frac{\kappa}{2}-\ell}}$ by $\mathscr{T}^{\frac{\kappa+1}{2}-\ell}_<$, see Appendix~\ref{app:indecomposables} for more details). The fact that BPS states live in indecomposable representations is typical for supergroup theories \cite{Gaberdiel:2011vf, Troost:2011fd}.

\smallskip

Finally, we discuss the moduli of the theory. Moduli of large $\mathcal{N}=4$ theories are superconformal descendants of $(\ell^+,\ell^-,u)=(\tfrac{1}{2},\tfrac{1}{2},0)$ BPS states \cite{Gukov:2004ym}.
%as reviewed in Appendix~\ref{subapp:N4 moduli}. 
These can come from the vacuum representation or the large $\mathcal{N}=4$  BPS representation labelled by $[h=\tfrac{1}{2},\ell^+=\tfrac{1}{2},\ell^-=\tfrac{1}{2},u=0]$. These states in turn come from the worldsheet representations 
\begin{align}
\sigma(\mathscr{F}^0_{\lambda_0})&\sim \mathscr{L} \oplus \sigma(\mathscr{G}^0_{>,+})\ , \\
\sigma^2(\mathscr{F}^{1/2}_{\lambda_{1/2}})&\sim \sigma(\mathscr{G}^0_{>,+}) \oplus \sigma^2(\mathscr{G}^{1/2}_{>,+})\ .
\end{align}
The module $\mathscr{L}$ contains a single physical state, namely the vacuum itself, which corresponds to the spacetime vacuum. The actual moduli therefore come from the representation $\sigma(\mathscr{G}^0_{>,+})$ which indeed appears twice. This reflects the situation in the dual CFT, where one of the moduli comes from the untwisted sector and changes the radius of $\mathrm{S}^1$, whereas the other modulus carries one away from the symmetric orbifold point. The two moduli in the theory are exactly on the same footing, in agreement with the fact that the geometry of the two-dimensional moduli space is the upper half plane \cite{Gukov:2004ym}.

\section{The spacetime DDF operators}\label{sec:DDF}

In the previous sections we have shown that the spacetime partition function of string theory on $\mathrm{AdS}_3 \times \mathrm{S}^3 \times \mathrm{S}^3 \times \mathrm{S}^1$ coincides with the partition function of the symmetric orbifold of $\mathcal{S}_\kappa$ if $k^+=1$. In this section we want to establish that also the algebraic structure of the two sides agree, thus extending the analysis of \cite{Eberhardt:2019qcl} to the present setting. Moreover, we show that the correspondence can be extended to the case of $k^+>1$, in which case the dual CFT becomes the symmetric orbifold of large $\mathcal{N}=4$ Liouville theory. Most of the arguments are very similar to what was done in \cite{Eberhardt:2019qcl}, and we shall therefore be rather brief. 

%In particular, we will construct the DDF operators following \cite{Giveon:1998ns,Elitzur:1998mm} and check that they satisfy the expected commutation relations. Much of what we will say applies in fact for generic $k^+$ and $k^-$, but we shall concentrate on $k^+=1$ in the following. 

%We have now established that string theory on $\mathrm{AdS}_3 \times \mathrm{S}^3 \times \mathrm{S}^3 \times \mathrm{S}^1$ is well-defined, even if $k^+=1$. Furthermore, we have shown that the  . In this Section, we want to show that also the algebraic structure of the two sides coincides. For this, we work again mostly in the RNS-formalism. Furthermore, almost everything what we say applies in fact for generic $k^+$ and $k^-$.

\subsection{Spacetime operators}

In \cite{Elitzur:1998mm} the DDF operators generating the large $\mathcal{N}=4$ superconformal algebra were constructed for the background $\mathrm{AdS}_3 \times \mathrm{S}^3 \times \mathrm{S}^3 \times \mathrm{S}^1$. The analysis was performed in the RNS formalism assuming that $k^\pm\geq 2$, and it is a priori not clear whether the construction continues to make sense also for $k^+=1$. Using similar arguments as in \cite{Eberhardt:2019qcl} (where the corresponding problem was studied for the case of $\mathrm{AdS}_3 \times \mathrm{S}^3 \times \mathbb{T}^4$), we have checked that the DDF operators of  \cite{Elitzur:1998mm} are also well-defined for $k^+=1$. 

Let us denote the large ${\cal N}=4$ spacetime algebra generators (our conventions follow \cite{Eberhardt:2017pty}) by 
\be 
\mathcal{L}_m\ ,\qquad \mathcal{G}^{\alpha\beta}_r\ , \qquad \mathcal{K}^{(\pm)a}_m\ , \qquad \mathcal{U}_m\ , \qquad \mathcal{Q}^{\alpha\beta}_r\ ,
\ee
where $\mathcal{L}_m$ are the modes of the spacetime energy momentum tensor, $ \mathcal{G}^{\alpha\beta}_r$ those of the spacetime supercharges, while $\mathcal{K}^{(\pm)a}_m$ and $\mathcal{U}_m$ define the R-symmetry generators. In addition, there are four free fermions that are denoted by $\mathcal{Q}^{\alpha\beta}_r$. 

As was explained in \cite{Eberhardt:2019qcl} --- the argument is essentially the same here ---  the modes of this algebra can take values in $\tfrac{1}{w}\mathds{Z}$ (or $\tfrac{1}{w}\big(\mathds{Z}+\tfrac{1}{2}\big)$ in the case of fermions). %This is reminiscent of the symmetric orbifold, where in the $w$-cycle twisted sector the algebra generators are  fractionally moded. 
By the same reasoning as in \cite{Eberhardt:2019qcl} this then suggests that the spacetime states that arise from the continuous representations on the worldsheet are in general  (i.e.\ for arbitrary $k^+$ and $k^-$) described by the symmetric product orbifold of
\be 
\text{large $\mathcal{N}=4$ Liouville theory with $(k^+,k^-)$}  \ .
\ee
We shall review the construction of large $\mathcal{N}=4$ Liouville theory in the following section, and explain the crucial steps in this derivation in Section~\ref{Liouvilleworld}. In Section~\ref{kp=1} we will then demonstrate that large ${\cal N}=4$ Liouville theory reduces, for $k^+=1$, to $\mathcal{S}_\kappa$. Furthermore, since for $k^+=1$ the entire worldsheet spectrum comes from the continuous representations, this is in fact a complete description of the theory.

\subsection{Large \texorpdfstring{$\mathcal{N}=4$}{N=4} Liouville theory} \label{subsec:large N4 Liouville}
Let us first discuss large $\mathcal{N}=4$ Liouville theory, which does not seem to be well-known. 
We shall first assume $k^\pm \ge 2$, and study the case of $k^+=1$ in Section~\ref{kp=1}. To motivate the construction of this theory, we consider a free boson coupled to the curvature of the worldsheet (i.e.\ with background charge), together with the $\mathfrak{su}(2)_{k^+-2} \oplus \mathfrak{su}(2)_{k^--2} \oplus \mathfrak{u}(1)$ R-symmetry and 8 free fermions. (This is basically the same field content as for the worldsheet 
theory in the RNS formalism, except that the $\mathfrak{sl}(2,\mathds{R})$ factor has been replaced by a boson with screening charge.) It was noticed in \cite{Ito:1992nq} that this theory supports large $\mathcal{N}=4$ supersymmetry with levels $k^+$ and $k^-$ for the two $\mathfrak{su}(2)$ currents. 
%; this is shown in Appendix~\ref{subapp:Liouville representation}. 
%It is the natural large $\mathcal{N}=4$ supersymmetric analogue of the construction of a Virasoro algebra out of a free boson with a screening charge. 
The free boson with screening charge $Q=\frac{(k-1)}{\sqrt{k}}$  leads to a continuous spectrum, whose gap above the vacuum equals
\be 
\Delta_\phi=\frac{c^\phi-1}{24}=\frac{(k-1)^2}{4k}\ .
\ee
We can combine this with arbitrary $\mathfrak{su}(2)_{k^\pm-2}$ and $\mathfrak{u}(1)$ representations, thus leading to the general formula for the gap 
\begin{align} 
\Delta_{\ell^+,\ell^-,u}&=\frac{(k-1)^2}{4k}+\frac{\ell^+(\ell^++1)}{k^+}+\frac{\ell^-(\ell^-+1)}{k^-}+\frac{u^2}{k^++k^-}\\
&=\frac{(\ell^++\frac{1}{2})^2}{k^+}+\frac{(\ell^-+\frac{1}{2})^2}{k^-}+\frac{k-2}{4}+\frac{u^2}{k^++k^-}\ ,
\end{align}
where we have used \eqref{eq:criticality}.
We should note that, generically, all the BPS representations lie below this gap since 
\be 
\Delta_{\ell^+,\ell^-,u}-h_\text{BPS}(\ell^+,\ell^-,u)=\frac{k}{4}\left(1-\frac{2\ell^-+1}{k^-} -\frac{2\ell^++1}{k^+}\right)^2\ge 0\ ,
\ee
where we have used the expression for the BPS bound, see e.g.\ eq.~(A.13) of \cite{Eberhardt:2017pty}. The only BPS states that appear in large ${\cal N}=4$ Liouville theory therefore arise if 
\be 
\frac{2\ell^-+1}{k^-} +\frac{2\ell^++1}{k^+}=1\ .
\ee
The fact that such solutions exist is related to the fact that also the continuous sector of the worldsheet theory of $\mathrm{AdS}_3 \times \mathrm{S}^3 \times \mathrm{S}^3 \times \mathrm{S}^1$ contributes to the BPS spectrum \cite{Eberhardt:2017pty}.

%Thus, this construction has the following spectrum of large $\mathcal{N}=4$ representations:
%\be 
%\bigoplus_{\ell^+=0}^{\frac{k^+-2}{2}} \bigoplus_{\ell^-=0}^{\frac{k^--2}{2}} \bigoplus_{(u,\bar{u}) \in \Gamma} \ \boldsymbol{\oplus}\hspace*{-.48cm}\int_{\Delta_{\ell^+,\ell^-,u}}^\infty \mathrm{d}h\ [h,\ell^+,\ell^-,u] \otimes \overline{[h,\ell^+,\ell^-,\bar{u}]}\ .
%\ee
%Of course, we could also choose non-diagonal modular invariants of the two $\mathfrak{su}(2)_{k^\pm-2}$ theories.

The full spectrum of large ${\cal N}=4$ Liouville theory is obtained by taking the diagonal modular invariant of all of the representations that lie above the gap (and have allowed $\mathfrak{su}(2)_{k^\pm-2}$ and $\mathfrak{u}(1)$ representations).  We should note that in large $\mathcal{N}=4$ Liouville, each representation appears precisely once, whereas in the free bosons realisation from above, each representation appears twice since opposite values of the momentum lead to the same Virasoro representation.

\subsection{The Liouville spectrum from the worldsheet}\label{Liouvilleworld}

Next, we want to reproduce this Liouville spectrum directly from the worldsheet. Solving the  mass-shell condition in the spectrally flowed sector leads to 
\be 
\frac{\frac{1}{4}+p^2}{k}-wh+\frac{k}{4}w^2+\frac{\ell^+(\ell^++1)}{k^+}+\frac{\ell^-(\ell^-+1)}{k^-}+\frac{u^2}{k^++k^-}+N=\frac{1}{2}\ ,
\ee
where $N$ is the contribution to the conformal weight from the oscillator part. Solving this equation for the conformal weight $h$ of the dual CFT yields
\be 
h=\frac{k}{4w}(w^2-1)+\frac{\Delta_{\ell^+,\ell^-,u}}{w}+ \frac{N}{w}+\frac{p^2}{kw} \ .
\ee
This matches exactly the form expected from the symmetric orbifold of large $\mathcal{N}=4$ Liouville: the first term is the universal ground state energy of the twisted sector, which equals $\tfrac{c}{24w}(w^2-1)$, where $c=6k$ is the central charge of the `seed theory', while 
the second term describes the gap in the $w$-cycle twisted sector. Since the modes are $\frac{1}{w}$-fractionally moded in the $w$-cycle twisted sector, the contribution of $N$ has to be divided by $w$. Finally, the term $\tfrac{p^2}{kw}$ leads to a continuum in the spectrum (since $p$ is any real number corresponding to the momentum of the long string). Furthermore, the representations belonging to $p$ and $-p$ are identified on the worldsheet --- they describe the same $\mathfrak{sl}(2,\mathds{R})$ representation --- and appear only once in the spectrum, as appropriate for $\mathcal{N}=4$ Liouville, see the comment at the end of the previous section.

In order to conclude from this that the complete spectrum matches we use again a character argument (as in \cite{Eberhardt:2019qcl}). To compute the relevant characters, we again make use of the free-field construction of \cite{Ito:1992nq}. Both the worldsheet theory as well as Liouville theory has 8 free fermions (after imposing the physical state conditions on the worldsheet). In addition also the bosonic degrees of freedom match: the $\mathfrak{su}(2)_{k^+-2} \oplus \mathfrak{su}(2)_{k^--2} \oplus \mathfrak{u}(1)$ algebra is the same on both sides and the $\mathfrak{sl}(2,\mathds{R})_k$ factor has the character of a free boson after imposing the physical state conditions. This reproduces the contribution of the Liouville boson.

Thus, we have matched the spectrum as well as the chiral algebras on both sides of the duality. Since Liouville theory is believed to be uniquely characterised by this data (and the same should be true for large ${\cal N}=4$ Liouville), this goes a long way towards proving the  duality in this case.

We should stress that the `symmetric orbifold of Liouville theory' contains single-particle states for which only one copy is in the ground state of Liouville theory, while the other copies are in the `vacuum' --- this is part of the spectrum as determined from the dual worldsheet analysis. (This is different from the naive definition of the symmetric orbifold where the `vacuum' would not be allowed for any copy.) As a consequence, the effective central charge scales as $6 N \frac{k^+ k^-}{k^+ +k^-}$, and the spectrum has the correct density at large conformal dimension.

\subsection{The case of \texorpdfstring{$k^+=1$}{k+=1}}\label{kp=1}
Upon setting $k^+=1$, the construction of $\mathcal{N}=4$ Liouville theory 
%of Appendix~\ref{subapp:Liouville representation} 
breaks down since the level of the corresponding bosonic algebra is $-1$, which makes the theory non-unitary. Instead, the superconformal algebra collapses to $\mathcal{S}_\kappa$, i.e.\ as chiral algebras we have the equivalence \cite{Goddard:1988wv}
\be 
A_\gamma(k^+=1,k^-=\kappa+1) =\mathfrak{su}(2)_\kappa\oplus \mathfrak{u}(1) \oplus \text{4 free fermions} \ , 
\ee
mirroring exactly what happens on the worldsheet. %This is explained in Appendix~\ref{app:large N4}. 
%Note that the appearance of the null fields in the spacetime algebra is mirrored on the worldsheet by the appearance of short representations.

Contrary to the $k^\pm\ge 2$ case, the $\mathcal{S}_\kappa$ theory (and hence also $A_\gamma$ at $k^+=1$) contains \textit{only} BPS representations.
%, again reflecting precisely our worldsheet analysis. 
This just follows from the fact that the conformal weight of a representation with $\mathfrak{su}(2)$ spin $\ell^-$ and $\mathfrak{u}(1)$-charge $u$ is
\be 
\Delta_{\ell^-,u}=\frac{\ell^-(\ell^-+1)}{\kappa+2}+\frac{u^2}{\kappa+2}=h_\text{BPS}(0,\ell^-,u)\ .
\ee
As a consequence, any large $\mathcal{N}=(4,4)$ theory at $k^+=1$ \textit{cannot} have a continuum (such as the one that appears in Liouville theory). Furthermore, the above DDF analysis predicts that the CFT dual of string theory on ${\rm AdS}_3\times {\rm S}^3 \times {\rm S}^3 \times {\rm S}^1$ must be a symmetric orbifold, whose seed theory has large ${\cal N}=4$ superconformal symmetry with levels $k^\pm=k^\pm_{\rm worldsheet}$. For $k^+=1$, the seed theory must therefore be $\mathcal{S}_\kappa$, thus inevitably leading to the proposal of \cite{Eberhardt:2017pty}  (for $k^+=1$). 

%Hence a large $\mathcal{N}=(4,4)$ theory at $k^+=1$ is necessarily equal to $\mathcal{S}_\kappa$ and thus \textit{cannot} have a continuum as Liouville theory did.
%This result fits together with our previous analysis, and shows that not only the spectrum agrees, but also that the DDF operators give rise to the spectrum generating fields of the symmetric orbifold of $\mathcal{S}_\kappa$ on the worldsheet. This therefore goes a long way towards proving the conjecture of for $k^+=1$.

%The existence of the spacetime DDF algebra then necessitates the appearance of the symmetric orbifold of $\mathcal{S}_\kappa$ and in fact we have shown above the agreement of partition functions, see \eqref{eq:full string partition function}. 
%
%Thus, we have shown the equivalence of strings on $\mathrm{AdS}_3 \times \mathrm{S}^3 \times\mathrm{S}^3 \times \mathrm{S}^1$ for $k^+=1$ with the symmetric orbifold of $\mathcal{S}_\kappa$. This essentially proves the conjecture made in \cite{Eberhardt:2017pty}.

\section{Discussion}\label{sec:discussion}

In this paper we have found a family of examples that relate a solvable worldsheet theory describing strings on AdS$_3$ to a solvable family of $2d$ CFTs. The relevant backgrounds describe string theory on 
${\rm AdS}_3 \times {\rm S}^3 \times {\rm S}^3 \times {\rm S}^1$ with pure NS-NS flux and minimal flux through one of the two ${\rm S}^3$'s, while the dual CFTs are symmetric orbifolds of the so-called ${\cal S}_\kappa$ theory, the simplest conformal field theory with large ${\cal N}=4$ superconformal symmetry \cite{Sevrin:1988ew,Gukov:2004ym}. We have shown that the spacetime spectrum of the worldsheet theory agrees precisely with the dual symmetric orbifold CFT in the large $N$ limit. We have furthermore shown that the spectrum generating fields on the worldsheet (the DDF operators) obey the same algebra as those of the symmetric orbifold. This gives strong support to the identification of the dual CFT that was proposed in \cite{Eberhardt:2017pty}, see also \cite{Elitzur:1998mm}. Our results are a natural generalisation of the results obtained for the $\mathbb{T}^4$ case in \cite{Eberhardt:2018ouy,Eberhardt:2019qcl}.

We have also analysed the situation where the NS-NS flux through both spheres is bigger than its minimal value ($k^\pm >1$), and in this case, our analysis suggests that the dual CFT is the symmetric orbifold of large ${\cal N}=4$ Liouville theory. In this case the spectrum of the symmetric orbifold is entirely accounted for in terms of the continuous representations on the worldsheet, while the role of the spacetimes states that originate from discrete representations on the worldsheet is less clear.\footnote{Note that if one of the levels takes the minimal value, say $k^+=1$, then the worldsheet spectrum does not contain any discrete representations.} Again, this mirrors precisely what was found for the $\mathbb{T}^4$ case in \cite{Eberhardt:2019qcl}.

It is suggestive that, apart from some small technical differences, the analysis (as well as the resulting picture) that we find here is quite similar to that obtained in the $\mathbb{T}^4$ case. This suggests that similar results may also hold for other backgrounds (say with less supersymmetry), and it would be interesting to explore this. It would also be interesting to probe these dual pairs in more detail, say, by comparing their 3-point functions, or by computing $1/N$ corrections (which should correspond to higher genus corrections from the worldsheet viewpoint). In any case, we feel that these three dimensional examples will provide a useful testing ground for various aspects of the AdS/CFT correspondence.

\section*{Acknowledgements}

We thank Andrea Dei, Rajesh Gopakumar, Wei Li and Ida Zadeh for useful discussions. We would also like to thank the Erwin Schr\"odinger Institute in Vienna, where this work was completed, for hospitality. LE is supported by the Swiss National Science Foundation, and the work of the group is more generally supported by the NCCR SwissMAP which is also funded by the Swiss National Science Foundation.

\appendix

\section{Conventions} \label{app:OPEs}

\subsection{The RNS formalism of strings on \texorpdfstring{$\mathrm{AdS}_3 \times \mathrm{S}^3 \times \mathrm{S}^3 \times \mathrm{S}^1$}{AdS3xS3xS3xS1}} \label{subapp:RNS formalism AdS3xS3xS3xS1}

The bosonic generators on the worldsheet give rise to $\mathfrak{sl}(2,\mathds{R})_{k+2} \oplus \mathfrak{su}(2)_{k^+-2}\oplus \mathfrak{su}(2)_{k^--2}$, together with one free boson. Their modes satisfy the commutation relations
\begin{subequations}
\begin{align}
[\mathscr{J}^3_m,\mathscr{J}^3_n]&=-\tfrac{k+2}{2}\, m \delta_{m+n,0}\ , \\
[\mathscr{J}^3_m,\mathscr{J}^\pm_n]&=\pm \mathscr{J}^\pm_{m+n}\ , \\
[\mathscr{J}^+_m,\mathscr{J}^-_n]&=(k-2)m \delta_{m+n,0}-2\,\mathscr{J}^3_{m+n,0}\ , \\
[\mathscr{K}^{(\pm)3}_m,\mathscr{K}^{(\pm)3}_n]&=\tfrac{k^\pm-2}{2}\, m \delta_{m+n,0}\ , \\
[\mathscr{K}^{(\pm)3}_m,\mathscr{K}^{(\pm)\pm}_n]&=\pm \mathscr{K}^{(\pm)\pm}_{m+n}\ , \\
[\mathscr{K}^{(\pm)+}_m,\mathscr{K}^{(\pm)-}_n]&=(k^\pm+2)m \delta_{m+n,0}+2\,\mathscr{K}^{(\pm)3}_{m+n,0}\ , \\
[\partial \Phi_m,\partial \Phi_n]&=m \delta_{m+n,0}\ .
\end{align}
\end{subequations}
There are moreover ten fermions on the worldsheet, which we denote by $\psi^a$, $\chi^{(\pm)a}$ and $\lambda$. We take them to have anticommutation relations
\begin{subequations}
\begin{align}
\{\psi^3_r,\psi^3_s\}&=-\tfrac{k}{2}\,\delta_{r+s,0}\ ,\\
\{\psi^+_r,\psi^-_s\}&=k\, \delta_{r+s,0}\ ,\\
\{\chi^{(\pm)3}_r,\chi^{(\pm)3}_s\}&=\tfrac{k^\pm}{2}\,\delta_{r+s,0}\ ,\\
\{\chi^{(\pm)+}_r,\chi^{(\pm)-}_s\}&=k^\pm\, \delta_{r+s,0}\ ,\\
\{\lambda_r,\lambda_s\}&=\delta_{r+s,0}\ .
\end{align}
\end{subequations}

Out of the bosonic currents at level $k+2$, $k^+-2$ and $k^--2$ and the free fermions, one can define `supersymmetric' currents at level $k$, $k^+$ and $k^-$, respectively
\begin{subequations}
\begin{align}
J^\pm&= \mathscr{J}^\pm\mp \frac{2}{k} (\psi^3 \psi^\pm)\ ,\label{eq:supersymmetric current a}\\
J^3&= \mathscr{J}^3+\frac{1}{k} (\psi^+\psi^-)\ ,\label{eq:supersymmetric current b} \\
K^{(\pm)\pm}&=\mathscr{K}^{(\pm)\pm} \pm \frac{2}{k^\pm} (\chi^{(\pm)3}\chi^{(\pm)\pm})\ ,\label{eq:supersymmetric current c}\\
K^{(\pm)3}&=\mathscr{K}^{(\pm)3}+\frac{1}{k^\pm} (\chi^{(\pm)+}\chi^{(\pm)-})\ ,\label{eq:supersymmetric current d}
\end{align}
\end{subequations}
whose zero modes correspond to the global (bosonic) generators of the spacetime supersymmetry algebra. Via picture changing, we can also write them in the canonical $(-1)$ picture, where they simply read
\be 
J^\pm=\psi^\pm \mathrm{e}^{-\phi}\ , \quad J^3=\psi^3 \mathrm{e}^{-\phi}\ , \quad K^{(\pm)\pm}=\chi^{(\pm)\pm} \mathrm{e}^{-\phi}\ , \quad K^{(\pm)3}=\chi^{(\pm)3} \mathrm{e}^{-\phi}\ .
\ee
\subsection{The \texorpdfstring{$\mathfrak{d}(2,1;\alpha)_k$}{d(2,1;alpha)k} algebra}\label{subapp:d21alpha commutation relations} 

We take the affine Kac-Moody algebra to be defined by
\begin{subequations}
\begin{align}
[J^3_m,J^3_n]&=-\tfrac{1}{2}km\delta_{m+n,0}\ ,   \label{eq:d21alpha commutation relations a}\\
[J^3_m,J^\pm_n]&=\pm J^\pm_{m+n}\ ,  \label{eq:d21alpha commutation relations b}\\
[J^+_m,J^-_n]&=km\delta_{m+n,0}-2J^3_{m+n}\ ,  \label{eq:d21alpha commutation relations c}\\
[K^{(\pm)3}_m,K^{(\pm)3}_n]&=\tfrac{1}{2}k^\pm m\delta_{m+n,0}\ ,  \label{eq:d21alpha commutation relations d}\\
[K^{(\pm)3}_m,K^{(\pm)\pm}_n]&=\pm K^{(\pm)\pm}_{m+n}\ ,  \label{eq:d21alpha commutation relations e}\\
[K^{(\pm)+}_m,K^{(\pm)-}_n]&=k^\pm m\delta_{m+n,0}+2K^{(\pm)3}_{m+n}\ , \label{eq:d21alpha commutation relations f}\\
[J^a_m,S^{\alpha\beta\gamma}_n]&=\tfrac{1}{2}\tensor{(\sigma^a)}{^\alpha_\mu} S^{\mu\beta\gamma}_{m+n}\ ,  \label{eq:d21alpha commutation relations g}\\
[K^{(+)a}_m,S^{\alpha\beta\gamma}_n]&=\tfrac{1}{2}\tensor{(\sigma^a)}{^\beta_\nu} S^{\alpha\nu\gamma}_{m+n}\ ,  \label{eq:d21alpha commutation relations h}\\
[K^{(-)a}_m,S^{\alpha\beta\gamma}_n]&=\tfrac{1}{2}\tensor{(\sigma^a)}{^\gamma_\rho} S^{\alpha\beta\rho}_{m+n}\ , \label{eq:d21alpha commutation relations i}\\
 \{S^{\alpha\beta\gamma}_m,S^{\mu\nu\rho}_n\}&=km \varepsilon^{\alpha\mu}\varepsilon^{\beta\nu}\varepsilon^{\gamma\rho}\delta_{m+n,0}-\varepsilon^{\beta\nu}\varepsilon^{\gamma\rho} \tensor{(\sigma_a)}{^{\alpha\mu}} J^a_{m+n}+\gamma\varepsilon^{\alpha\mu}\varepsilon^{\gamma\rho} \tensor{(\sigma_a)}{^{\beta\nu}} K^{(+)a}_{m+n}\nonumber\\
 &\qquad+(1-\gamma)\varepsilon^{\alpha\mu}\varepsilon^{\beta\nu} \tensor{(\sigma_a)}{^{\gamma\rho}} K^{(-)a}_{m+n}\ .
 \label{eq:d21alpha commutation relations j}
\end{align}
\end{subequations}
Here, $\alpha,\beta,\dots$ are spinor indices and take values in $\{+,-\}$. On the other hand, $a$ is an $\mathfrak{su}(2)$ adjoint index and takes values in $\{+,-,3\}$. It is raised and lowered by the standard $\mathfrak{su}(2)$-invariant form. Explicitly, we have
\begin{subequations}
\begin{align}
\tensor{(\sigma^-)}{^+_-}&=2\ , & \tensor{(\sigma^3)}{^-_-}&=-1\ , & \tensor{(\sigma^3)}{^+_+}&=1\ , & \tensor{(\sigma^+)}{^-_+}&=2\ , \\
\tensor{(\sigma_-)}{^{--}}&=1\ , & \tensor{(\sigma_3)}{^{-+}}&=1\ , & \tensor{(\sigma_3)}{^{+-}}&=1\ , & \tensor{(\sigma_+)}{^{++}}&=-1\ , \\
\tensor{(\sigma^-)}{_{--}}&=2\ , & \tensor{(\sigma^3)}{_{+-}}&=1\ , & \tensor{(\sigma^3)}{_{-+}}&=1\ , & \tensor{(\sigma^+)}{_{++}}&=-2\ .
\end{align}
\end{subequations}
$\epsilon^{\alpha\beta}$ is the epsilon symbol with $\epsilon^{+-}=1$.
Finally, $\gamma$, $k^+$ and $k^-$ are related to $\alpha$ and $k$ by
\be 
\gamma=\frac{\alpha}{1+\alpha}\, , \qquad k^+=\frac{(\alpha+1)k}{\alpha} \, , \qquad k^-=(\alpha+1)k\, .
\ee
We note that unitarity requires $k^+,k^- \in \mathds{Z}_{\ge 0}$.

\section{The Wakimoto representation of \texorpdfstring{$\boldsymbol{\mathfrak{d}(2,1;\alpha)_k}$}{d(2,1;alpha)k}}\label{app:Wakimoto}
In this Appendix, we explain the Wakimoto representation that is used in the derivation of the hybrid formalism for $\mathrm{AdS}_3 \times \mathrm{S}^3 \times \mathrm{S}^3 \times \mathrm{S}^1$ in the main body of the paper.

We start with four pairs of topologically twisted fermions (or $bc$ systems), satisfying
\be 
p^{\alpha\beta}(z) \, \theta^{\gamma\delta}(w) \sim \frac{\epsilon^{\alpha\gamma}\epsilon^{\beta\delta}}{z-w}\ .
\ee
Furthermore, we also have the bosonic $\mathfrak{sl}(2,\mathds{R})_{k+2} \oplus \mathfrak{su}(2)_{k^+-2} \oplus \mathfrak{su}(2)_{k^--2}$ currents $\mathscr{J}^a$ and $\mathscr{K}^{(\pm)a}$. 
%We recall that the 
%$p^{\alpha\beta}(z)$ carry charge $+\tfrac{1}{2}$ under $\mathfrak{su}(2)_{k^-}$, while that of 
%$\theta^{\alpha\beta}(z)$ is $-\tfrac{1}{2}$. Moreover, they are defined in picture number $(-\tfrac{1}{2})$ and $(\tfrac{1}{2})$, respectively. 
%In order to define a $\mathfrak{d}(2,1;\alpha)_k$ algebra, which closes off-shell, we therefore need to define a field carrying charge $q$ under $\mathfrak{su}(2)_{k^-}$ in the picture $(-q)$. 

\subsection{The root system of \texorpdfstring{$\mathfrak{d}(2,1;\alpha)$}{d(2,1;alpha)}} 

To continue systematically, let us recall the basic idea of the Wakimoto representation. Starting from the root system of a Lie (super)algebra, one first constructs a realisation of the (nilpotent) positive roots in terms of $\beta\gamma$ systems. Then one extends this construction to the positive Borel subalgebra by introducing as many free bosons as the rank of the Lie algebra. Finally, the generators for the negative roots are then uniquely fixed by requiring them to satisfy all the OPEs.
This procedure requires the breaking of some symmetries. For $\mathfrak{d}(2,1;\alpha)_k$, a minimal choice is to break the $\mathfrak{su}(2)_{k^-}$ symmetry and keep the rest of the bosonic subalgebra manifest.

In the context of $\mathfrak{d}(2,1;\alpha)_k$, we pick as Cartan subalgebra $J^3_0$, $K^{(+)3}_0$ and $K^{(-)3}_0$, and take the simply roots to be 
\be 
\alpha_1=(1,0,0)\ , \qquad \alpha_2=(0,1,0)\ , \qquad \alpha_3=\big(-\tfrac{1}{2},-\tfrac{1}{2},\tfrac{1}{2}\big)\ .
\ee
The first two roots are bosonic, while $\alpha_3$ is fermionic, so this corresponds to the distinguished choice of simple roots.\footnote{Recall that in Lie superalgebras, there is no unique choice of simple roots, see e.g.\ \cite{Frappat:1996pb}.} The step operators corresponding to the positive roots are then 
\be 
J^+\ , \qquad K^{(+)+}\ , \qquad K^{(-)+}\ , \qquad S^{\alpha\beta+}\ ,
\ee
for $\alpha$, $\beta\in \{+,-\}$.
\subsection{Constructing the Borel subalgebra}
We first explain how to construct $J^a$ and $K^{(+)a}$. The topologically twisted fermions lead to the generators 
of $\mathfrak{sl}(2,\mathds{R})_{-2} \oplus \mathfrak{su}(2)_2$
\begin{align} 
J^{\text{(f)}a}&=\frac{1}{2}c_a \tensor{(\sigma^a)}{_{\alpha\gamma}}\epsilon_{\beta\delta} (p^{\alpha\beta}\theta^{\gamma\delta})\ ,  \\
K^{\text{(f)}a}&=\frac{1}{2}\epsilon_{\alpha\gamma} \tensor{(\sigma^a)}{_{\beta\delta}}(p^{\alpha\beta}\theta^{\gamma\delta})\ ,
\end{align}
where the different constants were explained in Appendix~\ref{app:OPEs}. We then define
\be 
J^a=\mathscr{J}^a+J^{\text{(f)}a}\ , \qquad K^{(+)a}=\mathscr{K}^{(+)a}+K^{\text{(f)}(+)a}\ ,
\ee
which can be checked to agree with \eqref{eq:supersymmetric current a}--\eqref{eq:supersymmetric current d}. 
Next we introduce a Wakimoto representation for $\mathfrak{su}(2)_{k^--2}$ in terms of a $\beta\gamma$ system\footnote{In order to distinguish this from the $\beta\gamma$ system of the superconformal ghost, see eq.~(\ref{superconf}), we denote the relevant fields here with a hat.} together with a free boson $\partial\hat{\chi}$, see e.g.\ \cite{DiFrancesco:1997nk}. Then 
the remaining elements of the Borel subalgebra are 
\begin{align}
S^{\alpha\beta+}&=p^{\alpha\beta}-\frac{k^+}{2(k^++k^-)} (\theta^{\alpha\beta}\hat\beta)\ , \label{S-}\\
K^{(-)+}&=\hat\beta\ ,  \label{K-+}\\
K^{(-)3}&= \sqrt{\frac{k^-}{2}} \partial \hat\chi+(\hat\beta\hat\gamma)+\frac{1}{2} \epsilon_{\alpha\gamma}\epsilon_{\beta\delta} (p^{\alpha\beta}\theta^{\gamma\delta})\ , \label{K-3}
\end{align}
where the explicit form of $K^{(-)3}$ is obtained by demanding the OPEs of $\mathfrak{d}(2,1;\alpha)_k$.

\subsection{The complete algebra}
The remaining fields are much more complicated, but they can be found by a direct computation and are uniquely determined. Explicitly they are given as 
\begin{subequations}
\begin{align}
K^{(-)-}&=-(\hat\beta\hat\gamma\hat\gamma)-k^- \partial \hat\gamma-\sqrt{2k^-} (\partial \chi \hat\gamma)-\epsilon_{\alpha\gamma}\epsilon_{\beta\delta}(p^{\alpha\beta}\theta^{\gamma\delta} \hat\gamma)\nonumber\\
&\qquad-\frac{k^+(k^++2k^-)}{2(k^++k^-)^2} \left(\theta^{++}\theta^{+-}\theta^{-+}\theta^{--}\hat\beta\right)-\frac{k^+(k^-+1)}{2(k^++k^-)} \epsilon_{\alpha\gamma}\epsilon_{\beta\delta} \left(\theta^{\alpha\beta} \partial \theta^{\gamma\delta}\right)\nonumber\\
&\qquad+\frac{1}{2} c_a\tensor{(\sigma_a)}{_{\alpha\gamma}} \epsilon_{\beta\delta}\left(\theta^{\alpha\beta}\theta^{\gamma\delta} \left(\mathscr{J}^{(+)a}+\tfrac{1}{3}J^{\text{(f)}(+)a}\right)\right) \nonumber\\
&\qquad-\frac{k^-}{2(k^++k^-)} \epsilon_{\alpha\gamma} \tensor{(\sigma_a)}{_{\beta\delta}} \left(\theta^{\alpha\beta}\theta^{\gamma\delta} \left(\mathscr{K}^{(+)a}+\tfrac{1}{3}K^{\text{(f)}(+)a}\right)\right)\ ,
\label{K--} \\
S^{\alpha\beta-}&=\frac{k^+\sqrt{k^-}}{\sqrt{2}(k^++k^-)} \big( \theta^{\alpha\beta}\partial \hat\chi\big)+\frac{k^+}{2(k^++k^-)} \big(\theta^{\alpha\beta}\hat\beta\hat\gamma\big)-\big(p^{\alpha\beta}\hat\gamma\big) \nonumber\\
&\qquad+c_a \tensor{(\sigma_a)}{^\alpha_\gamma} \left(\theta^{\gamma\beta}\left(\mathscr{J}^a+\frac{3k^++2k^-}{4(k^++k^-)} J^{\text{(f)}a} \right)\right)\nonumber\\
&\qquad+\tensor{(\sigma_a)}{^\beta_\gamma} \left(\theta^{\alpha\gamma}\left(-\frac{k^-}{k^++k^-}\mathscr{K}^{(+)a}+\frac{k^+-2k^-}{4(k^++k^-)} K^{\text{(f)}(+)a} \right)\right)\nonumber\\
&\qquad+\frac{k^+(2k^-+1)}{2(k^++k^-)} \partial \theta^{\alpha\beta}-\frac{k^+(k^++2k^-)}{12(k^++k^-)^2} \epsilon_{\gamma\mu}\epsilon_{\delta\nu}\theta^{\alpha\gamma}\theta^{\delta\mu}\theta^{\nu\beta}\ .
\end{align}
\end{subequations}
The energy-momentum tensor becomes in terms of the defining fields
\begin{multline}
T=\frac{1}{k}\left(-\mathscr{J}^3\mathscr{J}^3+\tfrac{1}{2}\left(\mathscr{J}^+\mathscr{J}^-+\mathscr{J}^-\mathscr{J}^+\right)\right)+\frac{1}{2}(\partial \hat\chi\partial \hat\chi)+\frac{\partial^2 \hat\chi}{\sqrt{2k^-}}-(\hat\beta\partial \hat\gamma)\\
+\frac{1}{k^+}\left(\mathscr{K}^{(+)3}\mathscr{K}^{(+)3}+\tfrac{1}{2}\left(\mathscr{K}^{(+)+}\mathscr{K}^{(+)-}+\mathscr{K}^{(+)-}\mathscr{K}^{(+)+}\right)\right)-\epsilon_{\alpha\gamma}\epsilon_{\beta\delta} (p^{\alpha\beta}\theta^{\gamma\delta}) \ ,
\end{multline}
which is the standard energy-momentum tensor of $\mathfrak{sl}(2,\mathds{R})_{k+2} \oplus \mathfrak{su}(2)_{k^+-2} \oplus \mathfrak{su}(2)_{k^--2}$, together with the four pairs of topologically twisted fermions.

We should note that, in the limit $k^- \to \infty$, the above formulae lead to the construction for $\mathfrak{psu}(1,1|2)_k$ \cite{Bars:1990hx, Berkovits:1999im, Gotz:2006qp, Eberhardt:2019qcl}.

\section{The short representation of \texorpdfstring{$\boldsymbol{\mathfrak{d}(2,1;\alpha)}$}{d(2,1;alpha)}}
In this Appendix, we will display the short representation \eqref{eq:d21alpha short multiplet} explicitly. We will denote the states that appear by 
\begin{align} 
&\ket{m,m_+,m_-}\ , \qquad m \in \mathds{Z}+\lambda\ , \qquad m_+\in \{-\tfrac{1}{2},\tfrac{1}{2}\}\ ,\qquad m_- \in \{-\tfrac{\ell}{2}, \dots,\tfrac{\ell}{2}\}\ , \\
&\ket{m,0,m_-,\pm}\ , \qquad m \in \mathds{Z}+\tfrac{1}{2}+\lambda\ , \qquad m_- \in \{-\tfrac{\ell\pm 1}{2},\dots,\tfrac{\ell \pm 1}{2}\}\ .
\end{align}
For the action of the bosonic subalgebra we choose the conventions that for $\mathfrak{sl}(2,\mathds{R})$ we have 
\be
J^3_0 \ket{j,m} = m\, \ket{j,m} \qquad J^\pm_0 \ket{j,m}=\big(m\pm j\big)\ket{j,m\pm 1}\ , 
\ee
while for the spin $\ell$ representation of $\mathfrak{su}(2)$ we set
\be 
K^3_0 \ket{\ell,m}=m \ket{\ell,m}\ , \qquad K^\pm_0 \ket{ \ell, m}=(\ell\mp m)\ \ket{\ell, m\pm 1}\ .
\ee
The states are then not unit normalised, but this convention is nevertheless convenient. The action of the supercharges is 
\begin{subequations}
\begin{align}
S^{\alpha\beta\gamma}_0 \ket{m,m_+,m_-}&=\frac{\alpha \epsilon^{\beta,2m_+}}{\sqrt{2\ell+1}}\Big(-\ket{m+\tfrac{\alpha}{2},0,m_-+\tfrac{\gamma}{2},+}\nonumber\\
&\qquad+\big(m+\alpha j\big)\big(\gamma \ell-m_-\big) \ket{m+\tfrac{\alpha}{2},0,m_-+\tfrac{\gamma}{2},-}\Big)\ , \\
S^{\alpha\beta\gamma}_0 \ket{m,0,m_-,+}&=\frac{\alpha\left(m+\alpha\left(j-\frac{1}{2}\right)\right)\left(\gamma\left(\ell+\frac{1}{2}\right)-m_-\right)}{\sqrt{2\ell+1}} \ket{m+\tfrac{\alpha}{2},\tfrac{\beta}{2},m_-+\tfrac{\gamma}{2}}\ , \nonumber \\
S^{\alpha\beta\gamma}_0 \ket{m,0,m_-,-}&=\frac{\alpha}{\sqrt{2\ell+1}} \ket{m+\tfrac{\alpha}{2},\tfrac{\beta}{2},m_-+\tfrac{\gamma}{2}}\ .
\end{align}
\end{subequations}
One can check directly that this defines a representation of $\mathfrak{d}(2,1;\alpha)$ (in the conventions of eqs.\ \eqref{eq:d21alpha commutation relations a}--\eqref{eq:d21alpha commutation relations j}), provided that
\be 
j=(1-\gamma)\left(\ell+\tfrac{1}{2}\right)+\tfrac{1}{2}\ ,
\ee
see eq.~\eqref{eq:j shortening condition} in the main text.

Note that for $\ell=0$, the states $\ket{m,0,m_-,-}$ are never produced by the action of the generators, and hence can be decoupled from the multiplet, see eq.~(\ref{ultrashort}). Similarly, the states
\begin{align} 
&\ket{m,m_+,m_-}\ , & m &\in j+\mathds{Z}_{\ge 0}\ ,\\
&\ket{m,0,m_-,\pm}\ , & m &\in j \mp \tfrac{1}{2}+\mathds{Z}_{\ge 0}\ 
\end{align}
form a subrepresentation, which is the discrete representation on which $\mathscr{G}_{>,+}$ is based, while the states
\begin{align} 
&\ket{m,m_+,m_-}\ , & m &\in -j+\mathds{Z}_{\le 0}\ ,\\
&\ket{m,0,m_-,\pm}\ , & m &\in -j \pm \tfrac{1}{2}+\mathds{Z}_{\le 0}\ 
\end{align}
form the discrete subrepresentation which gives rise to $\mathscr{G}_{<,-}$ in the affine algebra. In order to obtain the other two discrete representations (which give rise to $\mathscr{G}_{>,-}$ and $\mathscr{G}_{<,+}$ in the affine algebra), one has to replace $j$ by $1-j$ in the continuous representations of $\mathfrak{sl}(2,\mathds{R})$.

\section{Characters and modular properties at \texorpdfstring{$\boldsymbol{k^+=1}$}{k+=1}} \label{app:characters}
In this Appendix, we determine the characters of $\mathfrak{d}(2,1;\alpha)$ for $k^+=1$. To do so, we exploit the conformal embedding (which only exists for $k^+=1$) \cite{Bowcock:1999uy, Creutzig:2017uxh}
\be
\mathfrak{sl}(2,\mathds{R})_k \oplus \mathfrak{su}(2)_{1} \oplus \mathfrak{su}(2)_{k^-} \subset \mathfrak{d}(2,1;\alpha)_k\ . \label{eq:d21alpha conformal embedding}
\ee
The modular properties of the characters will allow us to prove modular invariance of the full spectrum, see eq.~(\ref{eq:Hilbert space Grothendieck}). It will also allow us to compute the fusion rules via the Verlinde formula. We will use the conventions of the main text, so in particular
\be 
k^+=1\ , \quad k^-=\alpha=\kappa+1\ , \quad k=\gamma=\frac{\kappa+1}{\kappa+2}\ .
\ee

\subsection{Admissible \texorpdfstring{$\mathfrak{su}(2)$}{su(2)} WZW-models} \label{subsec:admissible su(2)}
We begin by discussing $\mathfrak{sl}(2,\mathds{R})_k$ at level $k=\tfrac{\kappa+1}{\kappa+2}$. On the level of the algebra (i.e.~disregarding the hermitian structure, which does not matter for the calculation of the characters), this algebra is isomorphic to 
\be 
\mathfrak{su}(2)_{-\frac{\kappa+1}{\kappa+2}}\ .
\ee
While the level of the $\mathfrak{su}(2)$ algebra is negative (and hence the model is non-unitary), the level is what is called admissible, see e.g.\ \cite{DiFrancesco:1997nk, Kac:1979fz}. (In the following we will mostly follow the notation of \cite{Creutzig:2013yca}.) To explain what this means we write 
\be 
-\frac{\kappa+1}{\kappa+2}+2 =\frac{\kappa+3}{\kappa+2}= \frac{p}{q}\ ,
\ee
where $p=\kappa+3$ and $q=\kappa+2$. Admissibility amounts to the condition that $\mathrm{gcd}(p,q)=1$ and $p \in \mathds{Z}_{\ge 2}$, $q \in \mathds{Z}_{\ge 1}$, all of which are obviously satisfied. The fact that the algebra is admissible means that the vacuum representation has a null-vector at level $(p-1)q=(\kappa+2)^2$.  This singular vector restricts the representation theory of the admissible $\mathfrak{su}(2)$ WZW-model significantly.
%Continuous representations and their spectrally flowed images are not conventional highest weight representations. The continuous representation is a so-called relaxed highest weight representation, meaning that the zero mode representation is not highest weight, but there still exists a state which is annihilated by all positive affine modes. 
The admissible irreducible representation of $\mathfrak{su}(2)$ at this level are \cite{Adamovic:1995aa}, see also \cite{Mukhi:1989bp}
\begin{align} 
&\mathscr{L}_{r,0}\ ,\qquad r\in \{1,\dots,\kappa+2\}\ , \\
&\mathscr{D}_{r,s}^\pm\ , \qquad r \in \{1,\dots,\kappa+2\}\ , \qquad s \in \{1,\dots,\kappa+1\}\ , \\
&\mathscr{E}_{r,s,\lambda}\ , \qquad r \in \{1,\dots,\kappa+2\}\ , \qquad s \in \{1,\dots,\kappa+1\}\ ,
\end{align}
where $\lambda \in [0,1)$ encodes the quantisation of the $J^3_0$-eigenvalue mod $\mathds{Z}$. We denote the conformal dimension of the ground states by $\Delta_{r,s}$, where
\be 
\Delta_{r,s}=\frac{((\kappa+2)r-(\kappa+3)s)^2-(\kappa+3)^2}{4(\kappa+2)(\kappa+3)}\ , \label{eq:admissible conformal weight}
\ee
and we have the field identification $\Delta_{r,s}=\Delta_{\kappa+3-r,\kappa+2-s}$. As a consequence $\mathscr{E}_{r,s,\lambda}$ and $\mathscr{E}_{\kappa+3-r,\kappa+2-s,\lambda}$ describe the same representation.

The characters of the representation $\mathscr{E}_{r,s,\lambda}$ were determined in \cite{Creutzig:2013yca} 
\be 
\mathrm{ch}[\mathscr{E}_{r,s,\lambda}](t;\tau)=\frac{\chi^{\text{Vir}}_{r,s}(\tau)}{\eta(\tau)^2}\sum_{m \in \mathds{Z}+\lambda} x^m\ ,\label{eq:sl2 cont rep characters}
\ee
where $x=\mathrm{e}^{2\pi i t}$ is the chemical potential of $\mathfrak{sl}(2,\mathds{R})$. Here, $\chi^{\text{Vir}}_{r,s}(\tau)$ is the character of the representation $(r,s)$ of the corresponding Virasoro minimal model of central charge
\be 
c^\text{Vir}=1-\frac{6}{(\kappa+2)(\kappa+3)}\ ,
\ee
which are explicitly \cite{DiFrancesco:1997nk}\footnote{The modular parameter $q=\mathrm{e}^{2\pi i \tau}$ should not be confused with the parameter $q$ of the Virasoro minimal model.}
\be 
\chi^{\text{Vir}}_{r,s}(\tau)=\frac{q^{\Delta_{r,s}^{\text{Vir}}-\frac{1}{24}(c^\text{Vir}-1)}}{\eta(\tau)} \sum_{\ell \in \mathds{Z}}\big( q^{\ell(pq\ell+q r-p s)}-q^{(p\ell-r)(q\ell-s)} \big)\ . \label{eq:Virasoro minimal characters}
\ee
The expression in (\ref{eq:sl2 cont rep characters}) is a bit formal because of the infinite sum over $m$, which converges nowhere (and will lead to a sum over delta functions as in \cite{Eberhardt:2018ouy}). 

The theory has again a spectral flow symmetry, which we shall denote by $\sigma$. It acts on the representations as \cite{Creutzig:2013yca}
\begin{subequations}
\begin{align} 
\sigma(\mathscr{L}_{r,0})&=\mathscr{D}^+_{\kappa+3-r,\kappa+1}\ , \label{eq:sl2R spectral flow identification a}\\
\sigma^{-1}(\mathscr{L}_{r,0})&=\mathscr{D}^-_{\kappa+3-r,\kappa+1}\ , \label{eq:sl2R spectral flow identification b}\\
\sigma(\mathscr{D}^-_{r,s})&=\mathscr{D}^+_{\kappa+3-r,\kappa+1-s}\ . \label{eq:sl2R spectral flow identification c}
\end{align}
\end{subequations}
Finally, there are short exact sequences analogous to \eqref{eq:SES 1}--\eqref{eq:SES 4}, which read
\begin{subequations}
\begin{align}
0 &\longrightarrow \mathscr{D}^+_{r,s} \longrightarrow \mathscr{E}_{r,s,\lambda_{r,s}} \longrightarrow \mathscr{D}^-_{\kappa+3-r,\kappa+2-s} \longrightarrow 0\ , \label{eq:sl2 SES 1}\\
0 &\longrightarrow \mathscr{D}^-_{r,s} \longrightarrow \mathscr{E}_{r,s,-\lambda_{r,s}} \longrightarrow \mathscr{D}^+_{\kappa+3-r,\kappa+2-s} \longrightarrow 0\ ,\label{eq:sl2 SES 2}
\end{align}
\end{subequations}
where
\be 
\lambda_{r,s}=\frac{r-1}{2}-\frac{\kappa+3}{2(\kappa+2)} s\ .
\ee
Hence, for $\lambda=\pm \lambda_{r,s}$, $\mathscr{E}_{r,s,\lambda}$ becomes indecomposable.

\subsection{The branching rules of \texorpdfstring{$\mathfrak{d}(2,1;\alpha=\kappa+1)_k$}{d(2,1;alpha)k} into its bosonic subalgebra} \label{subsec:branching rules}

After this interlude we now return to the case of  $\mathfrak{d}(2,1;\alpha=\kappa+1)_k$ with $k^+=1$. We want to understand the branching rules of the representations of $\mathfrak{d}(2,1;\alpha)_k$ under the conformal embedding (\ref{eq:d21alpha conformal embedding}). For the case of the vacuum representation of $\mathfrak{d}(2,1;\alpha)_k$ (and generic $\kappa \ne \mathds{Q}$), this was worked out in \cite{Creutzig:2017uxh}. This result can be generalised to $\kappa\in\mathds{Z}_{\geq 0}$, and we find 
\be 
\mathscr{L}\cong\bigoplus_{r=0}^{\kappa+2} \mathscr{L}_{r,0}\otimes\mathscr{M}_{r \bmod 2}^{(1)} \otimes \mathscr{M}_r^{(\kappa+1)}\ ,  \label{eq:d21alpha vacuum decomposition}
\ee
where $\mathscr{M}_{2\ell+1}^{(\kappa+1)}$ denotes the spin $\ell$ representation of $\mathfrak{su}(2)_{\kappa+1}$ and similarly for the level $1$ factor. 
%The left hand side of this equation denotes as in the main text the vacuum module of $\mathfrak{d}(2,1;\alpha)_k$ at $k^+=1$, whereas on the right hand side the modules of the three terms in the conformal embedding \eqref{eq:d21alpha conformal embedding} appear. We have denoted the $\mathfrak{su}(2)_{\kappa+1}$ (and $\mathfrak{su}(2)_1$) by the dimensionality of the ground state representation, i.e.\ $\mathscr{M}_{2\ell+1}^{\kappa+1}$ denotes the spin $\ell$ representation.

By exploiting the spectral flow rules \eqref{eq:spectral flow identification a}--\eqref{eq:spectral flow identification d} and \eqref{eq:sl2R spectral flow identification a}--\eqref{eq:sl2R spectral flow identification c} together with the short exact sequences \eqref{eq:SES 1}--\eqref{eq:SES 4} and \eqref{eq:sl2 SES 1}-- \eqref{eq:sl2 SES 2}, we can read off from this the branching rules of all modules,
\begin{subequations}
\begin{align}
\mathscr{G}^\ell_{<,+}&\cong \bigoplus_{r=1}^{\kappa+2} \mathscr{D}^+_{\kappa+3-r,\kappa+1-2\ell} \otimes \mathscr{M}_{r+2\ell+1 \bmod 2}^{(1)} \otimes \mathscr{M}_r^{(\kappa+1)}\ , \\
\mathscr{G}^\ell_{<,-}&\cong \bigoplus_{r=1}^{\kappa+2} \mathscr{D}^-_{r,2\ell+1} \otimes \mathscr{M}_{r+2\ell+1 \bmod 2}^{(1)} \otimes \mathscr{M}_r^{(\kappa+1)}\ , \\
\mathscr{G}^\ell_{>,+}&\cong \bigoplus_{r=1}^{\kappa+2} \mathscr{D}^+_{r,2\ell+1} \otimes \mathscr{M}_{r+2\ell+1 \bmod 2}^{(1)} \otimes \mathscr{M}_r^{(\kappa+1)}\ , \\
\mathscr{G}^\ell_{>,-}&\cong \bigoplus_{r=1}^{\kappa+2} \mathscr{D}^-_{\kappa+3-r,\kappa+1-2\ell} \otimes \mathscr{M}_{r+2\ell+1 \bmod 2}^{(1)} \otimes \mathscr{M}_r^{(\kappa+1)}\ , \\
\mathscr{F}^\ell_\lambda&\cong \bigoplus_{r=1}^{\kappa+2} \mathscr{E}_{r,2\ell+1,\lambda+\ell+\frac{r+1}{2}} \otimes \mathscr{M}_{r+2\ell+1 \bmod 2}^{(1)} \otimes \mathscr{M}_r^{(\kappa+1)}\ .
\end{align}
\end{subequations}

\subsection{The characters} \label{subapp:characters}

Given that we know the characters of the individual factors of the above branchings, it is now straightforward to compute the complete characters of $\mathfrak{d}(2,1;\alpha)_k$ for $k^+=1$. In particular, the character of the spin-$\ell$ representation of $\mathfrak{su}(2)_\kappa$ is explicitly given as 
\be 
\mathrm{ch}[\mathscr{M}^{(\kappa)}_{2\ell+1}](v;\tau)=\frac{\Theta_{2\ell+1}^{(\kappa+2)}(v;\tau)-\Theta_{-2\ell-1}^{(\kappa+2)}(v;\tau)}{\Theta_1^{(2)}(v;\tau)-\Theta_{-1}^{(2)}(v;\tau)}\ ,
\ee
where $v$ is the chemical potential of $\mathfrak{su}(2)_{k^-}$ with $z=\mathrm{e}^{2\pi i v}$, and the theta functions are explicitly 
\be 
\Theta_{m}^{(k)}(v;\tau)=\sum_{n \in \mathds{Z}+\frac{m}{2k}} q^{kn^2}z^{kn}\ .
\ee
We introduce similarly chemical potentials $t$ and $u$ for the subalgebras $\mathfrak{sl}(2,\mathds{R})$, and $\mathfrak{su}(2)_{k^+}$, respectively,\footnote{In particular, $t$ will play the role of the modular parameter in the dual CFT.} and define
\be 
x=\mathrm{e}^{2\pi i t}\ , \quad y=\mathrm{e}^{2\pi i u}\ , \quad z=\mathrm{e}^{2\pi i v}\ , \quad q=\mathrm{e}^{2\pi i \tau}\ ,
\ee
where $\tau$ is the modular parameter of the worldsheet. The characters we are interested in are 
\be 
\mathrm{ch}\big[\sigma^w\big(\mathscr{F}^\ell_\lambda\big)\big](t,u,v;\tau)\equiv\mathrm{tr}_{\sigma^w(\mathscr{F}^\ell_\lambda)}\Big(x^{J_0^3} y^{K^{(+)3}_0}z^{K^{(-)3}_0} q^{L_0-\frac{c}{24}}\Big)\ .
\ee
%We could define this of course also for the other representations $\mathscr{L}$, etc.\ but we shall not have need for it.
Because of the zero-modes, they contain in particular a sum of the form
\begin{align} 
\sum_{m \in \mathds{Z}+\lambda} \mathrm{e}^{2\pi i (t-w\tau) m}&=\mathrm{e}^{2\pi i (t-w\tau) \lambda} \sum_{m \in \mathds{Z}}\mathrm{e}^{2\pi i (t-w\tau) m} \\
&=\mathrm{e}^{2\pi i (t-w\tau) \lambda}  \sum_{n \in \mathds{Z}} \delta(t-n-w\tau)\\
&=\sum_{n \in \mathds{Z}} \mathrm{e}^{2\pi i n \lambda} \delta(t-n-w\tau)\ . \label{eq:convert infinite sum to delta function}
\end{align}
%Geometrically, this means that the modular parameter $t$ of the boundary torus is expressed as $t=n+w\tau$ and hence the worldsheet torus is mapped holomorphically on the boundary torus.
We want to show that the full character can be written as 
\begin{multline}
\mathrm{ch}[\sigma^w(\mathscr{F}^\ell_\lambda)]=q^{\frac{w^2}{4(\kappa+2)}}x^{\frac{\kappa+1}{2(\kappa+2)}w} y^{\frac{w}{2}}\sum_{n \in \mathds{Z}}\mathrm{e}^{2\pi i (\lambda+\frac{1}{2}) n} \delta(t-n - w\tau)\\
\times\frac{\vartheta_1\big(\frac{t+u+v}{2};\tau\big)\vartheta_1\big(\frac{t+u-v}{2};\tau\big)}{\eta(\tau)^4} \mathrm{ch}[\mathscr{M}^{(\kappa)}_{2\ell+1}](v;\tau)\ . \label{eq:character continuous representation}
\end{multline}
We should stress that it is, from this perspective, somewhat surprising that on the right-hand-side an $\mathfrak{su}(2)_\kappa$ character (rather than an $\mathfrak{su}(2)_{\kappa+1}$ character) appears. We should also mention that, with the exception of the additional $\mathfrak{su}(2)_\kappa$ character, this formula is almost identical to the $\mathfrak{psu}(1,1|2)_1$ characters computed in \cite{Eberhardt:2018ouy}.
\medskip

In order to prove (\ref{eq:character continuous representation}) it is sufficient to consider the 
unflowed sector, since the spectral flow \eqref{eq:spectral flow a}--\eqref{eq:spectral flow f} gives immediately the generalisation to any $w$. To start with we rewrite the characters of $\mathfrak{su}(2)_1$ as 
\be 
\mathrm{ch}[\mathscr{M}^{(1)}_m](u;\tau)=\frac{1}{\eta(\tau)} \sum_{n \in \mathds{Z}+\frac{m+1}{2}} q^{n^2} y^n=\frac{1}{\eta(\tau)}\vartheta\!\begin{bmatrix}
\frac{m+1}{2} \\ 0
\end{bmatrix}\!(u;2\tau)\ ,
\ee
where we have used that $\mathfrak{su}(2)_1$ is equivalent to a free boson at the self-dual radius. We also rewrite the Virasoro minimal model characters \eqref{eq:Virasoro minimal characters} as
\begin{align} 
\chi^{\text{Vir}}_{r,s}(\tau)&=\frac{1}{\eta(\tau)} \sum_{\ell \in \mathds{Z}} \Big( q^{\frac{(2\ell(\kappa+2)(\kappa+3)-s(\kappa+3)+r(\kappa+2))^2}{4(\kappa+2)(\kappa+3)}}-q^{\frac{(2\ell(\kappa+2)(\kappa+3)-s(\kappa+3)-r(\kappa+2))^2}{4(\kappa+2)(\kappa+3)}}\Big) \\
&=\frac{1}{\eta(\tau)} \sum_{\varepsilon=\pm} \quad\varepsilon\!\!\!\!\! \sum_{m \in \mathds{Z}+\frac{s}{2(\kappa+2)}-\frac{\varepsilon r}{2(\kappa+3)}} q^{(\kappa+2)(\kappa+3)m^2}\ .
\end{align}
It then follows from a direct computation that 
\begin{align}
\mathrm{ch}[\mathscr{F}^\ell_\lambda](t,u,v;\tau)&=\sum_{r=1}^{\kappa+2} \sum_{m \in \mathds{Z}+\lambda+\ell+\frac{r+1}{2}} \!\!\!\!\!\!\!\!\!\! \frac{x^m \chi^\text{Vir}_{r,2\ell+1}(\tau)\mathrm{ch}[\mathscr{M}^{(1)}_{r+2\ell+1 \bmod 2}](u;\tau)\mathrm{ch}[\mathscr{M}^{(\kappa+1)}_r](v;\tau)}{\eta(\tau)^2 } \nonumber \\
&=\sum_{r=1}^{\kappa+2}\frac{\mathrm{ch}[\mathscr{M}_{r+2\ell+1 \bmod 2}^{(1)}](u;\tau)}{\eta(\tau)^3\big(\Theta_1^{(2)}(v;\tau)-\Theta_{-1}^{(2)}(v;\tau)\big)}\sum_{m\in \mathds{Z}+\lambda+\ell+\frac{r+1}{2}} x^m  \nonumber\\
&\qquad\times\sum_{\varepsilon,\, \eta=\pm} \quad\varepsilon\eta\!\!\!\!\!\!\!\!\sum_{\genfrac{}{}{0pt}{}{n \in \mathds{Z}+\frac{\eta r}{2(\kappa+3)}}{p \in \mathds{Z}+\frac{2\ell+1}{2(\kappa+2)}-\frac{\varepsilon r}{2(\kappa+3)}}}\!\!\!\!\! q^{(\kappa+3)n^2+(\kappa+2)(\kappa+3)p^2} z^{(\kappa+3)n} \ . \label{eq:character proof 1}
\end{align}
The expression remains unchanged when extending the sum over $r$ from $1$ to $2\kappa+6$ and dividing the result by a factor of 2. Next we want to rewrite the sums over $n$ and $p$ by introducing 
\be 
a=\varepsilon \eta p+n\in \mathds{Z}+ \frac{\varepsilon\eta(2\ell+1)}{2(\kappa+2)}\ , \qquad b=n-\varepsilon\eta(\kappa+2)p \in \mathds{Z}+\ell+\frac{r+1}{2}\ .
\ee
The determinant of the matrix describing this change of variables is $\kappa+3$, which can be absorbed by restricting the summation over the $2(\kappa+3)$ values of $r$ to just $r=1,2$. Thus 
\eqref{eq:character proof 1} becomes
\begin{align}
\mathrm{ch}[\mathscr{F}^\ell_\lambda](t,u,v;\tau)&=\sum_{r=1,\,2} \frac{\mathrm{ch}[\mathscr{M}_{r+2\ell+1 \bmod 2}^{(1)}](u;\tau)}{2\eta(\tau)^3\big(\Theta_1^{(2)}(v;\tau)-\Theta_{-1}^{(2)}(v;\tau)\big)}\sum_{m\in \mathds{Z}+\lambda+\ell+\frac{r+1}{2}} x^m \nonumber\\
&\qquad\times \sum_{\varepsilon,\, \eta=\pm} \varepsilon \eta\sum_{\genfrac{}{}{0pt}{}{a \in \mathds{Z}+ \frac{\varepsilon\eta(2\ell+1)}{2(\kappa+2)}}{b \in \mathds{Z}+\ell+\frac{r+1}{2}}} \!\!\! q^{(\kappa+2)a^2+b^2} z^{(\kappa+2)a+b} \\
&=\sum_{r=1,\,2}\sum_{m\in \mathds{Z}+\lambda+\ell+\frac{r+1}{2}} x^m \ \frac{\Theta^{(\kappa+2)}_{2\ell+1}(v;\tau)-\Theta^{(\kappa+2)}_{-2\ell-1}(v;\tau)}{\eta(\tau)^2\big(\Theta_1^{(2)}(v;\tau)-\Theta_{-1}^{(2)}(v;\tau)\big)} \nonumber\\
&\qquad\times \mathrm{ch}[\mathscr{M}_{r+2\ell+1 \bmod 2}^{(1)}](u;\tau)\mathrm{ch}[\mathscr{M}_{r+2\ell \bmod 2}^{(1)}](v;\tau)\\
&=\frac{\mathrm{ch}[\mathscr{M}^{(\kappa)}_{2\ell+1}](v;\tau)}{\eta(\tau)^2} \bigg(\sum_{m \in \mathds{Z}+\lambda}  x^m \mathrm{ch}[\mathscr{M}_{2}^{(1)}](u;\tau)\mathrm{ch}[\mathscr{M}_{1}^{(1)}](v;\tau)\nonumber\\
&\qquad\qquad+\sum_{m \in \mathds{Z}+\lambda+\frac{1}{2}}  x^m \mathrm{ch}[\mathscr{M}_{1}^{(1)}](u;\tau)\mathrm{ch}[\mathscr{M}_{2}^{(1)}](v;\tau)\bigg) \ .
\end{align}
Here, we have first used the fact that the expression only depends on the product $\varepsilon \eta$ and hence we can trivially perform one of the two sums. In the final expression we have rewritten the result in terms of affine $\mathfrak{su}(2)_\kappa$ characters.

Next we rewrite the two $\mathfrak{su}(2)_1$ characters in terms of free fermion characters, i.e.\ we  use \eqref{eq:equivalence four fermions and su2su2 a} and \eqref{eq:equivalence four fermions and su2su2 b},  
%
%(or equivalently an $\mathfrak{so}(4)_1$ character) in the expression. Since we can give a free field realization of $\mathfrak{so}(4)_1$ in terms of four free fermions, it is possible to rewrite this combination as a free fermion character. Using \eqref{eq:equivalence four fermions and su2su2 a} and \eqref{eq:equivalence four fermions and su2su2 b}, we have
\begin{align} 
\mathrm{ch}[\mathscr{M}_{2}^{(1)}](u;\tau)\mathrm{ch}[\mathscr{M}_{1}^{(1)}](v;\tau)=\frac{\vartheta_2(\tfrac{u+v}{2};\tau)\vartheta_2(\tfrac{u-v}{2};\tau)-\vartheta_1(\tfrac{u+v}{2};\tau)\vartheta_1(\tfrac{u-v}{2};\tau)}{2\,\eta(\tau)^2}\ , \\
\mathrm{ch}[\mathscr{M}_{1}^{(1)}](u;\tau)\mathrm{ch}[\mathscr{M}_{2}^{(1)}](v;\tau)=\frac{\vartheta_2(\tfrac{u+v}{2};\tau)\vartheta_2(\tfrac{u-v}{2};\tau)+\vartheta_1(\tfrac{u+v}{2};\tau)\vartheta_1(\tfrac{u-v}{2};\tau)}{2\,\eta(\tau)^2}\ .
\end{align}
Thus we finally arrive the result 
\begin{multline}
\sum_{m \in \mathds{Z}+\lambda} x^m \mathrm{ch}[\mathscr{M}_{2}^{(1)}](u;\tau)\mathrm{ch}[\mathscr{M}_{1}^{(1)}](v;\tau)+\sum_{m\in \mathds{Z}+\lambda+\frac{1}{2}} x^m \mathrm{ch}[\mathscr{M}_{1}^{(1)}](u;\tau)\mathrm{ch}[\mathscr{M}_{2}^{(1)}](v;\tau) \\
=\sum_{m \in \mathds{Z}+\lambda+\frac{1}{2}}  x^m\ \frac{\vartheta_2\big(\frac{t+u+v}{2};\tau\big)\vartheta_2\big(\frac{t+u-v}{2};\tau\big)}{\eta(\tau)^2}\ ,
\end{multline}
which reproduces \eqref{eq:character continuous representation} upon turning the infinite sum over $m$ into a delta function as in \eqref{eq:convert infinite sum to delta function}.

\subsection{Modular properties} \label{subapp:modular properties}

Next we want to study the modular behaviour of the characters \eqref{eq:character continuous representation}. To obtain good modular properties, we insert a $(-1)^{\mathrm{F}}$ in the character, which amounts to the replacement $\vartheta_2 \longrightarrow \vartheta_1$. Moreover, to match the conventions of \cite{Eberhardt:2018ouy}, we include a $(-1)^w$ in the character. This merely defines what state in the representation is counted as being fermionic and which as bosonic. We have indicated these changes by a tilde in the character. %We stress that we are treating the character as a formal object and not as a meromorphic function. Hence the following manipulations will be somewhat formal. 
Our calculations follows \cite{Eberhardt:2018ouy} and are inspired by \cite{Maldacena:2000hw, Creutzig:2012sd}. 
For the $S$-modular transformation we find 
\begin{align}
& \tilde{\mathrm{ch}}[\sigma^w(\mathscr{F}^\ell_\lambda)](t,u,v;\tau)\rightarrow\ \mathrm{e}^{\frac{\pi i}{2\tau}(\frac{\kappa+1}{\kappa+2}t^2-u^2-(\kappa+1)v^2)}\tilde{\mathrm{ch}}[\sigma^w(\mathscr{F}^\ell_\lambda)]\big(\tfrac{t}{\tau},\tfrac{u}{\tau},\tfrac{v}{\tau};-\tfrac{1}{\tau}\big) \\[4pt]
& \quad\qquad =\frac{\mathrm{sgn}(\mathrm{Re}(\tau))}{i\tau} \mathrm{e}^{\frac{i \pi (t+w)}{2(\kappa+2)\tau}(-w+2u(\kappa+2)+t(2\kappa+3))}(-1)^w\sum_{m \in \mathds{Z}} \mathrm{e}^{2\pi i m (\lambda+\frac{1}{2})} \delta\Big(\frac{t+w-m \tau}{\tau}\Big)\nonumber\\
&\qquad\qquad \qquad\times\frac{\vartheta_1\big(\frac{t+u+v}{2};\tau\big)\vartheta_1\big(\frac{t+u-v}{2};\tau\big)}{\eta(\tau)^4} \sum_{\ell'=0}^{\frac{\kappa}{2}} S_{\ell\ell'}^{\mathfrak{su}(2)}\chi^{(\ell')}_\kappa(v;\tau)\\
&\quad\qquad =-i\,\mathrm{sgn}(\mathrm{Re}(\tau))\sum_{m \in \mathds{Z}}q^{\frac{m^2}{4(\kappa+2)}} x^{\frac{\kappa+1}{2(\kappa+2)}m} y^{\frac{m}{2}} \mathrm{e}^{-\frac{\pi i m w}{\kappa+2}} \mathrm{e}^{2\pi i m (\lambda+\frac{1}{2})}(-1)^w \delta(t+w-m \tau)\nonumber\\
&\qquad\qquad\qquad \times\frac{\vartheta_1\big(\frac{t+u+v}{2};\tau\big)\vartheta_1\big(\frac{t+u-v}{2};\tau\big)}{\eta(\tau)^4} \sum_{\ell'=0}^{\frac{\kappa}{2}} S_{\ell\ell'}^{\mathfrak{su}(2)}\chi^{(\ell')}_\kappa(v;\tau)\ . \label{eq:S modular transform character}
\end{align}
Here the prefactor $\mathrm{e}^{\frac{\pi i}{2\tau}(\frac{\kappa+1}{\kappa+2}t^2-y^2-(\kappa+1)z^2)}$ comes from the general transformation properties of weak Jacobi forms of index $-k=-\tfrac{\kappa+1}{\kappa+2}$, $k^+=1$ and $k^-=\kappa+1$, respectively, see e.g.\ \cite{Gaberdiel:2012yb}, and we have used the modular transformations of the theta-functions. We have also used the modular properties of the $\mathfrak{su}(2)_\kappa$ characters. In the final step we have set $t=m\tau -w$ (because of the $\delta$ function), and used that both $m$ and $w$ are integers. Finally, as in \cite{Eberhardt:2018ouy} we have inserted the formal identity
\be 
\delta\big(\tfrac{x}{\tau}\big)=\tau\, \mathrm{sgn}(\mathrm{Re}(\tau))\, \delta(x)\ , \label{eq:delta reparametrization}
\ee
which follows by writing
\be 
\delta(x)=\lim_{\epsilon \to 0} \frac{1}{\sqrt{2\pi \epsilon}} \mathrm{e}^{-\frac{x^2}{2\epsilon}}\ .
\ee
Here we have put the branch cut of the square root on the imaginary axis, which is the reason for the jump in  \eqref{eq:delta reparametrization} at this point. Note that the sign cancels out once we combine the left-movers with the right-movers.

The expression \eqref{eq:S modular transform character} can now be written as
\be 
\sum_{w'\in \mathds{Z}} \sum_{\ell'=0}^{\frac{\kappa}{2}}\int_0^1 \mathrm{d}\lambda' \, S_{(w,\lambda,\ell),(w',\lambda',\ell')}\, \, \tilde{\mathrm{ch}}[\sigma^{w'}(\mathscr{F}_{\lambda'})](t,u,v;\tau)\ ,
\ee
with
\be 
S_{(w,\lambda,\ell),(w',\lambda',\ell')}=-i\, \mathrm{sgn}(\mathrm{Re}(\tau)) \, \mathrm{e}^{2\pi i\big(w'\lambda+w\lambda' -\frac{\pi i w w'}{2(\kappa+2)}\big)} S^{\mathfrak{su}(2)}_{\ell \ell'}\ , \label{eq:S matrix}
\ee
thus obtaining the $S$-matrix (\ref{Smatrix}). As in \cite{Creutzig:2012sd}, it is not independent of $\tau$, but this dependence cancels out in physical calculations. The $S$-matrix is (formally) unitary, meaning
\begin{align} 
\sum_{w''\in \mathds{Z}} \sum_{\ell''=0}^{\frac{\kappa}{2}}\int_0^1 \mathrm{d}\lambda'' \, S^\dag_{(w,\lambda,\ell),(w'',\lambda'',\ell'')} S_{(w'',\lambda'',\ell''),(w',\lambda',\ell')}&=\delta_{w,w'}\, \delta(\lambda-\lambda'\bmod 1)\delta_{\ell,\ell'}\ .
\end{align}
Moreover, it is clearly symmetric. These properties suffice to deduce that the diagonal modular invariant is indeed modular invariant.

\subsection{The Verlinde formula} \label{subapp:verlinde}
We now use the formal S-matrix to derive the typical fusion rules using a continuum version of the Verlinde formula; the following derivation is parallel to Appendix~C.6 of \cite{Eberhardt:2018ouy}. For this, we also need the $S$-matrix element of the vacuum with a continuous representation. It follows from the 
exact sequences \eqref{eq:SES 1}--\eqref{eq:SES 4}, together with the identifications under spectral flow \eqref{eq:spectral flow identification a} -- \eqref{eq:spectral flow identification d}, that we have a resolution of the vacuum module as 
\begin{align} 
& &\cdots \longrightarrow &\sigma^{4\kappa+7-2\ell}\big(\mathscr{F}^\ell_{-\lambda_{\ell}}\big) \longrightarrow &\cdots \longrightarrow &\sigma^{3\kappa+7}\big(\mathscr{F}^{\frac{\kappa}{2}}_{-\lambda_{\frac{\kappa}{2}}}\big) \nonumber\\
\longrightarrow &\sigma^{3\kappa+5}\big(\mathscr{F}^{\frac{\kappa}{2}}_{\lambda_{\frac{\kappa}{2}}}\big)\longrightarrow &\cdots\longrightarrow &\sigma^{2\kappa+2\ell+5}\big(\mathscr{F}^\ell_{\lambda_\ell}\big)\longrightarrow &\cdots\longrightarrow &\sigma^{2\kappa+5}\big(\mathscr{F}_{\lambda_0}^0\big)\nonumber\\
\longrightarrow &\sigma^{2\kappa+3} \big(\mathscr{F}^0_{-\lambda_0}\big) \longrightarrow &\cdots \longrightarrow &\sigma^{2\kappa+3-2\ell}\big(\mathscr{F}^\ell_{-\lambda_{\ell}}\big) \longrightarrow &\cdots \longrightarrow &\sigma^{\kappa+3}\big(\mathscr{F}^{\frac{\kappa}{2}}_{-\lambda_{\frac{\kappa}{2}}}\big) \nonumber\\
\longrightarrow &\sigma^{\kappa+1}\big(\mathscr{F}^{\frac{\kappa}{2}}_{\lambda_{\frac{\kappa}{2}}}\big)\longrightarrow &\cdots\longrightarrow &\sigma^{2\ell+1}\big(\mathscr{F}^\ell_{\lambda_\ell}\big)\longrightarrow &\cdots\longrightarrow &\sigma\big(\mathscr{F}_{\lambda_0}^0\big)\longrightarrow \mathscr{L} \longrightarrow 0\ . \label{eq:resolution vacuum}
\end{align}
Thus we can write the vacuum character as 
\begin{multline} 
\mathrm{ch}[\mathscr{L}](t,u,v;\tau)=\sum_{m=0}^\infty \sum_{\ell=0}^{\frac{\kappa}{2}} (-1)^{2\ell} \Big(\mathrm{ch}\big[\sigma^{2\ell+2m(\kappa+2)+1}\big(\mathscr{F}^\ell_{\lambda_\ell}\big)\big](t,u,v;\tau) \\
-\mathrm{ch}\big[\sigma^{2(m+1)(\kappa+2)-2\ell-1}\big(\mathscr{F}^\ell_{-\lambda_\ell}\big)\big](t,u,v;\tau)\Big)\ ,
\end{multline}
and hence find for the $S$-matrix element of the vacuum and a continuous representation
\begin{align}
S_{\text{vac},(w,\lambda,\ell)}=\!\sum_{m=0}^\infty \sum_{j=0}^{\frac{\kappa}{2}}(-1)^{2j}\Big(S_{(2j+2m(\kappa+2)+1,\lambda_j,j),(w,\lambda,\ell)}-S_{(2(m+1)(\kappa+2)-2j-1,-\lambda_j,j),(w,\lambda,\ell)}\Big) .
\end{align}
By using the explicit form of the S-matrix \eqref{eq:S matrix} together with the $\mathfrak{su}(2)_\kappa$ S-matrix
\be 
S^{\mathfrak{su}(2)}_{\ell\ell'}=\sqrt{\frac{2}{\kappa+2}} \sin\Big(\frac{\pi(2\ell+1)(2\ell'+1)}{\kappa+2}\Big)\ ,
\ee
one finds after some algebra
\be 
S_{\text{vac},(w,\lambda,\ell)}=-  \frac{i(-1)^w \mathrm{sgn}(\mathrm{Re}(\tau)) S_{0\ell}^{\mathfrak{su}(2)}}{2\cos\big(\frac{\pi(2\ell+1)}{\kappa+2}\big)+2\cos(2\pi\lambda)}\ .
\ee
Thus the Verlinde formula becomes
\begin{align}
N^{(w_3,\lambda_3,\ell_3)}_{(w_1,\lambda_1,\ell_1)(w_2,\lambda_2,\ell_2)}&=\sum_{w \in \mathds{Z}} \sum_{\ell=0}^{\frac{\kappa}{2}} \int_0^1 \mathrm{d}\lambda \frac{S_{(w_1,\lambda_1,\ell_1)(w,\lambda,\ell)}S_{(w_2,\lambda_2,\ell_2)(w,\lambda,\ell)}S_{(w_3,\lambda_3,\ell_3)(w,\lambda,\ell)}^*}{S_{\text{vac},(w,\lambda,\ell)}}\\
&=\Big(\delta_{w_3,w_1+w_2+1} \delta\big(\lambda_3=\lambda_1+\lambda_2-\tfrac{\gamma}{2}\big)\nonumber\\
&\qquad\qquad\qquad+\delta_{w_3,w_1+w_2-1} \delta\big(\lambda_3=\lambda_1+\lambda_2+\tfrac{\gamma}{2}\big)\Big) N^{\ell_3}_{\ell_1\ell_2}\nonumber\\
&\qquad+\delta_{w_3,w_1+w_2} \delta\big(\lambda_3=\lambda_1+\lambda_2+\tfrac{1}{2}\big) \big(N^{\ell_3+\frac{1}{2}}_{\ell_1\ell_2}+N^{\ell_3-\frac{1}{2}}_{\ell_1\ell_2}\big)\ , \label{eq:fusion rules Grothendieck}
\end{align}
where, $N^{\ell_3}_{\ell_1\ell_2}$ are the $\mathfrak{su}(2)_\kappa$ rules, 
\be 
N^{\ell_3}_{\ell_1\ell_2}=\delta_\mathds{Z}(\ell_1+\ell_2+\ell_3)\begin{cases} 1 & |\ell_1-\ell_2| \le \ell_3 \le \min(\ell_1+\ell_2,\kappa-\ell_1-\ell_2) \\
0 & \text{otherwise}
\end{cases}
\ee
We take them by definition to be zero if one of the indices does not take values in $\{0,\tfrac{1}{2},\dots,\tfrac{\kappa}{2}\}$.

\section{The free field realization of \texorpdfstring{$\boldsymbol{\mathfrak{d}(2,1;\alpha)_k}$}{d(2,1;alpha)k} at \texorpdfstring{$\boldsymbol{k^+=k^-=1}$}{k+=k-=1}} \label{app:kappa=0}

\subsection{The symplectic boson theory} \label{subapp:symplectic boson}
Let us begin by explaining the free field realisation of $\mathfrak{sl}(2,\mathds{R})_{1/2}$ in terms of a single pair of symplectic bosons. This theory and its fusion rules were analysed in detail in \cite{Ridout:2008nh, Ridout:2010qx, Ridout:2010jk}, see also \cite{Eberhardt:2018ouy} for some background explanations.

The (pair of) symplectic bosons $\xi_m^\alpha$ with $\alpha=\pm$, satisfy the commutation relations 
\be
{}[ {\xi}_m^\alpha, \xi_n^\beta] = \epsilon^{\alpha\beta}\delta_{m,-n}\ . 
\ee
They give rise to an $\mathfrak{sl}(2,\mathds{R})_{1/2}$ affine algebra by setting
\be
J^{a}_m=-\tfrac{1}{4}c_a (\sigma^a)_{\alpha\beta} (\xi^\alpha\xi^\beta)_m \ .
\ee
Both $\xi^+_r$ and $\xi^-_r$ are spin-$\tfrac{1}{2}$ fields and possess therefore NS- and R-sector representations. The NS-sector highest weight representation is described by
\be 
\xi_r^\alpha\, \ket{0}=0\ ,  \qquad r \ge \tfrac{1}{2} \ , \qquad \alpha\in \{+,-\} \ ,
\ee
and gives the vacuum representation of the theory.
On the other hand, the R-sector representations of the symplectic boson pair have a zero-mode representation on the states $|m\rangle$ with action 
\be\label{xiaction}
\xi^+_0 \, |m\rangle = \sqrt{2}\, |m+\tfrac{1}{2} \rangle \ , \qquad 
\xi^-_0 \, |m\rangle = \sqrt{2} (m-\tfrac{1}{4})\,  |m-\tfrac{1}{2} \rangle \ , 
\ee
so that, in terms of the $\mathfrak{sl}(2,\mathds{R})$ generators we have, 
\be
J^3_0 \, |m\rangle = m \, |m\rangle \ , \qquad \mathcal{C}^{\mathfrak{sl}(2,\mathds{R})} \, |m\rangle = \frac{3}{16}\, |m\rangle \ . 
\ee
Thus the R-sector representations of the symplectic boson are labelled by $\lambda \in \mathds{R}/\tfrac{1}{2}\mathds{Z}$, describing the eigenvalues of $J^3_0$ mod $\tfrac{1}{2}\mathds{Z}$, see also Appendix~C.1 of \cite{Eberhardt:2018ouy}

Each symplectic boson representation decomposes into two $\mathfrak{sl}(2,\mathds{R})_{1/2}$ representations, since the $\mathfrak{sl}(2,\mathds{R})_{1/2}$ currents are bilinear in the symplectic bosons. The NS-sector representation decomposes into the two modules $\mathscr{K}_0$ and $\mathscr{K}_1$, which can be thought of as the vacuum and vector representation, respectively. Similarly, the R-sector representations decompose into the representations $\mathscr{E}_\lambda$ and $\mathscr{E}_{\lambda+1/2}$, where now $\lambda \in \mathds{R}/\mathds{Z}$ describes the eigenvalues of $J^3_0$ mod $\mathds{Z}$.  At $\lambda=\tfrac{1}{4}$, $\tfrac{3}{4}$, the modules become indecomposable, as can be seen from \eqref{xiaction}. 
The relevant modules that are required for the description of the full theory are in fact even bigger, and involve the indecomposable representations $\mathscr{S}$ and $\mathscr{S}'$ \cite{Ridout:2010jk}, whose composition series takes the form\footnote{Note that, unlike the situation discussed in Appendix~C.1 of \cite{Eberhardt:2018ouy}, we are considering here these modules as representations of $\mathfrak{sl}(2,\mathds{R})_{1/2}$, not as representations of the symplectic boson theory.}
\be 
\mathscr{S} : \quad 
\begin{tikzpicture}[baseline={([yshift=-.5ex]current bounding box.center)}]
\node (top) at (0,1.5) {$\mathscr{K}_0$};
\node (right) at (1.5,0) {$\sigma^2(\mathscr{K}_1)$};
\node (left) at (-1.5,0) {$\sigma^{-2}(\mathscr{K}_1)$};
\node (bottom) at (0,-1.5) {$\mathscr{K}_0$};
\draw[thick,->] (right) to (bottom);
\draw[thick,->] (left) to  (bottom);
\draw[thick,->] (top) to  (left);
\draw[thick,->] (top) to (right);
\end{tikzpicture}\ , \qquad
\mathscr{S}' : \quad 
\begin{tikzpicture}[baseline={([yshift=-.5ex]current bounding box.center)}]
\node (top) at (0,1.5) {$\mathscr{K}_1$};
\node (right) at (1.5,0) {$\sigma^2(\mathscr{K}_0)$};
\node (left) at (-1.5,0) {$\sigma^{-2}(\mathscr{K}_0)$};
\node (bottom) at (0,-1.5) {$\mathscr{K}_1$};
\draw[thick,->] (right) to (bottom);
\draw[thick,->] (left) to  (bottom);
\draw[thick,->] (top) to  (left);
\draw[thick,->] (top) to (right);
\end{tikzpicture}
\ee
%We describe the meaning of the composition series for $\mathscr{S}$ in more detail.
%Its bottom line is the irreducible vacuum representation $\mathscr{K}$, and it forms a proper subrepresentation of $\mathscr{S}$. Since $\mathscr{S}$ is indecomposable, the complement of $\mathscr{K}$ does {\it not} form another subrepresentation of $\mathscr{S}$. However, the quotient space $\mathscr{S}/\mathscr{K}$ contains  subrepresentations, namely the two irreducible representations described by the middle line of $\mathscr{S}$. Again, their complement is not another subrepresentation, so one needs to quotient again by the representations in the middle line. The resulting space is then the irreducible vacuum representation $\mathscr{K}$ appearing at the top of the diagram. 
%In this language, the direction of the arrows indicates the symplectic boson action: symplectic bosons can map from top to bottom, but not back.
%The top element of the composition series is called the ``head", whereas the bottom element is called the ``socle".

The representation $\mathscr{S}$ is closely related to $\mathscr{E}_{1/4}$ and $\mathscr{E}_{3/4}$ since, on the level of the Grothen\-dieck ring, we have
\be 
\mathscr{E}_{1/4} \sim \sigma(\mathscr{K}_0) \oplus \sigma^{-1}(\mathscr{K}_1)\quad \Longrightarrow \quad 
\mathscr{S} \sim \sigma (\mathscr{E}_{3/4}) \oplus \sigma^{-1}(\mathscr{E}_{1/4}) \ ,
\ee
while the analogous statement for $\mathscr{S}'$ is 
\be 
\mathscr{S}' \sim  \sigma (\mathscr{E}_{1/4}) \oplus \sigma^{-1}(\mathscr{E}_{3/4}) \ . 
\ee
Here $\sigma$ denotes the spectral flow of the symplectic boson theory which acts via
\be\label{sigmaxi}
\sigma (\xi_r^\alpha)=\xi_{r-\frac{\alpha}{2}}^\alpha\ . 
\ee
The fusion rules of this theory were worked out in \cite{Ridout:2010jk}, and are explicitly
\begin{subequations}
\begin{align}
\mathscr{E}_\lambda \times \mathscr{E}_\mu &\cong \begin{cases}
\sigma(\mathscr{E}_{\lambda+\mu-\frac{1}{4}}) \oplus \sigma^{-1}(\mathscr{E}_{\lambda+\mu+\frac{1}{4}})\ , \quad & \lambda+\mu \ne 0\ , \\
\mathscr{S}\ , \quad &\lambda+\mu=0\ , \\
\mathscr{S}'\ , \quad &\lambda+\mu=\frac{1}{2}\ , \\
\end{cases} \label{eq:symplectic boson fusion rules a}\\
\mathscr{E}_\lambda \times \mathscr{S} &\cong \sigma^{2}(\mathscr{E}_{\lambda+\frac{1}{2}}) \oplus 2\cdot \mathscr{E}_\lambda \oplus \sigma^{-2}(\mathscr{E}_{\lambda+\frac{1}{2}})\ , \label{eq:symplectic boson fusion rules b}\\
\mathscr{E}_\lambda \times \mathscr{S}' &\cong \sigma^{2}(\mathscr{E}_{\lambda}) \oplus 2\cdot \mathscr{E}_{\lambda+\frac{1}{2}} \oplus \sigma^{-2}(\mathscr{E}_{\lambda})\ , \label{eq:symplectic boson fusion rules c}\\
\mathscr{S} \times \mathscr{S}&\cong \mathscr{S}' \times \mathscr{S}'\cong\sigma^{2}(\mathscr{S}') \oplus 2\cdot \mathscr{S} \oplus \sigma^{-2}(\mathscr{S}')\ ,\label{eq:symplectic boson fusion rules d} \\
\mathscr{S} \times \mathscr{S}'&\cong\sigma^{2}(\mathscr{S}') \oplus 2\cdot \mathscr{S} \oplus \sigma^{-2}(\mathscr{S}')\ .\label{eq:symplectic boson fusion rules e}
\end{align}
\end{subequations}

\subsection{The explicit form of the free field representation} \label{subapp:free field realization explicit}
In order to describe the free field realisation of $\mathfrak{d}(2,1;\alpha)_k$ at $k^+=k^-=1$, we now combine a symplectic boson pair with four free fermions, which we take to satisfy the anticommutation relations
\be 
\{\psi^{\alpha\beta}_r,\psi^{\gamma\delta}_s\}=\epsilon^{\alpha\gamma}\epsilon^{\beta\delta}\delta_{r+s,0}\ .
\ee
The generators of $\mathfrak{d}(2,1;\alpha)_k$ are then given by
\begin{subequations}
\begin{align}
J^a_m&=-\frac{1}{4} c_a (\sigma^a)_{\alpha\beta} (\xi^\alpha\xi^\beta)_m\ , \\
K^{(+)a}_m&=\frac{1}{4} (\sigma^a)_{\alpha\gamma}\epsilon_{\beta\delta} (\psi^{\alpha\beta}\psi^{\beta\delta})_m\ ,\\
K^{(-)a}_m&=\frac{1}{4} \epsilon_{\alpha\gamma}(\sigma^a)_{\beta\delta} (\psi^{\alpha\beta}\psi^{\beta\delta})_m\ ,\\
S^{\alpha\beta\gamma}_m&=\frac{1}{\sqrt{2}} (\xi^\alpha \psi^{\beta\gamma})_m\ .
\end{align}
\end{subequations}
The spectral flow automorphism of $\mathfrak{d}(2,1;\alpha)_k$ acts on $\xi^\alpha_r$ as in (\ref{sigmaxi}), while on the fermions we have 
\be 
\sigma(\psi^{\alpha\beta}_r)=\psi^{\alpha\beta}_{r+\frac{\alpha}{2}}\ .
\ee
\subsection{The fusion rules} \label{subapp:kappa=0 fusion rules}
With this free field realisation at hand, we can evaluate the fusion rules directly in this case. Using the conformal embedding (\ref{eq:d21alpha conformal embedding}), the $\mathfrak{d}(2,1;\alpha)_k$ representations decompose as 
\begin{subequations}
\begin{align}
\mathscr{L}&=(\mathscr{K}_0,\mathbf{1},\mathbf{1}) \oplus (\mathscr{K}_1,\mathbf{2},\mathbf{2})\ , \label{eq:kappa=0 decomposition a}\\
\mathscr{L}'&=(\mathscr{K}_0,\mathbf{2},\mathbf{2}) \oplus (\mathscr{K}_1,\mathbf{1},\mathbf{1})\ , \label{eq:kappa=0 decomposition b}\\
\mathscr{F}^0_\lambda&=(\mathscr{E}_\lambda,\mathbf{2},\mathbf{1}) \oplus (\mathscr{E}_{\lambda+\frac{1}{2}},\mathbf{1},\mathbf{2})\ , \label{eq:kappa=0 decomposition c}\\
\mathscr{T}_>^{\frac{1}{2}}&=\sigma^{-1}(\mathscr{S},\mathbf{1},\mathbf{1}) \oplus \sigma^{-1}(\mathscr{S}',\mathbf{2},\mathbf{2})\ , \label{eq:kappa=0 decomposition d}\\
\mathscr{T}_<^{\frac{1}{2}}&=\sigma^{-1}(\mathscr{S}',\mathbf{1},\mathbf{1}) \oplus \sigma^{-1}(\mathscr{S},\mathbf{2},\mathbf{2})\ , \label{eq:kappa=0 decomposition e}
\end{align}
\end{subequations}
where we have denoted the $\mathfrak{su}(2)_1$ representations by the dimension of their ground state representations. Furthermore, $\mathscr{T}_>^{\frac{1}{2}}$ and $\mathscr{T}_<^{\frac{1}{2}}$ are indecomposable representations that will be introduced in Appendix~\ref{app:indecomposables}. If $\lambda+\mu \ne 0$, $\tfrac{1}{2}$, the fusion rules are then 
\begin{align}
\mathscr{F}_\lambda^0 \times \mathscr{F}_\mu^0&=\big((\mathscr{E}_\lambda,\mathbf{2},\mathbf{1}) \oplus (\mathscr{E}_{\lambda+\frac{1}{2}},\mathbf{1},\mathbf{2})\big) \times (\mathscr{E}_\mu,\mathbf{2},\mathbf{1}) \\
&=\sigma(\mathscr{E}_{\lambda+\mu-\frac{1}{2}},\mathbf{2},\mathbf{1}) \oplus \sigma^{-1}(\mathscr{E}_{\lambda+\mu+\frac{1}{4}},\mathbf{2},\mathbf{1}) \nonumber \\
& \qquad \qquad \oplus \sigma(\mathscr{E}_{\lambda+\mu+\frac{1}{4}},\mathbf{1},\mathbf{2}) \oplus \sigma^{-1}(\mathscr{E}_{\lambda+\mu-\frac{1}{4}},\mathbf{1},\mathbf{2}) \nonumber \\
&=\sigma(\mathscr{F}_{\lambda+\mu-\frac{1}{4}}) \oplus \sigma^{-1}(\mathscr{F}_{\lambda+\mu+\frac{1}{4}})\ .
\end{align}
The other cases work similarly, and the complete fusion rules are therefore
\begin{subequations}
\begin{align}
\mathscr{F}^0_\lambda \times \mathscr{F}^0_\mu &=\begin{cases}
\sigma(\mathscr{F}_{\lambda+\mu-\frac{1}{4}}) \oplus \sigma^{-1}(\mathscr{F}_{\lambda+\mu+\frac{1}{4}})\ , \quad & \lambda+\mu \ne 0\ , \\
\sigma(\mathscr{T}_>^{\frac{1}{2}})\ , \quad &\lambda+\mu=0\ , \\
\sigma(\mathscr{T}_<^{\frac{1}{2}})\ , \quad &\lambda+\mu=\frac{1}{2}\ , \\
\end{cases} \\
\mathscr{F}^0_\lambda \times \mathscr{T}_>^{\frac{1}{2}}&=\sigma^{-1}(\mathscr{F}^0_{\lambda+\frac{1}{2}}) \oplus 2 \sigma^{-1}(\mathscr{F}^0_\lambda) \oplus \sigma^{-3}(\mathscr{F}^0_{\lambda+\frac{1}{2}})\ , \\
\mathscr{F}^0_\lambda \times \mathscr{T}_<^{\frac{1}{2}}&=\sigma^{-1}(\mathscr{F}^0_{\lambda}) \oplus 2 \sigma^{-1}(\mathscr{F}^0_{\lambda+\frac{1}{2}}) \oplus \sigma^{-3}(\mathscr{F}^0_\lambda)\ , \\
\mathscr{T}_>^{\frac{1}{2}} \times\mathscr{T}_>^{\frac{1}{2}} &\cong \mathscr{T}_<^{\frac{1}{2}} \times \mathscr{T}_<^{\frac{1}{2}}\cong \sigma(\mathscr{T}_<^{\frac{1}{2}})\oplus 2\, \sigma^{-1}(\mathscr{T}_>^{\frac{1}{2}})\oplus \sigma^{-3}(\mathscr{T}_<^{\frac{1}{2}})\ , \\
\mathscr{T}_>^{\frac{1}{2}} \times\mathscr{T}_<^{\frac{1}{2}} &\cong  \sigma(\mathscr{T}_>^{\frac{1}{2}})\oplus 2\, \sigma^{-1}(\mathscr{T}_<^{\frac{1}{2}})\oplus \sigma^{-3}(\mathscr{T}_>^{\frac{1}{2}})\ .
\end{align}
\end{subequations}

\subsection{The characters} \label{subapp:kappa=0 characters}
Finally, it is straightforward to compute the characters using this free field realisation. Let us demonstrate how to do this for $\mathrm{ch}[\mathscr{F}^0_\lambda](t,u,v;\tau)$. The symplectic boson R-sector representation has the character
\be 
\sum_{m \in \frac{1}{2}\mathds{Z}+\lambda} \frac{x^m}{q^{\frac{1}{12}}\prod_{n=1}^\infty(1-x^{\frac{1}{2}}q^n)(1-x^{-\frac{1}{2}}q^n)}=\sum_{m \in \frac{1}{2}\mathds{Z}+\lambda} \frac{x^m}{\eta(\tau)^2}\ ,
\ee
where we have used that the chemical potentials of the oscillators can be absorbed into the zero modes, which allows us to rewrite the denominator in terms of the eta function. For the character of $\mathfrak{sl}(2,\mathds{R})_{1/2}$, we have to keep every second state, and thus obtain
\be 
\mathrm{ch}[\mathscr{E}_\lambda](t;\tau)= \sum_{m \in \mathds{Z}+\lambda} \frac{x^m}{\eta(\tau)^2}\ .
\ee
On the other hand, the character of the four free fermions equals
\begin{subequations}
\begin{align} 
\mathrm{ch}[(\mathbf{2},\mathbf{1})](u,v;\tau)&=\frac{\vartheta_2\big(\tfrac{u+v}{2};\tau\big)\vartheta_2\big(\tfrac{u-v}{2};\tau\big)-\vartheta_1\big(\tfrac{u+v}{2};\tau\big)\vartheta_1\big(\tfrac{u-v}{2};\tau\big)}{2\eta(\tau)^2}\ , \\
\mathrm{ch}[(\mathbf{1},\mathbf{2})](u,v;\tau)&=\frac{\vartheta_2\big(\tfrac{u+v}{2};\tau\big)\vartheta_2\big(\tfrac{u-v}{2};\tau\big)+\vartheta_1\big(\tfrac{u+v}{2};\tau\big)\vartheta_1\big(\tfrac{u-v}{2};\tau\big)}{2\eta(\tau)^2}\ .
\end{align}
\end{subequations}
Combining these ingredients according to \eqref{eq:kappa=0 decomposition c}, we finally obtain
\be 
\mathrm{ch}[\mathscr{F}^0_\lambda](t,u,v;\tau)=\sum_{m \in \mathds{Z}+\lambda+\frac{1}{2}} \frac{x^m}{\eta(\tau)^4} \vartheta_2\big(\tfrac{t+u+v}{2};\tau\big)\vartheta_2\big(\tfrac{t+u-v}{2};\tau\big)\ ,
\ee
which matches with the general formula \eqref{eq:character continuous representation}.

\section{The indecomposable modules} \label{app:indecomposables}

In this Appendix, we discuss the atypical modules appearing in the $\mathfrak{d}(2,1;\alpha)_k$ WZW-model at $k^+=1$. We make an educated guess for their structure, which passes many non-trivial tests.

\subsection{The indecomposable modules}
One strategy to determine the possible indecomposable modules is to study the representations that appear in fusion products. In the typical case we have, see (\ref{fusionrules})
\be 
\mathscr{F}^0_{\lambda_1} \times \mathscr{F}^\ell_{\lambda_2}\cong \sigma\big(\mathscr{F}^\ell_{\lambda_1+\lambda_2-\frac{\gamma}{2}}\big)\oplus\mathscr{F}^{\ell+\frac{1}{2}}_{\lambda_1+\lambda_2+\frac{1}{2}}\oplus \mathscr{F}^{\ell-\frac{1}{2}}_{\lambda_1+\lambda_2+\frac{1}{2}}\oplus 
\sigma^{-1}\big(\mathscr{F}^\ell_{\lambda_1+\lambda_2+\frac{\gamma}{2}}\big)\ .
\ee
If several modules on the right hand side of the fusion rules become indecomposable, we expect them to join to form one big indecomposable module. This happens when
\be 
\lambda_1+\lambda_2+\tfrac{1}{2}\in \big\{\pm\lambda_{\ell+\frac{1}{2}},\pm\lambda_{\ell-\frac{1}{2}}\big\}\ ,
\ee
since we have the exact short sequences of modules\footnote{The notation here is a bit cavalier since the 
modules $\mathscr{F}^\ell_{\lambda_\ell}$ that appear as the middle term in the first two lines have different indecomposable structures (since one contains a discrete highest weight and the other a discrete lowest weight representation).} 
\begin{subequations}
\begin{align} 
0 &\rightarrow \mathscr{G}_{>,+}^\ell \longrightarrow \mathscr{F}^{\ell}_{\lambda_\ell} \longrightarrow \mathscr{G}_{>,-}^\ell \rightarrow 0 \ , \label{eq:SES 1}\\
0 &\rightarrow \mathscr{G}_{>,-}^\ell \longrightarrow \mathscr{F}^{\ell}_{\lambda_\ell} \longrightarrow \mathscr{G}_{>,+}^\ell \rightarrow 0 \ , \label{eq:SES 2}\\
0 &\rightarrow \mathscr{G}_{<,+}^\ell \longrightarrow \mathscr{F}^{\ell}_{-\lambda_\ell} \longrightarrow \mathscr{G}_{<,-}^\ell \rightarrow 0 \ , \label{eq:SES 3}\\
0 &\rightarrow \mathscr{G}_{<,-}^\ell \longrightarrow \mathscr{F}^{\ell}_{-\lambda_\ell} \longrightarrow \mathscr{G}_{<,+}^\ell \rightarrow 0 \ . \label{eq:SES 4}
\end{align}
\end{subequations}
The cases $\ell=0$ and $\ell=\tfrac{\kappa}{2}$ are special, since then two of these values are simultaneously attained, but some modules are not present. One finds that the following modules can join up to form bigger indecomposable modules:
\begin{subequations}
\begin{align}
\mathscr{T}_>^{\frac{1}{2}}&\sim\mathscr{F}^0_{\lambda_0} \oplus \sigma^{-2}\big(\mathscr{F}^0_{-\lambda_0}\big)\ , \label{inde1} \\
\mathscr{T}_>^{\ell}&\sim \mathscr{F}^{\ell-\frac{1}{2}}_{\lambda_{\ell-\frac{1}{2}}} \oplus
\sigma^{-1}\big(\mathscr{F}^{\ell-1}_{\lambda_{\ell-1}}\big)  \ , \quad \ell\in \big\{1,\tfrac{3}{2},\dots,\tfrac{\kappa+1}{2}\big\}\ , \\
\mathscr{T}_<^{\frac{\kappa+1}{2}}&\sim\mathscr{F}^{\frac{\kappa}{2}}_{-\lambda_{\frac{\kappa}{2}}}\oplus \sigma^{-2}\big(\mathscr{F}^{\frac{\kappa}{2}}_{\lambda_{\frac{\kappa}{2}}}\big)\ , \\
\mathscr{T}_<^{\ell}&\sim\mathscr{F}^{\ell-\frac{1}{2}}_{-\lambda_{\ell-\frac{1}{2}}} \oplus \sigma^{-1}\big(\mathscr{F}^\ell_{-\lambda_\ell}\big)\ , \quad \ell\in \big\{\tfrac{1}{2},1,\dots,\tfrac{\kappa}{2}\big\}\ . \label{inde4}
\end{align}
\end{subequations}
%We thus have an indecomposable module $\mathscr{T}_\pm^\ell$ for any $\ell\in \tfrac{1}{2}\mathds{N} \setminus \tfrac{\kappa+2}{2}\mathds{Z}$, where we cyclically identify $\mathscr{T}^\ell \equiv \mathscr{T}^{\ell+\kappa+2}$. 

In order to describe the precise structures of these indecomposables, we first use the short exact sequences \eqref{eq:SES 1} -- \eqref{eq:SES 4} to decompose the modules in the Grothendieck ring as 
\begin{align} 
\mathscr{T}_>^{\frac{1}{2}}&\sim \mathscr{G}^0_{>,+} \oplus \mathscr{G}^0_{>,-} \oplus \sigma^{-2}\big(\mathscr{G}^0_{<,+}\big)\oplus \sigma^{-2}\big(\mathscr{G}^0_{<,-}\big)\ , \\
\mathscr{T}_>^{\ell}&\sim  \mathscr{G}^{\ell-\frac{1}{2}}_{>,+}\oplus \mathscr{G}^{\ell-\frac{1}{2}}_{>,-} \oplus 
\sigma^{-1}\big(\mathscr{G}^{\ell-1}_{>,+}\big)\oplus \sigma^{-1}\big(\mathscr{G}^{\ell-1}_{>,-}\big) \ , \\
\mathscr{T}_<^{\frac{\kappa+1}{2}}&\sim \mathscr{G}^{\frac{\kappa}{2}}_{<,+} \oplus \mathscr{G}^{\frac{\kappa}{2}}_{<,-}\oplus \sigma^{-2}\big(\mathscr{G}^{\frac{\kappa}{2}}_{>,+}\big)\oplus \sigma^{-2}\big(\mathscr{G}^{\frac{\kappa}{2}}_{>,-}\big)\ , \\
\mathscr{T}_<^{\ell}&\sim\mathscr{G}^{\ell-\frac{1}{2}}_{<,+}\oplus \mathscr{G}^{\ell-\frac{1}{2}}_{<,-}\oplus \sigma^{-1}\big(\mathscr{G}^{\ell}_{<,+}\big)\oplus \sigma^{-1}\big(\mathscr{G}^{\ell}_{<,-}\big)\ .
\end{align}
Because of the identifications under spectral flow, see eqs.~\eqref{eq:spectral flow identification b} -- \eqref{eq:spectral flow identification d}, the right hand side always contains two isomorphic modules, and hence the indecomposable structure is
\begin{subequations}
\begin{align}
\mathscr{T}_>^{\frac{1}{2}}&:\quad
\begin{tikzpicture}[baseline={([yshift=-.5ex]current bounding box.center)}]
\node (top) at (0,1.5) {$\mathscr{G}^0_{>,-}$};
\node (right) at (1.5,0) {$\mathscr{G}^0_{>,+}$};
\node (left) at (-1.5,0) {$\sigma^{-2}\big(\mathscr{G}^0_{<,-}\big)$};
\node (bottom) at (0,-1.5) {$\sigma^{-2}\big(\mathscr{G}^0_{<,+}\big)$};
\draw[thick,->] (right) to (bottom);
\draw[thick,->] (left) to  (bottom);
\draw[thick,->] (top) to  (left);
\draw[thick,->] (top) to (right);
\end{tikzpicture}\ , &
\mathscr{T}_>^{\ell}&:\quad
\begin{tikzpicture}[baseline={([yshift=-.5ex]current bounding box.center)}]
\node (top) at (0,1.5) {$\mathscr{G}^{\ell-\frac{1}{2}}_{>,-}$};
\node (right) at (1.5,0) {$\mathscr{G}^{\ell-\frac{1}{2}}_{>,+}$};
\node (left) at (-1.5,0) {$\sigma^{-1}\big(\mathscr{G}^{\ell-1}_{>,-}\big)$};
\node (bottom) at (0,-1.5) {$\sigma^{-1}\big(\mathscr{G}^{\ell-1}_{>,+}\big)$};
\draw[thick,->] (right) to (bottom);
\draw[thick,->] (left) to  (bottom);
\draw[thick,->] (top) to  (left);
\draw[thick,->] (top) to (right);
\end{tikzpicture}\ , \label{eq:indecomposable structure a}\\
\mathscr{T}_<^{\frac{\kappa+1}{2}}&:\quad
\begin{tikzpicture}[baseline={([yshift=-.5ex]current bounding box.center)}]
\node (top) at (0,1.5) {$\mathscr{G}^{\frac{\kappa}{2}}_{<,-}$};
\node (right) at (1.5,0) {$\mathscr{G}^{\frac{\kappa}{2}}_{<,+}$};
\node (left) at (-1.5,0) {$\sigma^{-2}\big(\mathscr{G}^{\frac{\kappa}{2}}_{>,-}\big)$};
\node (bottom) at (0,-1.5) {$\sigma^{-2}\big(\mathscr{G}^{\frac{\kappa}{2}}_{>,+}\big)$};
\draw[thick,->] (right) to (bottom);
\draw[thick,->] (left) to  (bottom);
\draw[thick,->] (top) to  (left);
\draw[thick,->] (top) to (right);
\end{tikzpicture}\ , &
\mathscr{T}_<^{\ell}&:\quad
\begin{tikzpicture}[baseline={([yshift=-.5ex]current bounding box.center)}]
\node (top) at (0,1.5) {$\mathscr{G}^{\ell-\frac{1}{2}}_{<,-}$};
\node (right) at (1.5,0) {$\mathscr{G}^{\ell-\frac{1}{2}}_{<,+}$};
\node (left) at (-1.5,0) {$\sigma^{-1}\big(\mathscr{G}^{\ell}_{<,-}\big)$};
\node (bottom) at (0,-1.5) {$\sigma^{-1}\big(\mathscr{G}^{\ell}_{<,+}\big)$};
\draw[thick,->] (right) to (bottom);
\draw[thick,->] (left) to  (bottom);
\draw[thick,->] (top) to  (left);
\draw[thick,->] (top) to (right);
\end{tikzpicture}\ .\label{eq:indecomposable structure b}
\end{align}
\end{subequations}
Here, we have used again composition diagrams to display the indecomposable structure, see also Appendix~\ref{app:kappa=0}. Note that the bottom and top modules are always identical via the spectral flow identifications.

\subsection{The atypical fusion rules}
%We will now assume that the complete spectrum of representations in the $\mathfrak{d}(2,1;\alpha)_k$ WZW model is given by
%\begin{align}
%&\sigma^w\big(\mathscr{F}_\lambda^\ell\big)\ , \qquad \ell \in \big\{0,\tfrac{1}{2},\dots,\tfrac{\kappa}{2}\big\}\ ,\qquad w \in \mathds{Z}\ , \qquad \lambda \in \mathds{R}/\mathds{Z}\ ,\qquad\lambda \ne \pm \lambda_\ell\ , \\
%&\sigma^w\big(\mathscr{T}^\ell\big)\ , \qquad \ell \in \tfrac{1}{2}\mathds{Z} \setminus \tfrac{\kappa+2}{2}\mathds{Z}\ ,\qquad w \in \mathds{Z}\ .
%\end{align}
%The indecomposable representations fill precisely the hole in the typical representations. 

A heuristic way to determine the fusion rules of these indecomposable representations consists of writing them in terms of their summands as in (\ref{inde1}) -- (\ref{inde4}). We then apply the naive generalisations of the typical fusion rules (\ref{fusionrules}), and finally reassemble the result, so that the only representations that appear are the typical representations we considered before, together with those given in (\ref{inde1}) -- (\ref{inde4}).

Unfortunately, the general formula is rather clumsy, so let us just work out one example to illustrate the idea. We consider 
\begin{align}
\mathscr{T}_>^{\frac{1}{2}} \times  \mathscr{T}_>^{\frac{1}{2}} &\sim \big( \mathscr{F}^0_{\lambda_0} \oplus \sigma^{-2}\big( \mathscr{F}^0_{-\lambda_0}\big)\big)\times \big( \mathscr{F}^0_{\lambda_0} \oplus \sigma^{-2}\big( \mathscr{F}^0_{-\lambda_0}\big)\big) \\
&\cong \sigma \big(\mathscr{F}^0_{\lambda_1}\big) \oplus \mathscr{F}^{\frac{1}{2}}_{\lambda_{\frac{1}{2}}} \oplus \sigma^{-1} \big(\mathscr{F}^0_{\lambda_0}\big) \oplus 2\, \sigma^{-1} \big(\mathscr{F}^0_{\lambda_0}\big) \oplus 2\, \sigma^{-2} \big(\mathscr{F}^{\frac{1}{2}}_{\frac{1}{2}}\big) \nonumber\\
&\qquad\oplus  2\, \sigma^{-3} \big(\mathscr{F}^0_{-\lambda_0}\big) \oplus \sigma^{-3} \big(\mathscr{F}^0_{-\lambda_0}\big) \oplus \sigma^{-4} \big(\mathscr{F}^{\frac{1}{2}}_{-\lambda_{\frac{1}{2}}}\big) \oplus \sigma^{-5} \big(\mathscr{F}^0_{-\lambda_1}\big) \\
&\sim\mathscr{T}_>^1 \oplus 2\, \sigma^{-1}\big(\mathscr{T}_>^{\frac{1}{2}}\big) \oplus \sigma^{-3}\big(\mathscr{T}_<^{\frac{1}{2}}\big) \oplus \sigma \big(\mathscr{F}^0_{\lambda_1}\big) \oplus 2\, \sigma^{-2} \big(\mathscr{F}^{\frac{1}{2}}_{\frac{1}{2}}\big) \oplus \sigma^{-5} \big(\mathscr{F}^0_{-\lambda_1}\big) \ . 
\end{align}
The other cases work similarly, and the resulting products define an associative ring, which should therefore agree with the fusion ring.

\subsection{The atypical Hilbert space}
Finally, we discuss the structure of the atypical Hilbert space. Naively, we would construct an atypical Hilbert space as
\be 
\mathcal{H}^{\text{naive}}_{\text{atyp}}=\bigoplus_{w \in \mathds{Z}}\,  \bigoplus_{\ell = \frac{1}{2}}^{\frac{\kappa+1}{2}}\, \Bigl[ \sigma^w\big(\mathscr{T}_>^\ell\big) \otimes \overline{\sigma^w\big(\mathscr{T}_>^\ell\big)} \,\, \oplus \,\, \sigma^w\big(\mathscr{T}_<^\ell\big) \otimes \overline{\sigma^w\big(\mathscr{T}_<^\ell\big) 
} \Bigr]\ .\label{eq:Hilbert space naive}
\ee
While this contains now only modules which close under fusion, there are two problems with this proposal. First, locality requires that $L_0-\bar{L}_0$ acts diagonalisably, since otherwise the complete correlation functions would be multi-valued. In addition, \eqref{eq:Hilbert space naive} does not agree with \eqref{eq:Hilbert space Grothendieck} on the level of the Grothendieck ring, and hence would not be modular invariant. As explained in \cite{Quella:2007hr,Gaberdiel:2007jv}, the true Hilbert space is obtained by quotienting out an ideal $\mathcal{I} \subset \mathcal{H}_\text{atyp}^\text{naive}$ from $\mathcal{H}_\text{atyp}^\text{naive}$.

To construct this ideal, we note that there are natural long exact sequences
\begin{align}
&\underset{s_-}{\overset{s_+}{\longleftrightarrow}}\sigma^{w+1}\big(\mathscr{T}_>^{\frac{1}{2}}\big) \underset{s_-}{\overset{s_+}{\longleftrightarrow}}\cdots \underset{s_-}{\overset{s_+}{\longleftrightarrow}}\sigma^{w+2\ell}\big(\mathscr{T}_>^{\ell}\big) \underset{s_-}{\overset{s_+}{\longleftrightarrow}} \cdots \underset{s_-}{\overset{s_+}{\longleftrightarrow}} \sigma^{w+\kappa+1} \big(\mathscr{T}_>^{\frac{\kappa+1}{2}}\big) \nonumber\\
&\underset{s_-}{\overset{s_+}{\longleftrightarrow}}\sigma^{w+\kappa+3}\big(\mathscr{T}_<^{\frac{\kappa+1}{2}}\big) \underset{s_-}{\overset{s_+}{\longleftrightarrow}} \cdots \underset{s_-}{\overset{s_+}{\longleftrightarrow}}\sigma^{w+2(\kappa+2)+2\ell}\big(\mathscr{T}_<^{\ell}\big) \underset{s_-}{\overset{s_+}{\longleftrightarrow}} \cdots \underset{s_-}{\overset{s_+}{\longleftrightarrow}} \sigma^{w+2\kappa+3} \big(\mathscr{T}_<^{\frac{1}{2}}\big) \nonumber\\
&\underset{s_-}{\overset{s_+}{\longleftrightarrow}}\sigma^{w+2\kappa+5}\big(\mathscr{T}_>^{\frac{1}{2}}\big) \underset{s_-}{\overset{s_+}{\longleftrightarrow}}\cdots \underset{s_-}{\overset{s_+}{\longleftrightarrow}}\sigma^{w+2(\kappa+2)+2\ell}\big(\mathscr{T}_>^{\ell}\big) \underset{s_-}{\overset{s_+}{\longleftrightarrow}} \cdots \underset{s_-}{\overset{s_+}{\longleftrightarrow}} \sigma^{w+3\kappa+5} \big(\mathscr{T}_>^{\frac{\kappa+1}{2}}\big) \ ,
\end{align}
where $s_+$ maps to the right and $s_-$ to the left. 
%This can be seen from the structure \eqref{eq:indecomposable structure a} and \eqref{eq:indecomposable structure b}. 
The map $s_+$ maps the two upper right elements of the composition diagram of \eqref{eq:indecomposable structure a} or \eqref{eq:indecomposable structure b} to the two lower left elements of the next term in the sequence. Similarly $s_-$ maps the two upper left elements of the composition diagram to the two lower right elements of the previous term in the sequence. There are in fact $2(\kappa+2)$ such sequences, that are characterised by $w \in \{0,1,\dots,2\kappa+3\}$. For each such sequence, 
%We note that the terms in this sequence are in fact all of the form $\sigma^{w+2m(\kappa+2)+2\ell}\big(\mathscr{T}^\ell\big)$. The maps $s_\pm$ jump simply over the module $\sigma^{w+2m(\kappa+2)+2\ell}\big(\mathscr{T}^\ell\big)$ with $\ell\in \big\{0,\tfrac{\kappa+2}{2}\big\}$, since they are nonexistent. 
let us denote the $n$-th term in the sequence by $\sigma^w\big(\mathscr{X}^n\big)$, so that 
%. Thus, the long exact sequence can be simply written as
\be 
s_\pm: \sigma^w\big(\mathscr{X}^n\big) \longrightarrow \big(\mathscr{X}^{n\pm 1}\big)
\ee
for $n \in \mathds{Z}$. Furthermore, we denote the elements of the indecomposable modules by $\mathscr{X}^n[\varepsilon_1,\varepsilon_2]$, where $\varepsilon_i \in \{0,1\}$. $[\varepsilon_1,\varepsilon_2]=[0,0]$ denotes the top element, $[\varepsilon_1,\varepsilon_2]=[1,0]$ the left element, $[\varepsilon_1,\varepsilon_2]=[0,1]$ the right element and $[\varepsilon_1,\varepsilon_2]=[1,1]$ the bottom element. We thus have, cf.\ \cite{Eberhardt:2018ouy}
\begin{align}
s_+ \sigma^w\big(\mathscr{X}^n\big)[\varepsilon_1,\varepsilon_2] &= \sigma^w\big(\mathscr{X}^{n+1}\big)[\varepsilon_1+1,\varepsilon_2]\ , \\
s_- \sigma^w\big(\mathscr{X}^n\big)[\varepsilon_1,\varepsilon_2] &= \sigma^w\big(\mathscr{X}^{n-1}\big)[\varepsilon_1,\varepsilon_2+1]\ .
\end{align}
The ideal $\mathcal{I}$ by which we have to quotient out is then generated by
\be 
\mathcal{I}_\pm \equiv\bigoplus_{w=0}^{2\kappa+3}\bigoplus_{n \in \mathds{Z}}\big(s_\pm \otimes \overline{\mathds{1}}-\mathds{1} \otimes \overline{s_\mp}\big)\big(\sigma^w\big(\mathscr{X}^n\big) \otimes \sigma^w\big(\mathscr{X}^{n\pm 1}\big)\big)\ .
\ee
This leads to the following identifications in the naive atypical Hilbert space,
\begin{align}
\sigma^w\big(\mathscr{X}^n\big)[\varepsilon_1+1,\varepsilon_2] \otimes \overline{\sigma^w\big(\mathscr{X}^n\big)[\bar{\varepsilon}_1,\bar{\varepsilon}_2]} &\sim \sigma^w\big(\mathscr{X}^{n-1}\big)[\varepsilon_1,\varepsilon_2] \otimes \overline{\sigma^w\big(\mathscr{X}^{n-1}\big)[\bar{\varepsilon}_1,\bar{\varepsilon}_2+1]}\ , \\
\sigma^w\big(\mathscr{X}^n\big)[\varepsilon_1,\varepsilon_2+1] \otimes \overline{\sigma^w\big(\mathscr{X}^n\big)[\bar{\varepsilon}_1,\bar{\varepsilon}_2]} &\sim \sigma^w\big(\mathscr{X}^{n+1}\big)[\varepsilon_1,\varepsilon_2] \otimes \overline{\sigma^w\big(\mathscr{X}^{n+1}\big)[\bar{\varepsilon}_1+1,\bar{\varepsilon}_2]}\ .
\end{align}
This `gauge freedom' allows us, for example, to set $\varepsilon_1=\bar{\varepsilon_1}=0$, so that 
%. With this gauge choice, we have
\be 
\big(\mathscr{T}_>^\ell\big)^\text{gauge fixed} \sim \mathscr{F}^{\ell-\frac{1}{2}}_{\lambda_{\ell-\frac{1}{2}}}\ , \quad\big(\mathscr{T}_<^{\ell}\big)^\text{gauge fixed} \sim \mathscr{F}^{\ell-\frac{1}{2}}_{-\lambda_{\ell-\frac{1}{2}}}\ ,\qquad \ell\in \{\tfrac{1}{2},1,\dots,\tfrac{\kappa+1}{2}\} \ . 
\ee
Hence, the indecomposable modules become after gauge-fixing the moral analogue of the continuous representations for $\lambda=\pm \lambda_\ell$. The resulting space of states has then essentially the same form as \eqref{eq:Hilbert space Grothendieck}, and hence is in particular modular invariant.

\section{Theta functions}\label{app:theta}

Our conventions for the theta functions follow \cite{Blumenhagen:2013fgp} 
\begin{align}
\vartheta\!\begin{bmatrix}
  \alpha \\
  \beta
  \end{bmatrix}(z;\tau)&\equiv\sum_{n\in\mathds Z}\mathrm{e}^{\pi i(n+\alpha)^2\tau+2\pi i(n+\alpha)(z+\beta)}\\
  &=\mathrm{e}^{2\pi i\alpha (z+\beta)} q^{\frac{\alpha^2}{2}} \prod_{n=1}^\infty \big(1-q^n \big) \big(1+q^{n+\alpha-\frac{1}{2}} \mathrm{e}^{2\pi i(z+\beta)} \big)\big(1+q^{n-\alpha-\frac{1}{2}} \mathrm{e}^{-2\pi i(z+\beta)} \big)\ .
\end{align}
In this language the four Jacobi theta functions are 
\be
\vartheta_1\equiv\vartheta\!\begin{bmatrix}
  \frac{1}{2} \\
  \frac{1}{2}
  \end{bmatrix}\ ,\qquad
\vartheta_2\equiv\vartheta\!\begin{bmatrix}
  \frac{1}{2} \\
  0
  \end{bmatrix}\ ,\qquad
\vartheta_3\equiv\vartheta\!\begin{bmatrix}
  0 \\
  0
  \end{bmatrix}\ ,\qquad
\vartheta_4\equiv\vartheta\!\begin{bmatrix}
  0 \\
  \frac{1}{2}
  \end{bmatrix}\ ,
\ee
and we have the identity 
\begin{align} 
\vartheta_2\big(\tfrac{u+v}{2};\tau)\vartheta_2\big(\tfrac{u-v}{2};\tau\big)&=\vartheta_2(u;2\tau)\vartheta_3(v;2\tau)+\vartheta_3(u;2\tau)\vartheta_2(v;2\tau)\ , \label{eq:equivalence four fermions and su2su2 a} \\
\vartheta_1\big(\tfrac{u+v}{2};\tau)\vartheta_1\big(\tfrac{u-v}{2};\tau\big)&=-\vartheta_2(u;2\tau)\vartheta_3(v;2\tau)+\vartheta_3(u;2\tau)\vartheta_2(v;2\tau)\ , \label{eq:equivalence four fermions and su2su2 b}
\end{align}
which comes from the equivalence of four free fermions to $\mathfrak{su}(2)_1 \oplus \mathfrak{su}(2)_1$.

%We also need the modular properties of $\vartheta_1(z,\tau)$ and $\eta(\tau)$, which are explicitly 
%\be 
%\vartheta_1\left(\frac{z}{\tau};-\frac{1}{\tau}\right)=-i \sqrt{-i \tau} \, \mathrm{e}^{\frac{\pi i z^2}{\tau}} \vartheta_1(z;\tau)\  , \qquad 
%\eta\left(-\frac{1}{\tau}\right) =  \sqrt{-i \tau} \, \eta(\tau) \ .
% \label{eq:theta 1 transformation}
%\ee

\end{document}